\documentclass[prd,twocolumn,showkeys,superscriptaddress]{revtex4}
\usepackage{graphicx} 
\usepackage{amssymb} 
\usepackage{amsmath}
\usepackage{epsfig}
\usepackage{slashed}
\usepackage{psfrag}

\newcommand{\one}{1\!\!1}
\newcommand{\jet}{\ensuremath{J}}
\newcommand{\pdf}{\ensuremath{F}}
\newcommand{\soft}{\ensuremath{S}}
\newcommand{\softbub}{\ensuremath{\mathcal{B}}}
\newcommand{\jetbub}{\ensuremath{\mathcal{J}}}
\newcommand{\pdfbub}{\ensuremath{\Phi}}
\newcommand{\hardbub}{\ensuremath{\mathcal{H}}}
\newcommand{\hardfact}{\ensuremath{H}}
\newcommand{\hardco}{\ensuremath{C}}
\newcommand{\tran}{\ensuremath{t}}
\newcommand\3[1]{{\bf #1}}

\begin{document}
\title{Fully Unintegrated Parton Correlation Functions 
and 
Factorization in 
Lowest Order Hard Scattering} 
\author{J.C. Collins}
\author{T.C. Rogers}
\affiliation{Department of
Physics, Pennsylvania State University,\\ University Park, PA  16802,
USA}
\author{A.M. Sta\'sto}
\affiliation{Department of
Physics, Pennsylvania State University,\\ University Park, PA  16802,
USA}
\affiliation{H. Niewodnicza\'nski Institute of Nuclear Physics, Krak\'ow, Poland}

\date{\today}

\begin{abstract}
  Motivated by the need to correct the potentially large kinematic
  errors in approximations used in the standard formulation of
  perturbative QCD, we reformulate deeply inelastic lepton-proton
  scattering in terms of gauge invariant, universal parton correlation
  functions which depend on all components of parton four-momentum.
  Currently, different hard QCD processes are described by very
  different perturbative formalisms, each relying on its own set of
  kinematical approximations.  In this paper we show how to set up
  formalism that avoids approximations on final-state momenta, and
  thus has a very general domain of applicability.  The use of exact
  kinematics introduces a number of significant conceptual shifts
  already at leading order, and tightly constrains the formalism.  We
  show how to define parton correlation functions that generalize the
  concepts of parton density, fragmentation function, and soft factor.
  After setting up a general subtraction formalism, we obtain a
  factorization theorem.  To avoid complications with Ward identities
  the full derivation is restricted to Abelian gauge theories; even so
  the resulting structure is highly suggestive of a similar treatment
  for non-Abelian gauge theories.
\end{abstract}


\maketitle

\section{Introduction}
\label{sec:intro}

The standard leading twist formalism for calculating deeply inelastic
lepton-proton scattering (DIS) cross sections provides a foundation
for understanding parton distribution functions (PDFs) and
perturbative Quantum Chromodynamics (pQCD) in general.  At issue in
the present paper are the various approximations to parton kinematics
that are fundamental to the standard approach, as we will review in
Sec.~\ref{sec:breakdown}.  These approximations are appropriate
for very inclusive cross sections, and they result in a number of
appealing conceptual and practical simplifications.  The resulting
factorization theorems involve standard (fully integrated) PDFs that
have rigorous definitions and depend only on the longitudinal
component of the parton momentum~\cite{Efremov:1980kz,Collins:1981uw}.

But when the true final states are studied in more detail, problems
arise.  Although various resummation methods are used to overcome the
problems, it has become increasingly clear
\cite{MRW1,MRW2,Collins:2005uv,CZ} that it is the standard kinematic
approximations that should be questioned.  The problems arise because
the approximations change momenta of particles in the final state,
typically resulting in a final state that does not obey 4-momentum
conservation.  For a fully inclusive cross section, this is not a
critical issue.  Furthermore, if one is only concerned 
with obtaining a
leading-logarithm approximation (improved only by the use of an
appropriate running coupling), it is legitimate to be imprecise about
the details of the kinematic approximations.

But to be able to make systematic improvements by 
including
higher order
corrections to the hard scattering, and to the showering and evolution
kernels, etc, a precise formulation of the approximations is needed.
Of course, one must expect approximations to be needed if one is to
obtain tractable factorization results.  But for the formalism to be
generally applicable, it is necessary to define the approximations in
such a way that final-state momenta are unaltered.  Attempts to remedy
the situation are now stymied by the use of conventional ``integrated
parton densities''; there is a mismatch
\cite{MRW1,MRW2,Collins:2005uv,CZ} between the definitions of
integrated parton densities and the imperatives of factorization with
correct final-state kinematics.  The mismatch extends equally to
fragmentation functions.

As explained in \cite{MRW1,MRW2,Collins:2005uv,CZ}, when we use
methods that treat final-state kinematics exactly, we are led to the
replacement of conventional parton densities by more general
quantities.  
The range of methods, from standard ones with
approximated final-state kinematics to the improved ones that are the
subject of this paper, can be characterized by the kind of parton
densities, or generalization, that are used:
\begin{itemize}
\item 
{\bf Conventional integrated parton densities:} These are the usual
PDFs; they depend only on a longitudinal momentum fraction variable,
$x$, and the hard scale, $Q^2$.  All other components of parton
momentum are integrated over.  Correspondingly, the intrinsic
external parton transverse momentum $k_\tran$
and virtuality are neglected in the hard scattering.  These parton
densities have consistent operator definitions \cite{Collins:1981uw} used in the
classic proofs of factorization.

\item {\bf Unintegrated parton densities:} These are the $k_\tran$-dependent
PDFs obtained when the integral over parton transverse momentum is
left undone, while the minus component (or virtuality) is still
integrated over.  The concept of an unintegrated parton density
appeared quite early \cite{Soper:1977xd}, necessary for the treatment
of the transverse momentum distribution in the Drell-Yan process. 
These quantities are also called transverse-momentum-dependent (TMD)
parton densities.  There are also TMD fragmentation functions.

Unintegrated parton distribution functions
also appear naturally in the high energy limit of QCD  \cite{BFKL} where 
one of the key 
features
is the lack of the transverse momentum ordering
of subsequent gluon emissions.   
Small-$x$ resummation then provides an
evolution equation in a rapidity variable for this unintegrated gluon
distribution function.
In this approach the integrated gluon density is
defined simply as an integral over the 
transverse momentum up to a hard scale.
However, the definitions of TMD densities that we refer to 
in this work may not, in general, agree with other definitions
appearing in the literature.

As for a definition, the obvious and natural one is 
given
by the hadron
expectation value of the parton number operator in light-front
quantization, equivalent to a standard simple expression as an
expectation value of a bilocal field operator, in light-front gauge.
However, this definition suffers \cite{Collins:1981uw,TMD} from what we will call
rapidity divergences, from where the rapidity of internal gluons goes
to infinity.  These divergences occur even when ultraviolet and
infra-red divergences are cut off, as reviewed in \cite{C03}.

So a correct definition requires an explicit cutoff for the rapidity
divergences.  Collins and Soper \cite{Soper:1979fq,Collins:1981uw,TMD} used a
non-light-like axial gauge for this purpose.  Certain improvements are
probably necessary, as we will see.  Other work suffers from an
imprecision in the definitions.  For example, in
\cite{Kimber:2001sc,Kimber:2000bg}, we do not see the operator
definitions at all, and the issues concerning rapidity divergences are
completely hidden, probably in some application of a leading logarithm
approximation, with cutoffs on both real and virtual lines in Feynman
graphs.
However, it is important to understand the differences in the definitions
of TMDs within different theoretical treatments.  For example, in a
recent analysis of dijet correlations at the Relativistic Heavy Ion Collider (RHIC)~\cite{Szczurek:2007bt}, 
it is shown that there is a large variation between the predictions obtained
using different TMDs from the literature. 
Therefore it is necessary to construct
a precise and unique operator  definition of the unintegrated parton distribution functions.
 Then, hopefully,  the  different cases discussed above (for example 
those 
which emerge in the context of 
 BFKL \cite{BFKL} or angular ordered  \cite{CCFM,Kimber:2000bg} evolutions)  will follow  naturally as 
 particular limits of the general definition,
or as changes of factorization scheme.

For the concept of a parton density to be applicable in the real world
of hadron physics, it needs to include all the relevant
non-perturbative phenomena concerning the actual state of the parent
hadron.
In particular,  
the need for TMD densities arises when we treat final states in enough
detail to be sensitive to the transverse momenta of partons relative
to their parent hadrons, etc.
More recent work devoted to providing precise definitions to 
the TMD densities was done in Ref.~\cite{Hautmann:2007cx}.

\item {\bf Parton Correlation Functions:} An even closer examination of
final-state kinematics, as was done for small-$x$ physics by Watt,
Martin, and Ryskin \cite{MRW1,MRW2}, indicates that ``doubly
unintegrated'' PDFs (differential in all components of parton
momentum) are more appropriate in many situations.  These can also be
called ``fully unintegrated parton densities'', but in this paper we
will simply call them ``parton correlation functions'' (PCFs).
Collins and Jung \cite{Collins:2005uv} showed very generally that for
differential distributions in final-states the use of parton
correlation functions is necessary.  Collins and Zu \cite{CZ} set up a
complete formalism in a model field theory, suitable for Monte-Carlo
event generator implementation, with the possibility of incorporating
arbitrarily non-leading-order corrections.

Under the label of parton correlation function we will include both
the fully unintegrated generalization of parton densities, and related
objects for fragmentation, and for soft factors in factorization
theorems.

\end{itemize}

The need in the general case for making no approximations on
final-state kinematics impels us to a formalism that uses PCFs.
Unfortunately, a complete treatment and derivation of factorization
using PCFs does not yet exist for QCD.  Therefore, our aim in this
paper is to initiate the construction of such a formalism.  The new
formalism should handle non-leading-order corrections as generally as
the standard formalism.  Thus it will be as good, if not better, in
cases where more traditional approaches are applicable.  But it will
also apply to more general situations.

\begin{figure*}
\centering
  \begin{tabular}{c@{\hspace*{5mm}}c}
    \includegraphics[scale=0.5]{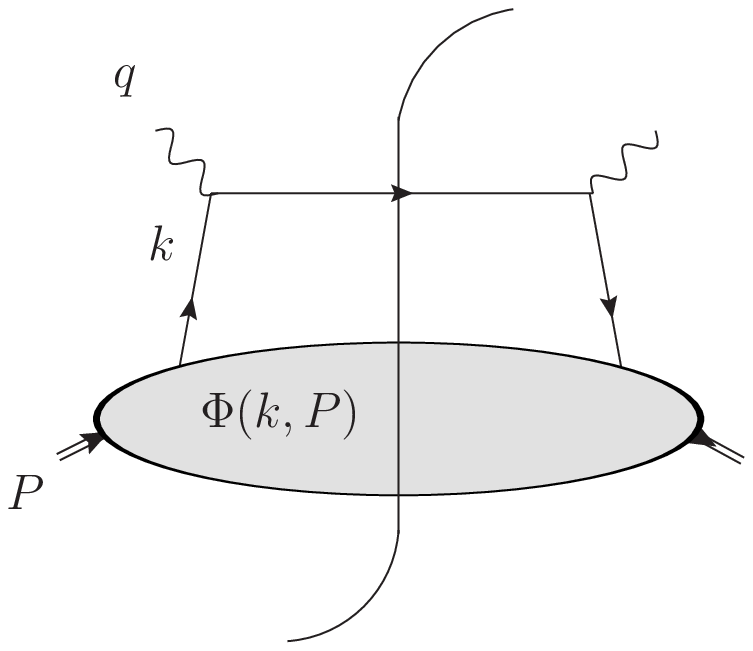}
    &
    \includegraphics[scale=0.5]{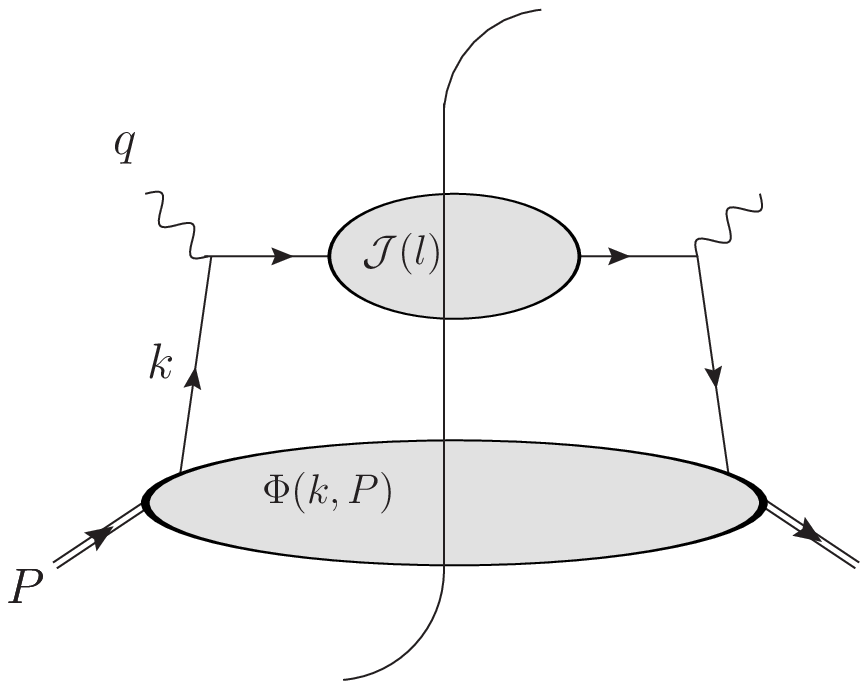}
  \\
  (a) & (b)
  \\[3mm]
  \multicolumn{2}{c}{
    \includegraphics[scale=0.5]{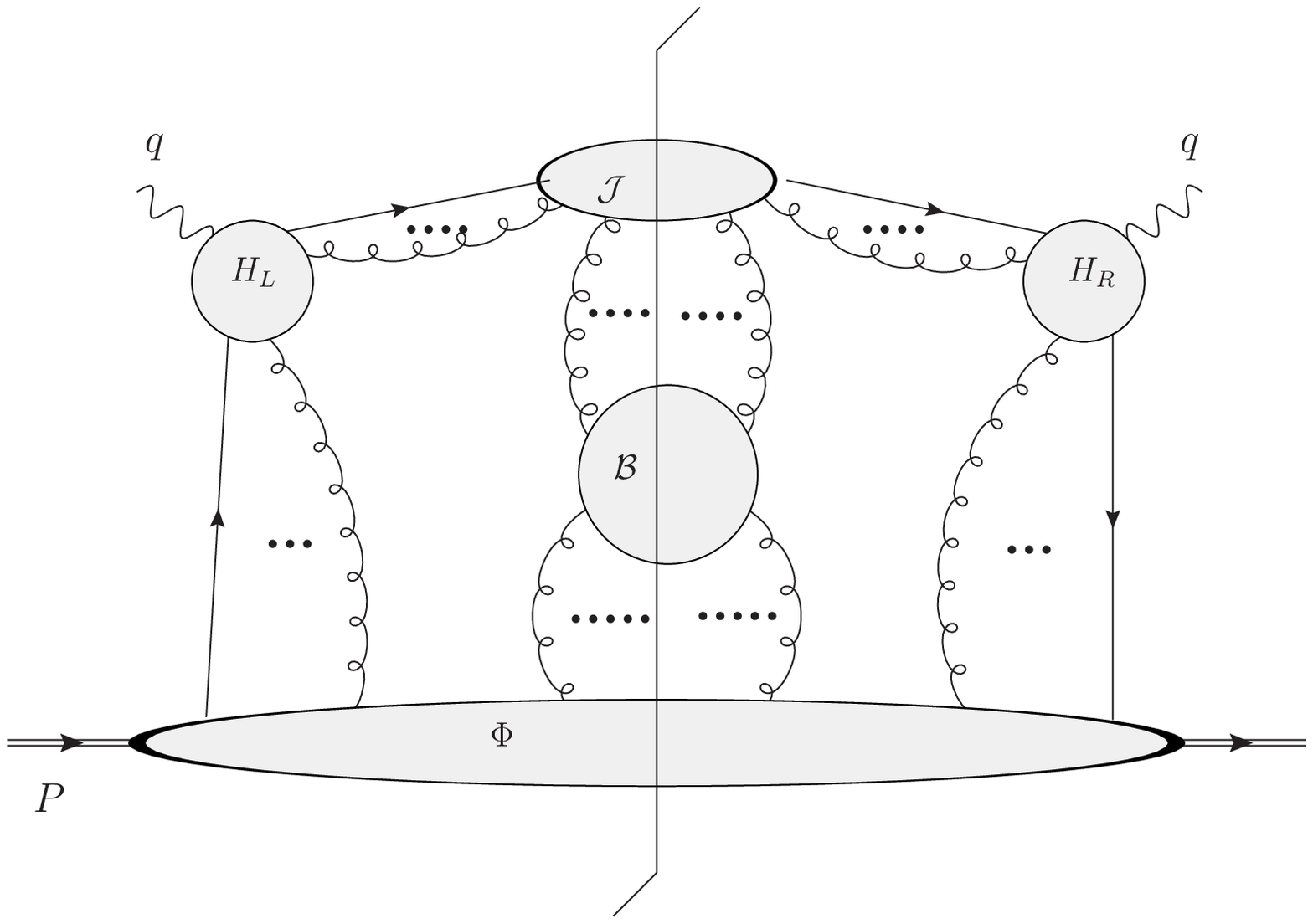}
  }
  \\
  \multicolumn{2}{c}{(c)}
  \end{tabular}
\caption{(a)  Parton model handbag diagram for DIS. 
(b) Handbag diagram with outgoing jet factor.
(c) The general reduced diagram for full QCD with one jet.
Note that in graphs like (c), the target collinear gluons may pass through 
the cut inside the target bubble.}
\label{LOdiags}
\end{figure*}

To treat kinematics correctly, we need to go back to the foundations,
and we will see that the basic structure of the derivation in fact
needs to be significantly modified \emph{even at the lowest-order,
  parton-model level} for DIS.  Normally, one starts with the handbag
diagram of Fig.~\ref{LOdiags}(a), where the outgoing struck quark is
exactly massless and on-shell.  But even without any sophisticated
treatment of QCD effects, we know that the quark must turn into a jet.
So a minimum logical foundation starts from Fig.~\ref{LOdiags}(b),
where the outgoing quark line now has acquired a jet subgraph.
In fact, examining the kinematics of such graphs gives the motivation 
\cite{MRW1,MRW2,Collins:2005uv,CZ} for searching for methods that use
PCFs. 

However, in real QCD, this is not sufficient.  There must certainly be
non-perturbative interactions at late times to neutralize the color of
the outgoing quark.  Furthermore, even in perturbation theory, and
even \emph{without going to the conventional domain for NLO
  corrections where extra high transverse momentum jets are produced},
it is necessary to consider all regions of the form symbolized by the
graphs of Fig.~\ref{LOdiags}(c).  There are arbitrarily many collinear
gluons exchanged between the jet and spectator subgraphs and the hard
scattering, and there are arbitrarily many 
soft (interacting) gluons
connecting the jet and spectator subgraphs.  We do not have
topological factorization \cite{CSS.fact} of the subgraphs.  Only
after an application of Ward identities to give a coherent sum over
different graphs does one obtain factorization for the cross section.
A consequence is 
that, to obtain a correct factorization formula, one
needs to make a  
careful selection of an appropriate gauge-invariant
definition of the parton correlation functions.  So the key issues
center around 
what such a definition is and why.

These issues are in fact present in conventional factorization, although most
treatments do not emphasize them.  The reasons why we now need to
examine them in much closer and exact detail are coupled to 
our current
aims of treating final states less inclusively.  The Ward identities
used to derive factorization apply only after the application of
certain approximations.  In the most obvious way of setting them up,
the approximations are only valid in regions of appropriately strongly
ordered kinematics.  Moreover, there is a general tendency in
constructing derivations to neglect power-suppressed corrections
whenever it assists the ease of derivation.  The Ward identity
arguments are therefore typically applied
to amplitudes where cutoffs are
imposed on the momenta of internal lines of the graphs.  But the Ward
identities are only exact when applied to amplitudes in the exact
theory, i.e., when no cutoffs whatever are applied to the internal
momenta other than, possibly, a conventional gauge invariant cutoff
for UV divergences.

There is therefore a conflict between the cutoffs used in discussing
why factorization occurs and the lack of cutoffs needed for the
validity of the Ward identities.  Once we go beyond the leading
logarithm approximation we cannot restrict the kinematics to strongly
ordered regions.  
A careful and precise formulation of approximations
and Ward identities is needed, otherwise there are likely to be
present uncontrolled correction terms that can readily violate
factorization.  
A subtraction formalism, such as we will use, avoids the problems with
direct cutoffs on the momenta of lines of Feynman graphs.
Examination of the literature --- e.g.,
\cite{CSS.fact,Nayak:2005rt} --- shows that the Ward identities for
proving factorization are typically stated and proved too inexplicitly
for our purposes.

We stress that the treatment in this paper is 
limited to lowest order in the hard scattering, where the final 
states are rather simple.  What we discuss in this paper 
may be thought of as a formal generalization of the parton model
to cases where the details of over-all kinematics are important.
This is a critical step toward a complete formalism 
because higher order calculations rely on the subtraction formalism to avoid
double counting the zeroth order contribution.  Knowing exactly 
what is subtracted requires a complete understanding of the zeroth order
contribution. 

If it turned out that factorization were universally true, these
issues would not be pressing.  However, recent work indicates
otherwise.  For example, 
it has been shown
\cite{bomhof_04,Bacchetta:2005rm,Bomhof:2006dp,Pijlman:2006tq,collins.qiu},  
rather unambiguously \cite{collins.qiu} 
that factorization fails for
production of hadrons of high transverse momentum in hadron-hadron
collisions.  This is in the case that the hadrons are close to
back-to-back azimuthally, when unintegrated parton densities are an
appropriate tool.  See also the recent paper by Bauer and Tackmann
\cite{Bauer:2007ad} for a closely related argument.

In other situations, model calculations can appear to show that
factorization fails, but the culprit is an inappropriate definition of
parton densities.  An example is the transverse single-spin-asymmetry
(SSA) in semi-inclusive deep-inelastic scattering (SIDIS), which
Brodsky, Hwang and Schmidt \cite{BHS} found to be non-factorizing.
Collins \cite{jcc.ssa} showed that factorization does actually hold,
but only if suitable Wilson line operators are used in the definition
of the unintegrated parton densities.

These motivations lead us to the detailed treatment in the present
paper.  The issue that makes the work of Collins and Zu \cite{CZ}
insufficient is the presence of extra gluonic connections between the
subgraphs of Fig.~\ref{LOdiags}(c).  Further difficulties are
associated with the masslessness of the gluon in QCD, and with the
complicated nature of Ward identities in a non-Abelian gauge theory.
It is useful to solve the difficulties one-by-one.  So the full
technical results in this paper are restricted to the case of an
Abelian gauge theory with a massive gluon.  However, much of our
treatment applies more generally and includes QCD.  Furthermore we
will obtain a factorization stated in terms of PCFs with 
precise
definitions as certain operator matrix elements.  The statement of
factorization is equally applicable to QCD.

In Sec.~\ref{sec:breakdown} we will deconstruct the standard parton
model/handbag diagram in the context of pQCD.  Then we will discuss in
Sec.~\ref{sec:requirements} what requirements are needed for a
formalism that gives a good treatment of parton kinematics.  In
Sec.~\ref{sec:basic.approx}, we discuss the projection operators which enable us to make 
 approximations  suitable for obtaining the  standard  parton model. In Sec.~\ref{sec:defns}
we list a set of candidate definitions
for the PCFs, and discuss the reasoning for our choices. Sec.~\ref{sec:sub} 
is devoted to   a subtraction procedure and in 
Sec.~\ref{sec:kin_approx} we discuss the kinematic approximations and the rapidity differences. In
Secs.~\ref{sec:real} and ~\ref{sec:virt}, we 
present calculations to support the
consistency of the structure outlined in Sec.~\ref{sec:requirements},
Finally in Sec.~\ref{sec:factorization} we present all-orders proofs
of the the factorization
formula with the PCFs, including the necessary Ward identities, but at
this point restricted to an Abelian theory.

\section{The Breakdown of Standard Kinematical Approximations}
\label{sec:breakdown}

\subsection{The parton-model approximation}
\label{sec:parton.model}

In this section we carefully analyze the derivation of the 
parton-model approximation for DIS, paying 
special attention to the
analysis of parton kinematics.

Although Fig.~\ref{LOdiags}(a), with its single on-shell final-state
struck quark, is the usual starting point, we actually need to start
with Fig.~\ref{LOdiags}(b), where the quark fragments into a group of
final-state particles.  We will show in what sense the sum over all
graphs of the form of Fig.~\ref{LOdiags}(b) is approximated by
Fig.~\ref{LOdiags}(a), i.e., we will analyze how it is possible to
neglect all higher-order corrections to the final-state bubble
$\jetbub(k+q)$.  For this part of the discussion, we will restrict our
attention to graphs without the extra gluonic attachments shown in
Fig.~\ref{LOdiags}(c).

In Fig.~\ref{LOdiags}(b), $q$ is the incoming virtual photon momentum 
and $P$ is the  momentum of the  target (proton).  As usual, $Q^2=-q^2$ is 
the photon virtuality and $x_{\rm Bj} \equiv Q^2/(2 P \cdot q)$ is 
the Bjorken scaling variable.  The incoming momenta may be expressed as
\begin{equation}
P = \left( P^+,\frac{M_p^2}{2P^+},{\bf 0}_\tran \right) \; , 
\qquad
q = \left( -xP^+,\frac{Q^2}{2xP^+},{\bf 0}_\tran \right) \; .
\end{equation}
Here, to provide a simple formula we use the Nachtmann variable $x$
instead of the Bjorken variable.
Since we want to start with \emph{no} approximation on kinematics, we
note the exact relation between the two variables:
\begin{equation}
  x = \frac{2 \, x_{\rm Bj}}{1+\sqrt{ 1 + 4\, \frac{M_p^2}{Q^2}\, 
x_{\rm Bj}^2}};
\end{equation}
they are equal up to a power-suppressed correction, as $Q\to\infty$.

The contribution to the hadronic tensor from Fig.~\ref{LOdiags}(b) is
\begin{equation}
\label{eq:f2model}
W^{\mu \nu}(q,P) \; = \; \sum_{j} \frac{e_{j}^{2}}{4 \pi} \int
\frac{d^4 k}{(2\pi)^4} {\rm Tr}\! \left[\gamma^{\mu} \jetbub(k+q) 
\gamma^{\nu} \pdfbub(k,P) \right] .
\end{equation}
Here, the sum is over quark flavors, and $e_j$ is the electric charge
of quark $j$ in units of the size of the electron's charge.  We
will leave implicit the
dependence on $j$
of the target and jet factors
$\jetbub$ and $\pdfbub$.  We will sometimes write the outgoing quark momentum as
$l \equiv k + q$.  The kinematic constraints, that the final states in
$\jetbub(k+q)$ and $\pdfbub(k,P)$ 
have positive energy and positive invariant mass,
impose limits on the values of $k$ where the integrand is
non-vanishing.

To obtain the standard LO DIS expression, we apply a number of
approximations valid at the leading power of $Q$.  First, we expand
both the upper bubble, $\jetbub(k+q)$, and the lower bubble, $\pdfbub(k,P)$, in a
basis of Dirac matrices:
\begin{equation}
\begin{aligned}
\pdfbub(k,P) & =  
    \pdfbub_S
    + \gamma^{\mu} \pdfbub_{\mu} +\sigma^{\mu\nu} \pdfbub_{\mu\nu}+\gamma_5 \pdfbub_{5}
    + \gamma^{\mu} \gamma_5\pdfbub_{5\mu} \, , 
\\
\jetbub(l)  &=
    \jetbub_S
    + \gamma^{\mu} \jetbub_{\mu} +\sigma^{\mu\nu} \jetbub_{\mu\nu}+\gamma_5 \jetbub_{5}+ \gamma^{\mu} \gamma_5\jetbub_{\mu 5} \,. 
\end{aligned}
\label{eq:clifford}  
\end{equation}
Then we observe that in the Breit frame, the trace in
Eq.~(\ref{eq:f2model}) is dominated by terms containing $\gamma^-$ in $\pdfbub$
and $\gamma^+$ in $\jetbub$, i.e., $\gamma^-\pdfbub^+$ and $\gamma^+\jetbub^-$, together with
terms only relevant for polarized scattering.  These terms dominate
because the coefficients $\pdfbub^+$ and $\jetbub^-$ are the large ones after a
boost from the rest frames of the final state of each of the bubbles.
Therefore, up to power-suppressed corrections, 
\begin{multline}
{\rm Tr}\! \left[\gamma^{\mu} \jetbub(k+q) \gamma^{\nu} \pdfbub(k,P) \right]
\\
\simeq 
{\rm Tr}\! \left[\gamma^{\mu} \gamma^+  
\gamma^{\nu} \gamma^{-} \right] \jetbub^-(k+q) \pdfbub^+(k,P) .
\end{multline}

Next, we focus attention on values of the
quark momentum where parton model kinematics are good. 
First, for each parton line, we define corresponding massless momenta:
\begin{equation}
\label{eq:parmodapprox}
\hat{k} \equiv (x_{\rm Bj} P^{+},0,{\bf 0}_\tran) 
\qquad
\hat{l} \equiv 
\left( 0,\frac{Q^{2}}{2 x_{\rm Bj} P^{+}},{\bf 0}_\tran \right).
\end{equation}
These are the parton momenta that normally appear in LO calculations. 
The parton-model region is where
the transverse and minus components of $k$ are small relative to 
$Q$, and where $k$ and $l$ both have small virtualities relative to
$Q^{2}$.  Then we can treat $\hat{k}$ and $\hat{l}$ as good
approximations to $k$ and $l$:
\begin{equation}
\label{pmapprox}
k \; \simeq \;  \hat{k}  \; ,  
\qquad
l \; \simeq \; \hat{l}  \;,
\end{equation}
with the errors being small compared with the large components of $k$
and $l$.  (For example the transverse momentum might be of order $M$,
compared with order $Q$ for the large components.)

The standard parton-model approximation is obtained by neglecting the
small momentum components, $k^-$, $k_\tran$, $l^+$, $l_\tran$, i.e., by
replacing $k$ and $l$ by $\hat{k}$ and $\hat{l}$. However, 
the replacement is only applicable in the hard scattering, where we
can neglect the small momentum components with respect to 
$Q$.  It is \emph{incorrect} to replace $k$ by $\hat{k}$ in
$\pdfbub(k,P)$ and $l$ by $\hat{l}$ in $\jetbub(l)$ 
because internal virtualities in $\pdfbub$ and $\jetbub$ may be small.  But it is
valid to replace $k^{+}$ by the fixed value $x_{\rm Bj} P^{+}$ inside
the lower bubble, and to replace $k^{-} + q^{-}$ by $q^{-}$ inside the
upper bubble.  These give only small fractional shifts in the large
components $k^+$ and $k^-+q^-$, and they give small fractional shifts
in the lines' virtualities.  Furthermore, we can perform a small
Lorentz transformation to set to zero the transverse momentum entering
$\jetbub(k+q)$.  After this we change variables for the $k$ integral to
$k^-$, ${\bf k}_\tran$ and $l^+$ to obtain a factorized approximation to
$W^{\mu \nu}(q,P)$:
\begin{widetext}
\begin{equation}
\label{eq:f2parmodapprox}
T_{\rm PM} W^{\mu \nu}(q,P) \; 
= \sum_{j} 
\frac{e_{j}^{2}}{4\pi} 
\left[ \int \frac{dk^- d^2 {\bf k}_\tran}
   {(2\pi)^4} \, \pdfbub^{+}(x_{\rm Bj} P^+,k^-,{\bf k}_\tran;P) \right]
\left[ \int dl^+ {\rm Tr}\! \left( \gamma^{\mu} \gamma^+ \gamma^{\nu} \gamma^- \right) 
\jetbub^-(l^+,q^-,{\bf 0}_\tran)\right] \;.
\end{equation}
\end{widetext}
The symbol $T_{\rm PM}$, which we call 
the ``parton model
approximator'', represents the operation of replacing the integrand in
Eq.~(\ref{eq:f2model}) by the integrand in
Eq.~(\ref{eq:f2parmodapprox}). 

At this stage, we should emphasize a distinction important for
a more detailed treatment of final states.  While the 
approximations, 
Eqs.~(\ref{pmapprox}),
are clearly good in the hard scattering calculation if $k^-$ 
and $k_\tran$ are small, the shift
in integration variables needed to get Eq.~(\ref{eq:f2parmodapprox})
introduces errors in the evaluation of $\pdfbub(k,P)$ and $\jetbub(l)$
that need to be 
examined
more carefully. 
Within the parton-model region of collinear
quark momentum, the integrand in (\ref{eq:f2parmodapprox}) is a good
numerical approximation to the original integrand,
if it is a smooth enough function.  
However, because it
involves replacing final-state momenta by somewhat different momenta,
the approximation will change certain kinds of cross 
sections differential in the final state. 
 
Even when we only treat inclusive cross sections, 
the changes in the kinematics affect the positions of
thresholds.  
Indeed, the approximated nonperturbative factors $\pdfbub^{+}(x_{\rm Bj}
P^+,k^-,{\bf k}_\tran;P)$ and $\jetbub^-(l^+,q^-,{\bf 0}_\tran)$ no longer restrict
$k$ to the actual kinematically allowed values of the original
unapproximated integral. 
Therefore, the approximations leading to
Eq.~(\ref{eq:f2parmodapprox}) can lead to unphysical results \cite{BS},
particularly if one is
interested in the details of the final state.  

For purely inclusive DIS, the usual formalism includes higher order
corrections that provide extra large-transverse-momentum jets, with
Eq.~(\ref{eq:f2parmodapprox}) corresponding to the first term in a
perturbative expansion of the hard scattering.  Higher-order terms in
the hard scattering include terms 
that can compensate for kinematic approximations that are
particularly bad at large $k_\tran$.

\subsection{Parton density and Wilson lines}
We now use the approximation Eq.\ (\ref{eq:f2parmodapprox}) 
to explain a definition of a parton density.
First, we recognize that $\pdfbub^{+}(x_{\rm Bj} P^+,k^{-},{\bf k}_\tran,P)$ may be 
written as
\begin{equation}
\pdfbub^{+}(x_{\rm Bj} P^+,k^{-},{\bf k}_\tran,P) = \frac{1}{4} {\rm Tr} \!
\left[ \gamma^{+} \pdfbub(x_{\rm Bj} P^+,k^{-},{\bf k}_\tran,P) \right],
\end{equation}
which suggests the following definition for the quark PDF
\begin{equation}
\label{oldpdf}
f_{j}(x_{\rm Bj}) \stackrel{??}{=} 
\int \frac{dk^- d^2 {\bf k}_\tran}{(2\pi)^4} {\rm Tr} \!
\left[ \frac{\gamma^{+}}{2} \pdfbub_{j}(x_{\rm Bj} P^+,k^{-},{\bf k}_\tran;P) \right].
\end{equation}
The overall numerical factor of $1/2$ is the standard convention; it
ensures that the PDF has exactly the normalization of a number
density, at least in field theories where light-front quantization is
non-problematic.

Now, the integral on the right-hand side is UV-divergent in a
renormalizable theory like QCD so, as it stands, Eq.~(\ref{oldpdf}) is
ill-defined and needs to be replaced by something else.  
The divergence comes from regions of the integral where
$k^{-}$ and $k_\tran$ are large, i.e., from values of $k$ that are far
from parton kinematics.  This is the domain where higher-order
corrections to the hard scattering are important, so it is appropriate
to modify the definition while preserving its treatment of
the parton-model region.

One possibility is to place some sort of UV
cutoff on the integral in 
Eq.~(\ref{oldpdf}) near the hard scale, 
$k^{-},k_\tran \lesssim Q$~\cite{Brodsky:2000ii}.  While physically plausible,
such a definition has problems when one tries to make it gauge
invariant --- the same problems that we will have to solve in our
improved treatment with parton correlation functions.  

The solution that is in fact used for normal QCD factorization, and
that corresponds exactly to what is done with the operator product
expansion for DIS, is to apply UV renormalization to the bilocal
operator.  With the insertion of appropriate Wilson line operators,
which give gauge-invariance, we get the usual definition \cite{Collins:1981uw}
\begin{multline}
\label{oldpdfR}
f_{j}(x_{\rm Bj},\mu) = \int \frac{dw^-}{4 \pi} 
e^{-i x_{\rm Bj} p^{+} w^-}  
\\\times
\langle p | \bar{\psi}(0,w^-,{\bf 0}_\tran) V^{\dagger}_{w}(u_{\rm J})
\gamma^{+} V^{}_{0}(u_{\rm J}) \psi(0) | p \rangle_{R}.  
\end{multline}
Here, $\psi(w)$ is the field operator for quark $j$,
and $| p \rangle$ is the proton state vector.  The subscript, $R$,
indicates that the operator is renormalized using ordinary
UV-renormalization techniques.  This definition
reproduces the basic structure of the integral in
Eq.~(\ref{oldpdf}), but renormalization removes the UV divergence with
a renormalization scale, $\mu$.  
However, a derivation of
factorization must allow for graphs with extra gluon 
exchanges, as in Fig.~\ref{LOdiags}(c).  It is known that therefore
in the operator defining the parton density, there must be inserted a
path-ordered exponential of the gluon field
along the light-like direction joining the quark and antiquark fields,
as in Eq.~(\ref{oldpdfR}).  
This also makes the definition gauge-invariant.  
Deriving an
appropriate generalization for a parton correlation function, where
the separation of the two quark fields is no longer light-like, will
be an important part of the present paper.

We will find that we need a
Wilson line that goes out to infinity from the origin along one line,
not necessarily light-like, and returns along a nearby line to a point
$w$.  So as a general notation we define $V_w(n)$ to be a Wilson line
from $w$ to infinity in the direction, $n$:
\begin{equation}
V_w(n) =  P \exp \!\left( -ig \int_0^{\infty} d\lambda \,
                    n \cdot A(w+\lambda n) \right) \; .
\end{equation}
Here, the symbol $P$ is a path-ordering operator.  

In Eq.\ (\ref{oldpdfR}), we use a light-like direction,
i.e., we replace $n$ by the vector $u_{\rm J} = (0,1,{\bf 0}_\tran)$; the
separation of the fields, $w$, is in the same direction.  Thus in the 
combination $V^{\dagger}_{w}(u_{\rm J}) V_{0}(u_{\rm J})$ the segments between $w$ and $\infty$
cancel, so that
\begin{equation}
  V^{\dagger}_{w}(u_{\rm J}) V_{0}(u_{\rm J}) 
= 
  P \exp \!\left( -ig \int_0^{w^-} d\lambda \, u_{\rm J} \cdot A(\lambda u_{\rm J}) \right) \; .
\end{equation}
Thus the Wilson line is simply along the straight line joining the
quark and antiquark fields.

If we set $\mu\sim Q$, the hard scattering can, as is well known, be
usefully calculated as a power series in $\alpha_s(Q)$, which is small
because of the asymptotic freedom of QCD.  (However, the simple
use of the perturbation expansion breaks down at small and large $x$.)

\subsection{Quark fragmentation factor}
The last factor in Eq.~(\ref{eq:f2parmodapprox}) is an integral
over a cut propagator.  In the absence of UV problems a simple unitary
argument shows that the integral over all values of $l^+$ is equal to
the value obtained by integrating just the  lowest-order term
in the cut propagator.
Equivalently, we may replace the final-state jet 
bubble as follows,
\begin{align}
\label{jetapprox}
\int dl^+ {\rm Tr}\! \left( \gamma^\mu \gamma^+ \gamma^\nu \gamma^- \right)
     \jetbub^-\big(\tilde{l}\big)
\hspace*{-1cm}&
\nonumber\\
& \mapsto
 \int dl^+ {\rm Tr}\! \left( \gamma^\mu \slashed{\tilde{l}} \gamma^\nu 
                       \gamma^-
                \right)
        2 \pi \delta_{+}\big(\tilde{l}^{2}\big)
\nonumber\\
&= 
\frac{2\pi}{\displaystyle  \hat{l}^- \hat{k}^+}
 {\rm Tr}\! \left( \gamma^{\mu} \slashed{\hat{l}} \gamma^{\nu} 
                \frac{ \slashed{\hat{k}} }{ 2 } \right).
\end{align}
Here the approximated momentum in $\jetbub$ is
$\tilde{l}=(l^+,q^-,{\bf0}_\tran)$, and we have normalized the trace so
that it corresponds to the trace in the calculation of a partonic
cross section. 
Graphically, (\ref{jetapprox}) 
corresponds to replacing
Fig.~\ref{LOdiags}(b),
where the full final-state jet is included, by
Fig.~\ref{LOdiags}(a), with the lowest order quark propagator.
It is important to recognize that two assumptions are necessary for 
this identification to be justified --- that it is valid to use the 
approximate momentum variable, $\tilde{l}$, in the upper bubble, 
and that it is valid to allow the integrals in Eq.~(\ref{eq:f2parmodapprox}) 
to be unconstrained by kinematical requirements.  
These assumptions go beyond the
use of on-shell parton kinematics in the hard scattering.

There is a delta-function that forces $\tilde{l}^2=0$ and hence
$l^+=0$.  We use $\hat{l}=(0,q^-,{\bf 0}_\tran)$ to denote the resulting
light-like momentum, and then we have
\begin{equation}
\label{eq:f2parmodapprox2}
T_{\rm PM} W^{\mu \nu}(P,q) \; 
= \; \sum_{j} 
\frac{e_{j}^{2}}{\displaystyle 8 \hat{k}^+ \hat{l}^-} f_{j}(x_{\rm Bj},\mu) 
{\rm Tr} \! \left( \gamma^{\mu} \slashed{\hat{l}} \gamma^{\nu} \slashed{\hat{k}} 
\right).
\end{equation}
Projecting out the $F_{2}(x_{\rm Bj},Q^{2})$ 
component produces the familiar expression
\begin{equation}
\label{f2pm}
F_2(x_{\rm Bj},Q^2) \; \simeq \; \sum_j e_j^2 \, x_{\rm Bj} f_j(x_{\rm Bj})\;.
\end{equation}

The diagrammatic representation of Eq.~(\ref{eq:f2parmodapprox2}) is the 
familiar handbag diagram in Fig.~\ref{LOdiags}(a).  
In fact, Fig.~\ref{LOdiags}(a) is the typical starting point for most
pedagogical  
introductions to a pQCD treatment of DIS
(e.g., \cite{esw}), although we see now that the 
justification for using Fig.~\ref{LOdiags}(a) involves a number of 
non-trivial steps.  

For our further work it is important to emphasize the distinction
between the approximation that one restricts attention to the
generalized handbag formula, 
Eq.~(\ref{eq:f2model}), 
and the set of approximations that lead from it
to the standard parton-model formula Eq.~(\ref{eq:f2parmodapprox2}).  
In writing down Eq.~(\ref{eq:f2model}), the only approximation
is to restrict to certain topologies of graph, whereas to 
reproduce Eq.~(\ref{eq:f2parmodapprox2}), we made several
very non-trivial kinematic approximations \footnote{An analysis similar to the one 
given above was given by Landshoff and
Polkinghorne~\cite{Landshoff:1971xb}  
and actually predates QCD.}.  
It is these later approximations that we will find we need to avoid.

\subsection{The Limits of Standard Kinematical Approximations}
\label{sec:limits}
The approximations at issue
change the momenta of final
state particles.  This can give problems whenever cross sections are
investigated that are differential in final-state jets, for example by
producing final states that violate energy-momentum conservation \cite{BS}.
In particular, at large $x$, the true parton kinematics are strongly
restricted, whereas 
Eq.~(\ref{eq:f2parmodapprox}) has these restrictions removed.

\begin{figure}
\centering
\epsfig{file=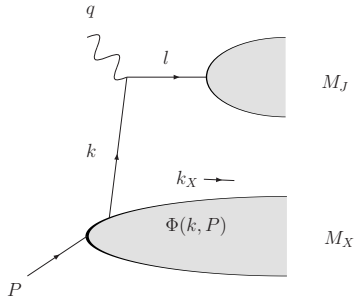,scale=0.5}
\caption{The amplitude for $\gamma^{\ast} p$ scattering into 
two jets with fixed masses.}
\label{lowx}
\end{figure}

To see this more explicitly, consider a particular final state
consisting of two outgoing jets, Fig.~\ref{lowx}, of momenta $l$ and
$k_X$.  Let the invariant masses of the struck quark jet and the
spectator jet be $M_{J}$ and $M_{X}$.  The Mandelstam variable, $s$,
is
\begin{equation}
\label{svar}
s = (1-x) M_{p}^{2} + \frac{Q^{2}}{x} (1-x).
\end{equation}  
In the center-of-mass frame, the 3-momenta of $p-k$ and $l$ are equal
and opposite, so that the internal parton transverse momentum obeys
$k_\tran^{2} = l_\tran^{2} = k_{T,X}^{2}$.  Thus 
\begin{multline}
\label{svar2}
s = (l^{0} + k_{X}^{0})^{2} 
\\
= \left( \sqrt{M_{J}^{2} + k_\tran^{2} 
+ l_{z}^{2}} + \sqrt{ M_{X}^{2} + k_\tran^{2} + k_{z,X}^{2}} \right)^{2}.
\end{multline} 
Since $M_{X}^{2},M_{J}^{2},l_{z}^{2},k_{z,X}^{2} > 0$, we have
$4 k_\tran^{2} < s$.
Thus, from Eq.~(\ref{svar}) we get a strict upper limit on the 
kinematically allowed values of $k_\tran^{2}$,
\begin{equation}
k_\tran^{2} < \frac{(1-x)}{4} M_{p}^{2} + \frac{Q^{2}}{4 x} (1-x).
\end{equation}
When $x$ is close to one, this limit is much less than $Q^2$.
This is in gross contradiction to 
Eq.~(\ref{eq:f2parmodapprox}) where the integral over 
$k$ is unrestricted.  Even if we apply renormalization at a scale $\mu\sim
Q$, this implies an effective cutoff of order $Q$, far above the
actual kinematic limit.  
However, it is not sufficient to set $\mu\sim Q\sqrt{1-x}$,
since corrections to the photon vertex still have an external
virtuality $Q^2$, for which the scale $\mu\sim Q$ is appropriate.  There is
a mismatch of scales. 
Therefore, we have an example where Eq.~(\ref{eq:f2parmodapprox}) 
is inappropriate even for a totally inclusive process.

\begin{figure}
\centering
\epsfig{file=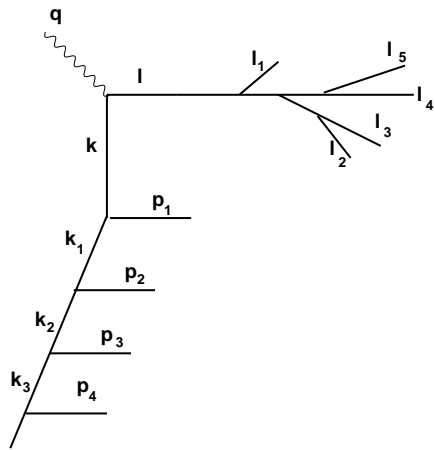,scale=0.4}
\caption{An event in which the collinear, on-shell matrix element 
--- the photon-quark vertex in this example --- is accompanied by initial 
and final state showers.
(In the showers, solid lines denote both quarks and gluons.) }
\label{fig:mceg}
\end{figure}

There are further problems with using the framework of standard
collinear factorization for observables that are differential in final 
state kinematics. 
This is illustrated in Fig.~\ref{fig:mceg}
where the photon-quark
vertex is now accompanied by initial and final state showers. 
This situation is appropriate not only for the discussion of jet cross
sections, but also for the theory of Monte-Carlo event generators (MCEGs).
Then the mass of the outgoing jet is given by
\begin{equation}
   m_J^2  =   (k+q)^2  =  2 (k^+-xP^+) \left( k^-+ \frac{Q^2}{2xP^+}
     \right)-k_\tran^2,
\end{equation}
so that
\begin{equation}
   k^+  =   xP^+ +  \frac{m_J^2+k_\tran^2}{2 (q^-+k^-)},
\end{equation}
which is strictly greater than $xP^+$.  (Note that $k^-$ is always
negative.)  This shows that away from collinear kinematics, there is a
substantial inconsistency between the value of the longitudinal
momentum used to evaluate the parton density, namely $k^+ = xP^+\simeq
x_{\rm Bj}P^+$ and the correct value of $k^+$.  Since we must allow
the transverse momentum and the parton virtualities to range up to
large values, this represents a substantial shift in $k^+$. 
The value of $k^+$ depends on both target-related and jet-related
variables, so particularly difficult problems arise in constructing a
systematic treatment of higher order corrections in a factorization
framework with conventional PDFs, as explained by Collins and Zu
\cite{CZ}.  Different numerical values for the same quantity are used
at different places in the formalism.

The important conclusion of this section is that the steps that allow one to replace
Fig.~\ref{LOdiags}(b) with Fig.~\ref{LOdiags}(a) introduce 
possibly large errors in certain types of calculation.  Since the  
kinematical approximations that allow us to replace Eq.~(\ref{eq:f2model}) 
by Eq.~(\ref{eq:f2parmodapprox}) are what normally allow us to 
replace the final state jet bubble in Fig.~\ref{LOdiags}(b) with the on-shell 
massless parton in Fig.~\ref{LOdiags}(a), then a unified treatment
must improve the approximations to avoid changing momenta in the final
state.

\section{What is Needed}
\label{sec:requirements}

In the last section, we argued that some of the
approximations that lead to the parton-model formula need to be
avoided, because they produce large 
kinematical errors that affect the measured final state.  
Furthermore, from the analysis of Libby and
Sterman~\cite{Libby:1978bx}, we know that 
for QCD the correct starting point is the sum of regions represented 
by Fig.~\ref{LOdiags}(c), and not just graphs (a) or even (b).  
In order to get a factorization formula and 
be able to perform
perturbative calculations, we need to 
rewrite Fig.~\ref{LOdiags}(c) in a useful approximation 
as the product of a hard part that 
can be calculated directly with ordinary Feynman graphs for on-shell 
external partons, and a collection of universal parton correlation 
functions to describe the non-perturbative physics.  Therefore, what 
is needed is a set of approximations and Ward identities
that reduce Fig.~\ref{LOdiags}(c) 
to a factorized form, but without the problems with 
kinematics that we have just discussed.

Each region of the form shown in Fig.~\ref{LOdiags}(c) has a set of
lines collinear to the target, subgraph $\pdfbub$, a set collinear to the
outgoing quark, subgraph $\jetbub$, a set of soft lines, $\softbub$, and two hard
subgraphs $\hardbub_L$ and $\hardbub_R$ (on the left and right of the final-state cut).  
(We use the symbol $\softbub$ for the soft bubble in a general graph.
The symbol $\soft$ is reserved for the soft PCF whose definition will 
arise when we discuss factorization.)
For each region, we will define
an approximator, in the same spirit as Eq.~(\ref{eq:f2parmodapprox}).
Our aim is to find definitions of approximators that simplify as much
as
possible the systematic application of Ward identities without
uncontrolled remainder terms.

\subsection{Requirements on approximators}

On the basis of the observations in the previous sections, we propose that the
approximators should obey the following: 
\begin{enumerate}

\item  The kinematics of the initial and final states 
must be kept exact.  Otherwise, large errors occur in certain types of
calculation.

\item The bubbles representing the sums over physical final states
  must be kept explicit.
  \\
  For example, we must take Fig.~\ref{LOdiags}(b) rather than
  Fig.~\ref{LOdiags}(a) as the starting point of the derivation of the
  parton model.  It can be argued that the integral over final-state
  bubbles such as $\jetbub$ is unity, as in Eq.~(\ref{jetapprox}).  But this
  involves a cancellation between final states of different invariant
  masses, and this violates the first requirement.

\item To avoid making kinematical approximations in the initial and
  final states, the non-perturbative factors need to be functions of
  \emph{all} components of parton four-momentum.
  \\
  Hence all of the non-perturbative factors are \emph{fully}
  unintegrated factors, rather than standard PDFs, i.e., they are
  PCFs.  In addition to the fully unintegrated PDF, we need to define
  a fully unintegrated soft factor and a fully unintegrated jet
  factor.

\item  The hard scattering matrix element should appear as an on-shell parton 
matrix element in the final factorization formula.  
\begin{itemize}
\item Setting on-shell the external partons of a hard-scattering
  subgraph involves no shift of the momenta of observable lines.  Thus
  it is a safe choice, since it merely involves changing the numerical
  value of the integrand.  
\item The use of on-shell and massless matrix elements allows the use
  of already existing Feynman graph calculations.  The only changes
  from the usual case concern the subtraction terms to remove the
  double counting of collinear and soft contributions.
\item This is the primary place where explicit higher-order
  calculations of Feynman graphs are actually used.  These
  calculations are much easier when on-shell and massless.
\item It is much easier to maintain gauge-invariance in on-shell
  amplitudes than in off-shell amplitudes.  For the PCFs, we make
  gauge invariant quantities with the aid of Wilson line factors in
  their definitions as matrix elements of operators.  But this is much
  harder to do in the hard-scattering coefficients unless their
  external parton lines are on-shell.
\end{itemize}

\item We must be able to apply
Ward identities exactly to the approximated graphs, in order to
convert the gluon exchanges 
in 
Fig.~\ref{LOdiags}(c) to a factorized form.
\begin{itemize}
\item Any approximation on momenta inside the hard scattering
  matrix element should be consistent with the use of Ward identities.
  In the process of factorizing soft and collinear gluons, it will be
  important to identify contributions to the PCFs.  The resulting
  constraints on the organization of the approximations will lead to
  corresponding constraints on the definitions of the PCFs.
\item It is easy to get a situation where Ward identities are
  applicable only with remainder terms that are of nonleading power in
  what we will term the core region of an approximator.  These are
  typically of leading power when the integrations are extended, as is
  always necessary, to a full range of kinematics.  As far as
  possible, therefore, the approximators should be arranged so that the remainder
  terms are \emph{exactly} zero.  Otherwise, explicit treatment of the
  remainder terms is needed to get factorization beyond a leading
  logarithm approximation. 
\end{itemize}

\item  Each approximator should give a good approximation in a particular 
region of momentum space, but,
for the purpose of proving factorization, should be well-defined for all momenta
for which it is used.  
\begin{itemize}
\item In the context of a systematic subtraction scheme 
  --- Sec.\ \ref{sec:sub} ---
  applied to all
  orders, an approximator $T_R$ for a region $R$ is used to provide a
  good approximation to a graph in region $R$, with errors suppressed
  by a power of $\Lambda/Q$ in the \emph{core of the region}.  
  ($\Lambda$ is a characteristic hadronic mass scale.)  
\item For a larger region $R_1$, we will apply its approximator
  $T_{R_1}$ to the graph with contributions from smaller regions
  subtracted, to compensate for double counting.
\item In order for this procedure to work, we need to take as the
  contribution of the region $R$ the integral of its approximation up
  to where the error becomes of order 100\%.  
\item Thus it must be possible to extend the formula for $T_R\Gamma$ beyond
  the core of the region $R$. 
  Therefore, its definition cannot assume
  the momenta are in the core of the region.
\end{itemize}

\end{enumerate}

In the next section, we will address 
the first four points
by returning to Fig.~\ref{LOdiags}(b) and demonstrating how the hard
scattering part of the graph may be approximated without violating
initial and final state kinematics.  To address the last three items,
we will discuss the definitions of the PCFs.  There we give a set of
candidate definitions for the PCFs.  Having completed this, we will be
in a position to approximate Fig.~\ref{LOdiags}(c) to produce a
factorization formula in terms of the lowest order on-shell-parton
hard scattering amplitude and the PCFs of Sec.~\ref{sec:defns}.

\subsection{Collinear Gluons in the Standard, Integrated Treatment}
\label{sec:std.coll}

\begin{figure}
\centering
\epsfig{file=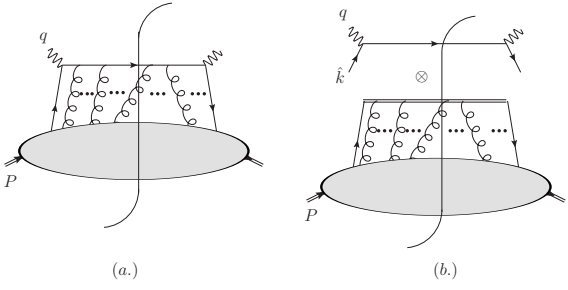,scale=0.45}
\caption{
Target collinear gluons explicit in the definition 
of the standard integrated PDF.}
\label{lessnaive}
\end{figure}

To clarify the strategy for dealing with soft and collinear gluons, it
is useful to recall relevant steps for obtaining the appropriate
gauge-invariant definition of the fully integrated PDF,
Eq.~(\ref{oldpdfR}), with LO factorization in the standard formalism.
Relevant graphs beyond the standard parton model graph
Fig.~\ref{LOdiags}(a) are those of the form Fig.~\ref{lessnaive}(a),
where we extend the handbag diagram
to allow for an arbitrary number of gluon
exchanges connecting the lower bubble with the outgoing quark.  
We
restrict the extra gluons to be collinear to the target.  To leading
power, the gluons are longitudinally polarized.  We thus have a
special case of Fig.~\ref{LOdiags}(c).  By the use of a simple Ward
identity, it is seen that the gluon attachments eikonalize, and may
be converted into a Wilson line factor, as illustrated in
Fig.~\ref{lessnaive}(b).  This leaves the convolution product of the
on-shell LO parton scattering amplitude with the gauge-invariant PDF
given in Eq.~(\ref{oldpdfR}).  In the light-cone gauge, $A^{+} = 0$,
the Wilson line operator is unity and we exactly reproduce the most
naive graph for lowest order hard scattering, Fig.~\ref{LOdiags}(a),
and graphs with extra collinear gluon exchanges are power suppressed.

As we will see, a number of complications arise in extending these
ideas to deal with more general cases.  Our aim is to make a precise,
general-purpose formalization suitable for the generalization.

\subsection{Soft and Collinear Gluons in the Generalized Formalism}
\label{sec:new.coll.soft}

Since we do not make any kinematical approximations on initial- or
final-state momentum variables, our generalization of the handbag
diagram will contain all final state bubbles.  In addition to the
initial-state-collinear gluons, we also need to show that
final-state-collinear and soft gluons also factorize into
appropriately defined gauge-invariant PCFs.  We will find that after
appropriate approximations, we can apply Ward identities that
disentangle the coupled subgraphs in Fig.~\ref{LOdiags}(c) to give the
factorized form shown diagrammatically
in Fig.~\ref{fig:WI.appl}(c) below, up 
to power suppressed corrections.  After some further
manipulations to compensate for double counting, we find that the
cross section, $\sigma$ (or a related object like a structure function), is a
convolution product of a hard factor, a PDF, a final state jet factor,
and a soft factor:
\begin{equation}
\label{factorization}
\sigma = \hardco \otimes \pdf \otimes \jet \otimes \soft 
+ \mathcal{O} \left( \left( \frac{\Lambda}{Q} \right)^{a} |\sigma | \right).
\end{equation}
Equation~(\ref{factorization}) establishes our notation 
for the PCFs --- $\pdf$ for the target PCF, $\jet$ for the jet PCF, 
and $\soft$ for the soft PCF.  The notation should be carefully distinguished 
from the notation ($\pdfbub$, $\jetbub$, and $\softbub$) that 
we have used so far in discussing the subgraph bubbles in a particular \emph{graph}.  
The last term in Eq.~(\ref{factorization}) indicates that errors should be 
suppressed by a power of $\Lambda/Q$ where
$\Lambda$ is a typical hadronic mass scale and the power is $a > 0$.

\section{The Basic Approximation}
\label{sec:basic.approx}

In this section we reexamine and reformulate the parton model
approximation, as appropriate for Fig.\ \ref{LOdiags}(b), in a form
suitable for our later work.  We arrange to use PCFs rather than
regular parton densities.  As regards parton kinematics, a suitable
definition was given by Collins and Zu \cite{CZ}.  We now extend this
to convenient projections onto appropriate two-dimensional spaces for
on-shell massless Dirac spinors.  Although the calculations are quite
elementary for tree graphs, a precise formal definition with a
convenient graphical notation will greatly assist later work with
higher order graphs.

Since we normally work with scalar structure functions, $F_1$ etc, we
define a projection tensor $P_{\mu\nu}$ for any chosen structure function.
Thus, the projection onto $F_1$ is done by
\begin{equation}
  \label{eq:F1.proj}
  P_{\mu\nu} 
  = \frac12 
         \left[
              -g_{\mu\nu}
              + \frac{ Q^2 P_\mu P_\nu }
                     { (P\cdot q)^2 +M_{p}^2 Q^2 }
         \right],
\end{equation}
so that $F_1= P_{\mu\nu}W^{\mu\nu}(q,P)$.

We now apply such a projection to Eq.~(\ref{eq:f2model}):
\begin{equation}
\begin{split}
\label{pm2}
\Gamma
= \frac{P_{\mu \nu}}{4\pi} 
          \int \frac{d^{4} k}{(2 \pi)^{4}}  
          e_{j}^{2} {\rm Tr}\bigg[\jetbub(l) \gamma^{\mu} \pdfbub(k,P) \gamma^{\nu} \bigg]. 
\end{split}
\end{equation}
{}From here on, a sum over quark flavors, $j$, is implicit.

The first step is to replace exact parton momentum variables with
approximated parton momentum variables (indicated with a hat)
\emph{inside the hard matrix element}:
\begin{equation}
\label{pmapp1}
k  \mapsto  \hat{k},\qquad
l  \mapsto  \hat{l}.
\end{equation}
These are the same as in our treatment of the conventional parton
model; but now we no longer use the other kinematic approximations in
the $\jetbub$ and $\pdfbub$ factors.  The approximated momenta, defined in Eq.\
(\ref{eq:parmodapprox}), form a particular case of the prescription in
Ref.~\cite{CZ}.  They are uniquely determined by the following
requirements.  First, the approximated momenta, $\hat{k}$ and
$\hat{l}$, describe a collinear struck parton and an on-shell final
state parton,
\begin{equation}
\label{onshell}
| {\bf \hat{k}}_\tran | =  \hat{k}^{-} = 0, \qquad
\hat{l}^{2} =  0.
\end{equation}
Then we require four-momentum conservation for both the exact and
approximated variables:
\begin{equation}
\label{momcons}
k + q =  l, \qquad
\hat{k} + q  =  \hat{l}.  
\end{equation}

Next we formalize the projection onto the leading power terms in the
trace over Dirac matrices by defining projection matrices ${\cal P}_{\rm T}$
and ${\cal P}_{\rm J}$:
\begin{equation}
{\cal P}_{\rm T} = \frac{1}{2} \gamma^- \gamma^+ \; ,  
\qquad
{\cal P}_{\rm J} = \frac{1}{2} \gamma^+ \gamma^- \; , 
\end{equation}
where
\begin{equation}
\gamma^-=\frac{1}{\sqrt{2}}(\gamma^0 -\gamma^3),\; \;\; \; \;\gamma^+=\frac{1}{\sqrt{2}}(\gamma^0+\gamma^3) \; .
\end{equation}
As elsewhere, we use the subscript T to denote target-related
quantities, and J to denote jet-related quantities.
Some properties of the projection matrices are listed in App.\
\ref{sec:projection}. 
It can be readily checked that ${\cal P}_{\rm T}$ projects a general
4-dimensional spinor onto the 2-dimensional subspace of spinors $u$
that obey the massless Dirac equation for momentum $\hat{k}$, i.e.,
$\hat{\slashed{k}} u = 0$.  Similarly, ${\cal P}_{\rm J}$ projects onto
spinors for momentum $\hat{l}$. 

We next use the  decompositions (\ref{eq:clifford}) of $\jetbub(l)$ and $\pdfbub(k,P)$ in a basis of Dirac matrices.
The terms that appear at the leading power in the trace can be
projected out by sandwiching $\pdfbub$ and $\jetbub$ between projection matrices:
\begin{align}
  {\cal P}_{\rm T} \, \pdfbub(k,P) \, \overline{{\cal P}_{\rm T}}
  & = {\cal P}_{\rm T} \, \pdfbub(k,P) \, {\cal P}_{\rm J}
\nonumber\\
  & =  \gamma^- \pdfbub^+ - \sigma^{-j} \pdfbub^{+j} + \gamma^- \gamma_5 \pdfbub_5^+ \, , 
\\
  {\cal P}_{\rm J} \, \jetbub(l) \, \overline{{\cal P}_{\rm J}}
  & = {\cal P}_{\rm J} \, \jetbub(l) \, {\cal P}_{\rm T} 
\nonumber\\
  &=
  \gamma^+ \jetbub^- - \sigma^{+j} \jetbub^{-j} + \gamma^+ \gamma_5 \jetbub_5^- \,. 
\end{align}
Using them, we replace Eq.~(\ref{pm2}) with its parton-model approximation
\begin{multline}
T_{\rm PM} \Gamma
=  \frac{P_{\mu \nu}}{4\pi} 
\int \frac{d k^{+} d k^{-} d^{2} {\bf k}_\tran}{(2 \pi)^{4}} \, 
\\\times 
e_{j}^{2} {\rm Tr}\bigg[\jetbub(l) {\cal P}_{\rm T} \, \gamma^{\mu} \,  
   {\cal P}_{\rm T}  \pdfbub(k,P) {\cal P}_{\rm J} \, \gamma^{\nu} \,  {\cal P}_{\rm J} \bigg], 
\label{pm3}
\end{multline}
which is changed from its previous definition.
The errors incurred by making this substitution are power
suppressed. 
We now restrict to the case of unpolarized scattering,
for simplicity, in which case only $\pdfbub^+$ and $\jetbub^-$ appear, so that
Eq.~(\ref{pm3}) can be rewritten as
\begin{multline}
T_{\rm PM} \Gamma[W] =
 \frac{P_{\mu \nu}}{4\pi}  \int \frac{d k^{+} d k^{-} d^{2} {\bf k}_\tran}
                          {(2\pi)^{4}}  
\\\times 
e_{j}^{2} {\rm Tr}\bigg[\gamma^{+} \jetbub^{-}(l) \, \gamma^{\mu} \,  \gamma^{-} \pdfbub^+(k,P)
\, \gamma^{\nu} \,  \bigg]. 
\label{pm4}
\end{multline}
Dividing and multiplying by $\hat{k}^+ = x_{\rm Bj} P^{+}$ and
$\hat{l}^- = q^{-}$, we obtain
\begin{multline}
T_{\rm PM} \Gamma
=
 \frac{P_{\mu \nu}}{4\pi} \int \frac{d^{4} k}{(2 \pi)^{4} \hat{k}^{+} \hat{l}^{-}}
\\\times
e_{j}^{2} {\rm Tr}\bigg[\hat{\slashed{l}} \, \gamma^{\mu} \,  \hat{\slashed{k}} \, \gamma^{\nu} \,  \bigg] \pdfbub^+(k,P)  
\jetbub^{-}(l). 
\label{pm5}
\end{multline}
This gives a good approximation to $\Gamma$ so long as 
the integral is dominated by the region where $k^{-},| {\bf k}_\tran | \ll k^{+}$.  
The lowest order hard matrix element squared
is immediately identifiable, and
we define it as
\begin{equation}
\label{eq:H0}
\big| \hardfact_0(q,\hat{k},\hat{l}) \big|^2 \equiv \; 
\frac{1}{2}e_{j}^2 \, P_{\mu \nu} 
  {\rm Tr}\!\left[ \slashed{\hat{l}} \, \gamma^{\mu} \,
                   \slashed{\hat{k}}   \, \gamma^{\nu} 
            \right]\;.
\end{equation}
The factor $1/2$ ensures that this is normalized just like the Born
graph for scattering on a spin-averaged massless quark.  This
definition then entails a factor of $1/ \hat{l}^- \hat{k}^+$ outside the hard scattering amplitude.
Thus, the hard scattering is evaluated with the on-shell parton
amplitude, while the momentum used to evaluate the PDF and the jet
factor remain exact.  We write the approximation as
\begin{multline}
T_{\rm PM} \Gamma
= 
  \frac{1}{2\pi} 
 \int \frac{d^{4} k}{(2 \pi)^{4} \hat{k}^{+} \hat{l}^{-}}
\\\times 
 \big| \hardfact_0(q,\hat{k},\hat{l}) \big|^2 \, \pdfbub^+(k,P)  \jetbub^{-}(l).
\label{pm6}
\end{multline}
In the hard matrix element in Eq.~(\ref{pm6}), the hatted approximate
variables should be regarded as functions of the exact variables.  All
the nonperturbative objects (the PCFs) are evaluated with unapproximated 
momentum variables.  Thus the only kinematical approximation is in the
evaluation of the hard matrix element.  

Finally, notice that at this order there is 
no ultraviolet divergence corresponding to $| k^{2} | \to \infty$ 
because of the kinematic constraints and positive energy condition on the 
final state.  This is to be contrasted with the situation in Eq.~(\ref{eq:f2parmodapprox}).

\begin{figure}
\centering
\epsfig{file=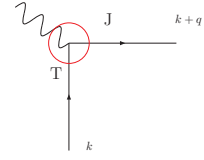,scale=0.45}
\caption{A schematic representation of the approximation performed in
  Eq.~(\ref{pm6}).} 
\label{fig:gamvertex_pro2}
\end{figure}

\begin{figure}
\centering
\epsfig{file=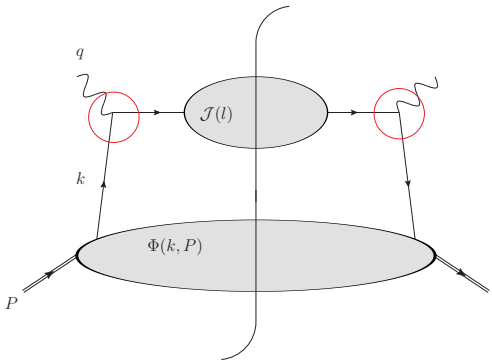,scale=0.45}
\caption{The approximation of Eq.~(\ref{pm6}).}
\label{fig:handapprox}
\end{figure}

The symbol, $T_{\rm PM}$, is the ``approximator'' which acts on
Eq.~(\ref{pm2}) to produce the approximation in Eq.~(\ref{pm6}).
Graphically, we depict the operation of $T_{\rm PM}$ by a circle
around the electromagnetic vertex, as in
Fig.~\ref{fig:gamvertex_pro2}.  It is defined as follows:
\begin{itemize}
\item Everything outside the circle is left unapproximated.  This is
  the essential change from the standard parton model.  
\item The 
  label T next to the circle symbolizes where approximations
  appropriate to
  the initial-state quark are made:
  \begin{itemize}
  \item Parton momentum $k$ is replaced \emph{inside} the circle by
    $\hat{k}$, thus projecting it onto the plus direction.
  \item The projection matrix ${\cal P}_{\rm T}$ is applied in Dirac spinor
    space.  
  \item 
    In the complex conjugate amplitude, to the right of
    the final-state cut, the projection matrix is $\overline{{\cal
        P}_{\rm T}} = {\cal P}_{\rm J}$.
  \end{itemize}

\item Similarly, the 
  label J symbolizes where
  approximations appropriate to outgoing quark are made:
  \begin{itemize}
  \item Parton momentum $l$ is replaced \emph{inside} the circle by
    $\hat{l}$, thus projecting it onto the minus direction.
  \item The projection matrix ${\cal P}_{\rm J}$ is applied in Dirac spinor
    space.  
  \item
    In the complex conjugate amplitude, to the right of
    the final-state cut, the projection matrix is $\overline{{\cal
        P}_{\rm J}} = {\cal P}_{\rm T}$.
  \end{itemize}

\end{itemize}
The approximations that lead to the factorized form in Eq.~(\ref{pm6})
are shown diagrammatically in Fig.~\ref{fig:handapprox}.  In contrast
to the standard formalism for DIS, the momenta used to evaluate
$\pdfbub^+(k,P)$ and $\jetbub^{-}(l)$ in Eq.~(\ref{pm6}) are exact.
Furthermore, \emph{the integrals over $k$ are constrained by the
  positive energy condition and relativistic kinematics,
in contrast to Eq.~(\ref{eq:f2parmodapprox}).}

\section{Defining Parton Correlation Functions}
\label{sec:defns}

A major complication in developing a generalized treatment of DIS is in
the difficulty of giving appropriate and consistent gauge invariant
definitions for the PCFs.  In the absence of gauge fields, the
definition of the PCFs is clear \cite{CZ}: they are just the obvious
matrix elements of quark fields, Fourier transformed into momentum
space.  

In a gauge theory, there are two sources of complication that are
intimately tied to each other: (a) There are important leading
contributions from regions like Fig.\ \ref{LOdiags}(c), where gluons
connect subgraphs that correspond to very different ranges of
rapidity.  (b) The operators in the definition of the PCFs must be
made gauge-invariant.  
As we will see in the later sections, the gluon
connections can be converted to a factorized form by applying certain
approximations, after which Ward identities are used to show that the
sum over the gluon exchanges corresponds to contributions associated
with Wilson lines inserted in the partonic operators in the
definitions of the PCFs and PDFs.  

The Wilson lines give gauge-invariant definitions, and there is, a
priori, a choice in the path used in the integral over the gluon field
in the Wilson lines.  However, only certain directions are suitable,
i.e., consistent with factorization.  Evidently, a complete discussion
can only be made within the context of a treatment of factorization of
soft and collinear gluons as will be described below in
Sec.~\ref{sec:real}.  But given that a definition has been
found or proposed, its properties can be discussed separately from the
motivation.

The characteristic difficulty is that the most obvious definitions
have divergences.  This applies not just to the PCFs but to PDFs as
well.  There are three basic sources of divergence:
\begin{enumerate}
\item {\bf UV divergences due to integration to infinite transverse
  momentum:}  These appear in graphs for ordinary integrated PDFs in
  all renormalizable theories.
  They also appear in virtual graphs contributing to unintegrated PDFs
  and PCFs.  In all cases they can be removed by renormalization
  counterterms, 
  beyond those needed to renormalize the interactions of the theory.

\item {\bf Divergences due to the masslessness of the gluon:}  These appear
  in Feynman graphs, but presumably are cut off by confinement effects
  in real QCD when the PCFs and PDFs are treated non-perturbatively in
  hadronic targets.  

  At this
  stage of our work, we will use a model Abelian gauge
  theory with a nonzero gluon mass, so that we separate the mass
  divergences from other issues. Given that color is confined in QCD,
  we can expect a real physical cutoff of these divergences, which
  however is difficult to discuss within pure perturbation theory.

\item {\bf Divergences due to the use of light-like Wilson lines:} We
  will call these rapidity divergences.  In simple Feynman-graph
  calculations, rapidity divergences are frequently confused with the
  divergences due to the masslessness of the gluon, since both arise
  from regions where the plus 
  or minus momentum of a gluon goes to zero.
\end{enumerate}

We will discuss the issues starting with fully integrated PDFs.  Then
we will examine the intermediate case of unintegrated PDFs, which are
differential in $k^+/P^+$ and $k_\tran$.  Finally we will examine the case
of the fully unintegrated functions, the PCFs, of all the three types
we will need: target related (like an ordinary PDF), jet related, and
soft factor.

\subsection{Integrated PDFs}

For the fully integrated PDFs, the definition \cite{Collins:1981uw} given in Eq.\
(\ref{oldpdfR}) is entirely satisfactory.  The primary parton fields
have a light-like separation in the minus direction
so that the Wilson line
can be taken along the line joining the parton fields,
as is needed to be consistent with factorization.  UV divergences are canceled by
renormalization, and there is a cancellation of all the 
rapidity divergences and of the final-state divergences associated with
masslessness of the gluon.  In a color singlet hadron with confinement
there should be no other divergences.

\subsection{Unintegrated PDFs}

\begin{figure*}
  \centering
  \psfrag{collA}{coll.\ targ.}
  \psfrag{collB}{coll.\ jet}
  \psfrag{soft}{soft}
  \psfrag{y_p}{$y_p$}
  \psfrag{y_Q}{$y_Q$}
  \psfrag{y_J}{$y_J$}
  \psfrag{infty}{$-\infty$}
  \begin{tabular}{c@{\hspace*{2cm}}c}
     \includegraphics{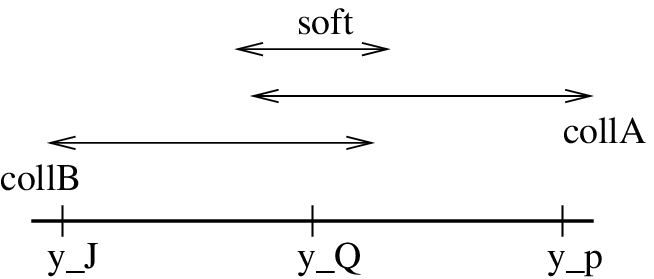}
     &\includegraphics{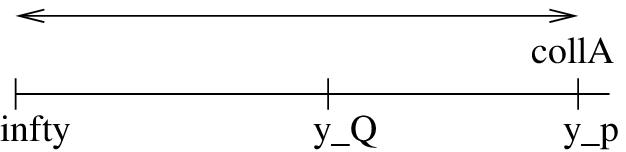}
  \\
        (a) & (b)      
    \end{tabular}
  \caption{(a) Range of rapidity for exchanged gluons in DIS.  Rightmost is the
    target region, and leftmost is the outgoing jet from the struck
    quark.
    (b) Rapidity for gluons in PDF 
    defined with light-like Wilson
    line (or for PDF defined in light-cone gauge without Wilson line).}
  \label{fig:rap.range}
\end{figure*}

As for a simple unintegrated PDF, a common approach is to define it as
a matrix element of parton fields without a Wilson line:
\begin{multline}
\label{pdf.lcg}
P(x,{\bf k}_\tran,\mu)  \stackrel{??}{=} 
    \int \frac{dw^- d{\bf w}_\tran}
           {16 \pi^{3}}
      e^{-i x p^{+} w^- + i {\bf k}_\tran \cdot {\bf w}_\tran} 
\\\times 
\langle p | \bar{\psi}(0,w^-,{\bf w}_\tran)
\gamma^{+} \psi(0) | p \rangle_{\rm lcg},
\end{multline}
but with the fields defined to be in the light-cone gauge $A^+=0$.  This
definition is natural, because if the rules of light-cone quantization
are applied naively, the definition becomes exactly the number density
of quarks.  However, as Collins and Soper \cite{Collins:1981uw} demonstrated,
this definition has a rapidity divergence
in the interacting theory.  

The rapidity divergence, from gluons whose rapidity goes to $-\infty$,
occurs 
\emph{even when all ultraviolet and mass divergences are cutoff}.
As illustrated in Fig.\ \ref{fig:rap.range}, the important range of
gluon rapidity for the actual cross section is between the rapidity of
the target and the rapidity of the jet.  Naturally gluons to the right
of the rapidity of the virtual photon can be considered as associated
with the PDF, and gluons to the left as associated with the jet's
fragmentation function, while gluons in the center belong to a soft
factor, in the Collins-Soper-Sterman (CSS)
factorization method \cite{TMD}.  But the use of
light-cone gauge gives contributions to the PDF from gluons not merely from the
positive rapidities that are naturally part of the PDF, but also from
rapidities running
all the way to $-\infty$.  Evidently, the parton density needs to be
redefined so that it has some kind of rapidity cutoff at around the
photon's rapidity.  This can be  accomplished by
Soper's \cite{Soper:1977xd} definition with a non-light-like
gauge-fixing vector $n$ of about the same rapidity as the virtual
photon.
Collins and Soper (CS) \cite{Collins:1981uw} derived
an equation for the derivative with respect to the direction of $n$; 
they used this equation to show very generally the existence of a
rapidity divergence in the light-cone-gauge limit $n^2/(n\cdot p)^2\to0$.

The definition in
Eq.~(\ref{pdf.lcg}) is readily converted to a gauge-invariant 
form by using the
fact that a Wilson line operator $V_w(n)$ in direction $n$ is unity in
the $n\cdot A=0$ gauge.  But a further complication is uncovered when the
definition is made exactly invariant:
\begin{multline}
\label{pdf1}
  P(x,{\bf k}_\tran,\mu)  
  = \int \frac{dw^- d{\bf w}_\tran}{16 \pi^{3}}
   e^{-i x p^{+} w^- + i {\bf k}_\tran \cdot {\bf w}_\tran}
\\\times 
 \langle p | \bar{\psi}(0,w^-,{\bf w}_\tran) V^{\dagger}_{w}(n) I_{n;w,0} 
     \gamma^{+} V_{0} (n) \psi(0) | p \rangle.
\end{multline}
Here Wilson lines go out from the parton fields to infinity in the
direction $n$.  With only these, we would get exact agreement with
Eq.\ (\ref{pdf.lcg}) in $n\cdot A=0$ gauge.  But as was pointed out by Belitsky et
al.\ \cite{Belitsky:2002sm}, strict gauge invariance requires that the
Wilson line be completed in the transverse direction by a segment
connecting the two points at infinity.  
This is accomplished in Eq.\ (\ref{pdf1}) by a factor of the following
operator
\begin{equation}
\label{pdfinf}
I_{n,w,0} = P \exp \!\left( -ig \int_C dz^{\mu} A_{\mu}(z) \right) \; ,
\end{equation}
where the contour $C$ is in the transverse direction and connects
$(0,\infty,{\bf w}_\tran)$ to $(0,\infty,{\bf 0}_\tran)$.  

Belitsky et al.\ \cite{Belitsky:2002sm} showed how this gauge link at
infinity is essential to get correct physics even in $n\cdot A=0$ gauge;
the naive definition Eq.\ (\ref{pdf.lcg}) is wrong.  
Their demonstration of the failure of Eq.~(\ref{pdf.lcg}) was done in
the context of one-gluon-exchange calculations of the Sivers function;
this is the single-spin-asymmetry (SSA) of the unintegrated parton
density. 
In Feynman gauge the Sivers function is obtained from from an
imaginary part associated with the usual part of the Wilson line, and
the gauge link at infinity does not contribute.  But in light-cone
gauge, the total contribution comes from the link at infinity.  The
naive definition Eq.\ (\ref{pdf.lcg}) gives zero for the Sivers
function, as observed by Brodsky, Hwang, and Schmidt \cite{BHS}.

For this particular calculation of the SSA, the problem with rapidity
divergences when $n$ is light-like did not appear,
so the calculation in \cite{Belitsky:2002sm} was done in light-like
gauge.  But the problem of rapidity divergences does appear 
when more gluons are exchanged, and it does appear with one-gluon
exchange in the unpolarized case, as is verified by explicit
calculation \cite{C03}.  Furthermore, couplings of multiple gluons to
the link at infinity should affect the unpolarized parton density
in axial gauge.

A satisfactory solution to all the difficulties is therefore to use
Eq.\ (\ref{pdf1}) as the definition of an unintegrated quark density,
with now a non-light-like vector $n$ for the Wilson lines' direction
and with a gauge-link at infinity.  Then $n$ being non-light-like cuts
off the rapidity divergence, and the CS evolution equation gives the
effect of changing the cutoff.  Apart from the gauge-link at infinity,
the definition is exactly that of CS
\cite{Collins:1981uw}.  Presumably the complications made evident by the
calculation of the Sivers function also infect the CS definition,
which makes no allowance for the gauge link at infinity. 

Unfortunately, the use of a non-light-like vector to set the gauge
condition or the directions of the Wilson lines produces some
practical complications: Feynman graph calculations are
algorithmically harder than when a light-like vector $n$ is used, and
the evolution equations have inhomogeneous terms that are difficult to
discuss explicitly.  The inhomogeneous terms are of non-leading power
in the hard scale $Q$, so that they are neglected in phenomenological
applications.  The appearance of the nonleading powers in $Q$,
however, suggests that the use of non-light-like Wilson lines might be
preferred in studies of higher-twist effects (see, for example, the
work of \cite{Bal}).  Collins, 
Hautmann and Metz \cite{C03,CM,CH} suggested
modified definitions.  They take the definition with light-like lines
as basic.  But to cancel rapidity divergences, it is
divided by an extra factor involving the vacuum expectation value of
certain Wilson lines.  The non-light-like vector needed to specify the
physically necessary rapidity cutoff is in this extra factor.  This
represents a kind of \emph{generalized} renormalization of the
operators whose matrix element is the parton density.  But we will not
take this route in the present paper.

\subsection{PCFs}

Parton correlation functions (PCFs) are defined like parton densities,
but without any of the integrals over $k_\tran$ or $k^-$.  The primary
issue in constructing a definition is the choice of directions of
Wilson lines.  So we start by discussing what leads us to our choice.
The issues are closely linked to those of rapidity divergences and the
important regions of gluon rapidity relevant for a process.

Compared with the case of PDFs (integrated or unintegrated) some
simplification occurs because we remove the integral over parton
virtuality (or $k^-$).  Any rapidity divergence occurs when the
rapidity of a gluon goes to $-\infty$.  At \emph{fixed} transverse
momentum, the divergence therefore occurs where the gluon's minus
momentum goes to infinity.  For emission of real gluons in a PCF, the
divergence is therefore cutoff, because the minus momentum on any one
gluon is restricted to $P^--k^-$ by the externally imposed $k^-$.
This occurs even if the gluon is ``dressed'' so that it decays to
multiple particles.  This contrasts with the case of an ordinary PDF,
integrated or unintegrated, with its integral over \emph{all} values
of $k^-$.

Although the use of exact parton kinematics in a PCF cuts off rapidity
divergences when the external transverse momentum is fixed, it does
nothing about virtual gluons.  Moreover, when we work with Feynman
graphs in a theory with massless gluons, the limit on rapidity expands
as the transverse momentum is reduced to zero.  So rapidity
divergences reappear as part of the infra-red divergences.  We regard
rapidity divergences as particularly dangerous because a derivation of
factorization associates the PCF or PDF with momenta that are related
to the target.  A rapidity divergence gives important contributions from
momenta that are infinitely far away, thereby at least endangering the
value of treating parton kinematics more exactly.

Therefore we define all our PCFs with non-light-like Wilson lines.
There are three types: (a) A PCF in a target, generalizing the
notion of parton density.  (b) A corresponding object, a jet PCF, for
fragmentation, generalizing the notion of fragmentation function.  (c)
A soft PCF factor.  These will be used to capture the physics
associated with partons whose momenta are respectively in the regions:
(a) target-collinear, (b) 
jet-collinear, (c) soft.

For the target and jet PCFs, the direction of the Wilson line is
chosen to be approximately at rest in the center-of-mass, so as to
correspond to a separation between left- and right-moving momenta, as
is natural for factorization.  
(For our purposes, we have generalized the notion of center-of-mass 
to any four-vector with zero rapidity so that it is applicable to 
space-like as well as time-like vectors.)
As for the soft factor, it concerns gluons central in rapidity, and is
defined in terms of a vacuum expectation of a suitable Wilson line
operator.  The Wilson line has segments representing the active
partons, and that therefore go in almost light-like directions.

There will be an apparently unfortunate profusion of directions of
Wilson lines.  The associated complications are manageable once one
recognizes that the directions provide rapidity cutoffs in each factor
to ensure it only concerns momenta appropriate for the given factor.  The
factors do acquire extra arguments corresponding to differences
between the rapidity cutoffs and the external momenta.  But then the
CS equation for the dependence on the Wilson line directions gives the
dependence on the extra arguments.  Thus the non-light-like Wilson
lines provide a tool for quantifying the evolution with respect to the
available rapidity range, and hence with respect to energy.  The
correspondence between rapidity and angle for a massless particle is
presumably the link that relates Collins-Soper evolution to the
well-known angular-ordering rule for coherent gluon emission in parton
showering.

We leave open the possibility that the definitions may be replaced
with definitions that use light-like Wilson lines with rapidity
divergences removed by gauge invariant factors as in the treatment of
the Sudakov form factor in Refs.~\cite{CM,CH2}, 
but that is left for later work.

The observations made so far in this section 
motivate our new definitions for the PCFs.  
In this section, we will simply state the definitions with 
qualitative remarks to motivate them.  The justification for the
definitions will come when we find we can set up a detailed set of
approximations which result in factorization with our given
definitions for the PCFs.  

First, we define light-like vectors corresponding to the directions of
the primary hadrons in the process:
\begin{equation}
\label{ref:u.def}
u_{\rm T}  =  (1,0,{\bf 0}_\tran),
\qquad
u_{\rm J}  =  (0,1,{\bf 0}_\tran).
\end{equation}
We also define slightly non-light-like vectors,
\begin{align}
\label{primedun}
n_{\rm T}  =  (1,-e^{-2y_{\rm T}},{\bf 0}_\tran),
\qquad
n_{\rm J}  =  (-e^{-|2y_{\rm J}|},1,{\bf 0}_\tran),
\end{align}
with $y_{\rm T}$ large and positive, $y_{\rm J}$ large and negative.  Our notation
is to use the letter $u$ for light-like vectors, and $n$ for
non-light-like vectors; the subscript indicates which direction of a
hadron or an active parton is approximated by the vector.  
To achieve factorization, we will also need a vector to 
characterize the boundary between  left and right-movers;
it should  
correspond to the gauge-fixing vector in the Collins-Soper
\cite{Collins:1981uw,TMD} formalism.  Therefore we define:
\begin{equation}
\label{eq:n.def}
  n_s = (-e^{y_s}, e^{-y_s}, {\bf 0}_\tran).
\end{equation}
In accordance with the results of Collins and Metz \cite{CM}, we use
space-like, not time-like, vectors.  They found that when virtual
gluon emission is included, space-like Wilson lines give the widest
and most universal factorization.

The above vectors will appear as directions of Wilson lines in the
definitions of PCFs, and they serve to provide cutoffs on the
rapidities of momenta in each PCF.  
Thus, each PCF is restricted to
rapidities appropriate to its function in a 
derivation of the factorization property.
During the derivation we will require that the rapidities $y_{\rm T}$,
$y_s$ and $y_{\rm J}$ correspond approximately to the target, a rest
vector in the center-of-mass, and the outgoing jet.  (The precise
values will not be relevant.)  After we have a factorization, we will
wish to exploit the universality of the PCFs to relate processes at
different energies.  This will involve, for example, boosting the
target PCF to change the target state from one energy to another.  The
boost will also apply to the vector $n_s$, thereby giving it an
inappropriate rapidity for proving factorization at the new energy.
The CSS equation gives the dependence on $y_s$, so that we can convert
the PCF to the one appropriate to the new energy.  Thus, although a
factorization proof can legitimately assume that $y_s=0$, i.e., that
$n_s$ can be considered at rest in the center-of-mass, we leave $y_s$ as
a parameter because we will need to exploit the $y_s$-dependence of
the PCFs.

In a subtraction scheme, we start the treatment from the smallest
region and successively generate terms for larger regions, with
subtractions to avoid double counting with previously encountered
regions.  Since the region of soft momenta is smaller than the regions
of collinear momenta, it is sensible to define the soft factor first,
but with some care to ensure compatibility with what 
later will be found to be
appropriate for the definition of the PDF and the jet factor.

\subsubsection{Soft factor}

Soft gluons couple to the target jet, with 
its large plus component of
momentum, and to the outgoing jet with 
its large minus component of
momentum.  This suggests that the soft factor is the vacuum
expectation value 
of Wilson lines that are nearly light-like in the plus
and minus directions.  In coordinate space we define the soft factor
by
\begin{multline}
  \tilde{\soft}(w,y_{\rm T},y_{\rm J},\mu) 
\\ 
=
  \langle 0| I^{\dagger}_{n_{\rm T};w,0}  
   V_{w}(n_{\rm T}) V^{\dagger}_{w}(n_{\rm J}) I_{n_{\rm J};w,0}V_{0}(n_{\rm J}) 
   V^{\dagger}_{0}(n_{\rm T})  | 0 \rangle_{R}.
\label{softdef}
\end{multline}
This expression has non-light-like Wilson lines going out in
approximately the plus and minus directions from a particular point in
spacetime which we may choose as the origin of our coordinate system,
times a conjugate amplitude with emission from a different point $w$.
Fourier transformation then gives a factor for the production of a
final state of a given momentum.  This represents emission from
outgoing eikonalized colored lines in directions appropriate to the
quarks $k$ and $k+q$ in the parton model.  That the Wilson lines are
not quite light-like restricts the states to those appropriate for a
finite energy process.  They also provide cutoffs on rapidity
divergences.

Now choosing the $n_{\rm J}$ Wilson line to be outgoing naturally matches
the idea that gluon radiation from this line concerns emission from
the actual $k+q$ line at the hard scattering.  This suggests that, to
match the incoming $k$ line, the direction for the $n_{\rm T}$ Wilson line
should be incoming from $-\infty$ rather than outgoing to $+\infty$.  The work
of Collins and Metz \cite{CM} shows otherwise: the choice of an
outgoing line (with color corresponding to an antiquark, if the $k$
line is an incoming quark) turns out to work better and to give
broader universality properties.

\begin{figure}
\centering
\psfrag{nJ}{$n_{\rm J}$}
\psfrag{nT}{$n_{\rm T}$}
\psfrag{I(nJ)}{$I(n_{\rm J})$}
\psfrag{I(nT)}{$I^\dag(n_{\rm T})$}
\psfrag{(w+,w-,wT)}{\hspace*{-3mm}$(w^+,w^-,\3{w}_\tran)$}
\psfrag{(0,0,0)}{$(0,0,\3{0}_\tran)$}
\epsfig{file=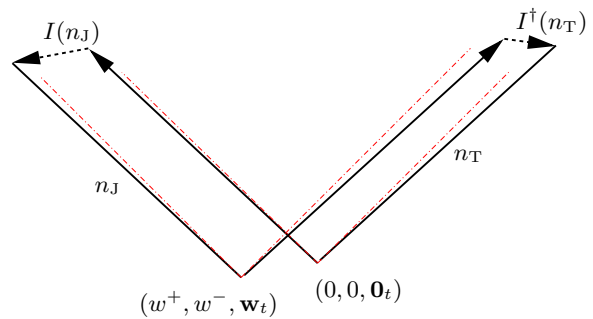,scale=0.4}
\caption{Structure of the Wilson loop that appears in  the 
definition of the soft factor (\ref{softdef}). Dot-dashed lines
indicate exactly light-like directions.} 
\label{fig:wilson_loop}
\end{figure}

The paths for these Wilson lines 
are illustrated in Fig.~\ref{fig:wilson_loop}, where we also indicate
the exactly light-like directions. 
The gauge
links,   $I^{\dagger}_{n_{\rm T};w,0}$ 
and  $I_{n_{\rm J};w,0}$,
at infinity are needed for strict gauge invariance.
It should be noted, of course, that the lines representing these gauge links in Fig.~\ref{fig:wilson_loop} 
should have components in the transverse direction (out of the page).  Thus, the soft 
factor is the vacuum expectation value of a closed Wilson loop.

Renormalized field operators 
are used in Eq.~(\ref{softdef}).
In a renormalizable theory, Eq.~(\ref{softdef}) will contain 
UV divergence, both from the divergences that appear in the Lagrangian, 
and from the fact that Eq.~(\ref{softdef}) involve Wilson lines 
meeting at a cusp.  Both types of divergence can be dealt with using standard
renormalization techniques (as indicated by the subscript, $R$). 
The renormalization scale is $\mu$.

\subsubsection{Target PCF}

\begin{figure}
\centering
\psfrag{n}{$n_s$}
\psfrag{'}{}
\psfrag{I(n)}{$I(n_s)$}
\psfrag{(w+,w-,wT)}{\hspace*{-3mm}$(w^+,w^-,\3{w}_\tran)$}
\psfrag{(0,0,0)}{$(0,0,\3{0}_\tran)$}
    \includegraphics[scale=0.35]{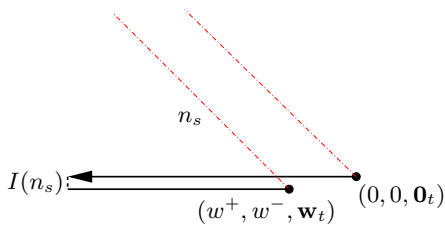}
\caption{Path of the Wilson line 
(in a highly boosted frame)
in the definition of the
  parton correlation function in the target. (The fully unintegrated PDF.)}
\label{fig:wilsonpdf}
\end{figure}

The target PCF for a quark should involve a gauge-invariant
expectation value of the quark fields inside the target.  Therefore, a
reasonable first attempt at a definition (in coordinate space) is
\begin{multline}
\tilde{F}(w,y_p,y_s,\mu)
\\
=
\langle p | \bar{\psi}(w) V^{\dagger}_{w}(n_s)  I_{n_s;w,0}  
           \dfrac{\gamma^{+}}{2}
V_{0}(n_s) \psi(0) | p \rangle_{R}.
\label{pcfguess}
\end{multline}
Here the Wilson lines are in direction $n_s$, and in proving
factorization we will assume that the rapidity $y_s = \frac12
\ln|n_s^+/n_s^-|$ is close to zero in the center-of-mass, i.e., that $n_s$
is approximately along the $-z$ direction in the center-of-mass.  That
$n_s$ is space-like is obtained from the results of Collins and Metz
\cite{CM}.  As we explained earlier, exhibiting the dependence on
$y_s$ allows the use of the CSS evolution of the PCFs.  The above
definition is compatible with the CS definition, where $n_s$ is the
gauge-fixing vector for the axial gauge $n_s \cdot A=0$.  For exact gauge
invariance we have also inserted a gauge link at infinity.  The path
for the complete Wilson line is shown in Fig.\ \ref{fig:wilsonpdf}.
The factor of $1/2$ with $\gamma^+$ is to keep the normalization the same
as for a PDF.

Although this is an excellent definition, 
we will find a different quantity arises when we first obtain a
factorization.  We will find gluonic effects that appear both in the
soft PCF factor and this definition of the target PCF.  To remove the
double counting, we will use a related definition which our
first proposal, Eq.~(\ref{pcfguess}), divided by a factor related to the
soft PCF: 
\begin{multline}
\tilde{F}_{\rm mod}(w,y_p,y_{\rm T},y_s,\mu) 
\\
=  
\frac{ \langle p | \bar{\psi}(w) V^{\dagger}_{w}(n_s)  I_{n_s;w,0}  
           \dfrac{\gamma^{+}}{2}
           V_{0}(n_s) \psi(0) | p \rangle_{R}
     }
     { \langle 0| I^{\dagger}_{n_{\rm T};w,0}  
            V_{w}(n_{\rm T}) V^{\dagger}_{w}(n_s) I_{n_s;w,0}
            V_{0}(n_s) V^{\dagger}_{0}(n_{\rm T})  | 0 \rangle_{R} 
     }.  
\label{pcf2}
\end{multline}
The denominator is the soft factor, but with the $n_{\rm J}$ Wilson line
changed to have direction $n_s$.  
The rapidity argument, $y_p$, is the exact rapidity of the target proton.
(This should be distinguished from $y_{\rm T}$ which parameterizes the 
direction of the target associated  Wilson line and points approximately in the target direction.)
Notice that this definition is given
in coordinate space.  When we Fourier transform to momentum space,
the division will be in the sense of a convolution product.

We understand the meaning of 
the definitions in Eq.~(\ref{softdef}) and Eq.~(\ref{pcf2})
as follows:
\begin{itemize}
\item The PCF for the soft factor treats central gluons accurately,
  the $n_{\rm T}$ and $n_{\rm J}$ Wilson lines providing accurate approximations
  to the quark lines and associated collinear subgraphs. 
\item As gluons approach the target rapidity, the accuracy of the
  $n_{\rm T}$ Wilson line 
  as an approximation for the incoming quark line is degraded
  in both the target and soft PCF.
\item The numerator in Eq.\ (\ref{pcf2}) provides a good approximation
  for gluons in the target range of rapidity, and the denominator
  accurately cancels the bad approximation for this same region in the
  soft factor.  
\item In the numerator of Eq.\ (\ref{pcf2}), central gluons, around the $n_s$ direction, are
  accurately given by an eikonal approximation of the form also
  appearing in the denominator, so we get a cancellation.
  Gluons of intermediate rapidity also cancel.
\item The non-lightlike Wilson lines in direction $n_s$ provide strong
  cut offs in Eq.\ (\ref{pcf2}) on gluon rapidities beyond the central region.
\item Similar ideas apply to negative rapidities and the
  correspondingly defined jet factor (next section).
\end{itemize}

The lack of question marks on the equal sign in Eq.~(\ref{pcf2})
implies that this definition is the main definition.  It has acquired
a second rapidity argument, which is quite undesirable, but this does
correspond to a definition by Idilbi, Ji, Ma and Yuan
\cite{Idilbi:2004vb,Ji:2004xq} for unintegrated PDFs in the context of
SIDIS.  However, various further reorganizations, together with an
application of the CS evolution equation can handle the dependence on
the extra arguments.

However, to match the derivation of factorization, an excellent choice
for $y_{\rm T}$ is to be close to the target rapidity.  In fact the simplest choice 
is to set $y_{\rm T}=y_p$.  In that case, the important dependence
for which we need the CS equation is on $y_s$.

\subsubsection{Jet PCF}
\begin{figure}
\centering
\psfrag{-n}{$-n_s$}
\psfrag{'}{}
\psfrag{I(-n)}{$I(-n_s)$}
\psfrag{(w+,w-,wT)}{\hspace*{-3mm}$(w^+,w^-,\3{w}_\tran)$}
\psfrag{(0,0,0)}{$(0,0,\3{0}_\tran)$}
    \includegraphics[scale=0.35]{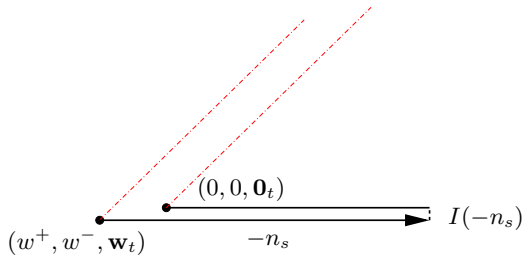}
    \caption{Path of the Wilson line in the definition of the
      fragmentation function.}
\label{fig:wilsonff}
\end{figure}

We define a jet PCF, to account for final-state-collinear behavior,
just like the target PCF, in versions without the denominator:
\begin{widetext}
\begin{equation}
\tilde{\jet}(w,jet direction,y_{\rm J},y_s,\mu)
 = \langle 0 | \bar{\psi}(w) V^{\dagger}_{w}(-n_s)  
          I_{-n_s;w,0}  \gamma^{-} V_{0}(-n_s) \psi(0)
          | 0 \rangle_{R} ,
\label{pcf3a}
\end{equation}
and with the denominator:
\begin{equation}
\tilde{\jet}_{\rm mod}(w,jet direction,y_{\rm J},y_s,\mu)
 = \frac{ \langle 0 | \bar{\psi}(w) V^{\dagger}_{w}(-n_s)  
          I_{-n_s;w,0}  \gamma^{-} V_{0}(-n_s) \psi(0)
          | 0 \rangle_{R}
        }
        { \langle 0| I^{\dagger}_{-n_s;w,0}  V_{w}(-n_s) V^{\dagger}_{w}(n_{\rm J}) I_{n_{\rm J};w,0}V_{0}(n_{\rm J}) 
               V^{\dagger}_{0}(-n_s)  | 0 \rangle_{R} 
        } .
\label{pcf3}
\end{equation}
\end{widetext}
The paths of the Wilson lines in this case are illustrated in 
Fig.~\ref{fig:wilsonff}. 
The arguments of the jet PDF will include a momentum variable,
analogous to the proton momentum in the target PCF, for the 
exact rapidity of the outgoing jet. 
We use a script $\jet$ as the symbol for the 
jet PCF rather than 
$\jetbub$ in order to distinguish the jet PCF 
from the jet bubble that appears  
in graphs like Fig.~\ref{LOdiags}.

A surprising feature is that in Eq.~(\ref{pcf3}) we use $-n_s$ for the
central-rapidity Wilson lines, which is the opposite of what we used in
Eq.~(\ref{pcf2}).  This arises from a correspondence with different
Wilson lines in the soft factor.  For the target PCF, the Wilson line
with direction 
$n_s$ is at negative rapidity relative to the target.  In
the \emph{rest} frame of the target $n_s$ has a large minus component,
and the Wilson line is associated with the outgoing $k+q$ quark.
However, $n_{\rm T}$ is the nearly light-like vector that characterizes
the momentum of the outgoing quark.
Therefore, the
minus component should be of the same 
sign as $n_{\rm J}$, i.e.,
positive, which is just what we have defined it to be.  
In contrast, in the jet PCF, Eq.~(\ref{pcf3}), the corresponding
Wilson line is to have 
the sign of
its \emph{plus} 
component
match that of the target
associated Wilson line, i.e., of $n_{\rm T}$.  Thus we use 
$-n_s$ not $+n_s$,
given that we use a space-like Wilson line.

\subsection{Connection to factorization}
The momentum-space PCFs are determined by the Fourier transforms of the 
above definitions.
For example, the momentum space PCF 
in the target is
\begin{multline}
\label{pcf4}
F_{\rm mod}(k,y_p,y_{\rm T},y_s,\mu) 
\\
= \int \frac{dw^+ dw^- d^{2}{\bf w}_\tran}{32 \pi^{4}}
   e^{-i k\cdot w} 
\tilde{F}_{\rm mod}(w,y_p,y_{\rm T},y_s,\mu).
\end{multline}
When we come to the issue of factorization it will be important to 
recall that the momentum space PCFs discussed in this section are 
well-defined for all values of momentum, even those that lie far
outside the range that is meant to be accurately described by the PCF.
For example, the momentum-space soft factor $\soft(l,y_{\rm T},y_{\rm J},\mu)$, obtained
by Fourier transforming Eq.~(\ref{softdef}), exists even for values 
of $l$ that are far from soft.  Of course, when the PCFs appear in 
the factorization formula for 
physical processes, they should be 
large only for appropriate values of momentum.  This is partly 
accomplished by the use of non-light-like Wilson lines in the definitions 
above, which, as we shall see in the next two sections, cut-off the 
contribution from light-like gluons.
The motivation for writing down Eqs.~(\ref{softdef}),~(\ref{pcf2}), 
and~(\ref{pcf3}) will become clearer 
in the next few sections where we will show explicitly 
what is needed for a factorization formula. 
This 
new factorization formula will be defined with 
exact over-all kinematics
that takes into account soft and 
collinear gluon emissions.  
We will show how the definitions for parton correlation functions 
listed above follow naturally from 
the factorization of soft and collinear gluons in DIS.

\section{Subtractions}
\label{sec:sub}

The overall approach we use is a subtractive approach --- e.g.,
\cite{JCC} --- generalized from the Bogoliubov approach to
renormalization.  Up to power-suppressed terms, each graph $\Gamma$ is
written 
as a sum over a contribution for each of its leading regions:
\begin{equation}
  \label{eq:Gamma}
  \Gamma = \sum_{R~\text{of}~\Gamma} C_R\Gamma + \text{power-suppressed}\;.
\end{equation}
A single graph typically has many different regions, each
corresponding to a different graphical decomposition of the form of
Fig.\ \ref{LOdiags}(c).  As explained below the definitions of the
terms $C_R\Gamma$ employ approximations and then subtractions to eliminate
double counting between regions.  We denote the chosen approximation
corresponding to a particular region by the action of an
``approximator'', $T_R$, as in Eq.~(\ref{pm3}) for the simple case of
parton model kinematics.

The different regions can be ranked according to their sizes, e.g. a
soft region corresponds to a smaller range of momentum than a
collinear region, and is therefore a smaller region in a
set-theoretic sense.  We define a region as minimal if there are no
smaller regions.  The contribution from a minimal region $R_0$ is
simply the action of the corresponding approximator on the
unapproximated graph,
\begin{equation}
\label{eq:smallest}
C_{R_{0}} \Gamma = T_{R_{0}} \Gamma.
\end{equation}

For the contributions from larger regions, we avoid double counting by
performing subtractions for the contributions from smaller regions.
So we define
\begin{equation}
  \label{eq:CR}
  C_R \Gamma = T_R\Bigl( \Gamma - \sum_{R'<R} C_{R'}\Gamma \Bigr) .
\end{equation}
For a minimal region, Eq.\ (\ref{eq:CR}) reduces to
Eq.~(\ref{eq:smallest}), so that it gives a valid recursive
definition of $C_R\Gamma$ with the terms being constructed sequentially
starting from the minimal region(s).

As exhibited in Fig.\ \ref{LOdiags}(c), a complication is that the
graphical representation of the regions does not directly correspond
to factorization, because of multiple gluon connections between the
different factors.  This contrasts with the case of Fig.\
\ref{LOdiags}(b), where we have topological factorization.  Therefore
an important constraint on choosing the definition of $T_R$ out of the
range of possibilities is that (if possible) the graphical
factorization results in an actual factorization after a sum over
graphs and regions:
\begin{equation}
  \label{eq:total}
  \sum_{R,\Gamma} C_R\Gamma = \text{factorized form}\;.
\end{equation}

\section{Kinematic approximations and rapidity differences}
\label{sec:kin_approx}
The kinematic approximations that enable factorization to be derived
utilize certain properties of Minkowski-space momenta.  We now review
them with a view to systematizing our later work.

Corresponding approximations in a Euclidean space are much more
trivial.  Thus if $\3p$ and $\3q$ are two spatial momenta with
$|\3p|\ll|\3q|$, then we can approximate $(\3p+\3q)^2$ by $\3q^2$, up to
a power-law correction.  This is simply because angles are bounded in
a Euclidean space.

In Minkowski space, we have to deal with unbounded rapidity variables
instead of angles.  (Rapidity is useful to us because our process has
a preferred axis.)  Suppose we have two 4-momenta $k_1$ and $k_2$ with
rapidities defined by $y_i = \frac12\ln|k_i^+/k_i^-|$, and such that
$|k_i^+k_i^-|$ is comparable to or bigger than $k_{i,\tran}^2$, as would
be the case for an on-shell momentum.  
These could be, for example, the four-momenta of  
two internal gluon lines in a graph, $\Gamma$.
We write the orders of
magnitude of the $({+},{-},T)$ components of each momentum as
\begin{equation}
  (k_i^+, k_i^-, k_{i,\tran}) = O\big( Me^{y_i}, Me^{-y_i}, M\big),
\end{equation}
where $M$ is an appropriate mass scale.

The interesting case will be where the rapidities are quite different,
let us say $e^{y_1-y_2}\gg 1$.  Then $k_1\cdot k_2$ is dominated by one
term:
\begin{equation}
\label{eq:mom.approx}
  k_1\cdot k_2 = k_1^+k_2^- \left[ 1 + O( e^{-(y_1-y_2)} )\right].
\end{equation}
We will be able to apply this in Fig.\ \ref{LOdiags}(c), for the
numerators of the attachments of the gluons from $\softbub$ to the collinear
subgraphs $\jetbub$ and $\pdfbub$, and for the attachments of the gluons from the
collinear subgraphs to the hard subgraphs.

There is an interesting region, where Eq.\ (\ref{eq:mom.approx}) fails
because one momentum variable has particularly small longitudinal
momentum components, i.e., $|k_i^+k_i^-| \ll k_{i,\tran}^2$.  This is called
the Glauber region, and it is a natural case to examine since it
corresponds to a virtual particle exchanged in small-angle elastic
scattering, as in a final-state interaction.  An important part of
factorization proofs is to arrange for a contour deformation to get
out of the Glauber region, when possible.  See Collins and Metz
\cite{CM} for a recent treatment of issues relevant to our discussion
in later sections; for the DIS reactions treated in this paper, a
contour deformation out of the Glauber region is possible.

It is worth observing that the word ``region'' has two slightly
different connotations in our discussions.  One refers to a locality
in the space of 4-momenta, as in the explanation of the Glauber region
in the previous paragraph.  The other connotation is as a locality in
the multi-dimensional space of loop or line momenta for a whole graph,
as in Eq.\ (\ref{eq:CR}).  For a graph with a single gluon exchange,
we often use the direct correspondence between the regions of the
graph and the regions for the gluon's momentum.

A propagator denominator
\begin{equation}
  (k_1+k_2)^2-m^2,
\end{equation}
needs a bit more care than a simple product $k_1\cdot k_2$, since the
appropriate approximation depends also on the relative virtualities of
$k_1$ and $k_2$.  We again assume that we have deformed out of any
Glauber region, and that $e^{y_1-y_2}\gg1$.  Then:
\begin{enumerate}

\item If the virtualities of $k_1$ and $k_2$ are comparable to each
  other, and both are comparable to or bigger than $m^2$, then the
  denominator is dominated by $k_1^+$ and $k_2^-$:
  \begin{equation}
    (k_1+k_2)^2-m^2 \simeq  2 k_1^+k_2^-. 
  \end{equation}
  An elementary application is to the virtual photon in Fig.\
  \ref{LOdiags}(a) where $Q^2\simeq 2k^+l^-$ when the initial quark is
  collinear to the target.

\item But if we also have to treat the case that one momentum, $k_2$
  say, has virtuality much less than that of the other, then although
  we can neglect $k_2^2$ with respect to $k_1^+k_2^-$, we cannot
  necessarily neglect $k_1^2$ or $m^2$.  Thus we can only write
  \begin{equation}
    (k_1+k_2)^2-m^2 \simeq k_1^2-m^2 + 2 k_1^+k_2^-.
  \end{equation}
  This amounts to replacing $k_2$ by its minus component.  This
  approximation will be the primary tool in deriving factorization.

\end{enumerate}

\section{Real Gluon Emission}
\label{sec:real}

\begin{figure}
\centering
\epsfig{file=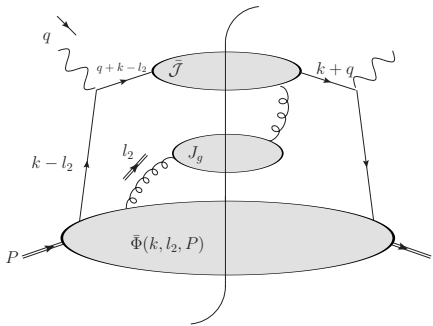,scale=0.45}
\caption{A graph that yields a contribution to the soft and collinear regions.}
\label{soft}
\end{figure}

In this section we examine the simplest case of the gluonic
corrections to the parton model that were summarized in Fig.\
\ref{LOdiags}(c).  This is given by the emission of one real dressed
gluon~\footnote{By ``dressed'' gluon, we mean the outgoing gluon with
  the final state bubble which represents all possible radiative
  corrections.}  --- Fig.~\ref{soft}.  We will consider virtual gluon
radiation in a later section.

\subsection{Regions for gluon exchange}

The gluon, of momentum $l_2$, evolves into a final state represented
by the bubble, $J_{g;\lambda\rho}(l_{2})$.  It attaches to a jet-associated
bubble denoted by $\bar{\jetbub}^\lambda(k+q,l_{2})$, and a target-associated
bubble $\bar{\pdfbub}^\rho(k,l_2,P)$.  Here $\lambda$ and $\rho$ are Lorentz indices,
and the subscript, $g$ indicates that this is the bubble associated
with an outgoing gluon. 

We will restrict attention to the case that there is no production of
extra jets of high transverse momentum.  Therefore, by standard
power-counting arguments, we only need to consider the contribution
from three regions of the gluon's momentum: soft, target-collinear,
and jet-collinear.  These are fundamentally distinguished by the
rapidity of the gluon relative to the target and the jet --- Fig.\
\ref{fig:rap.range}.  Each of these three regions corresponds to a
particular decomposition of the form of Fig.\ \ref{LOdiags}(c), and
for the moment we assume that no other regions are relevant.

A complication is that there are two ways to characterize the regions.
One is by the Libby-Sterman \cite{Libby:1978bx} analysis in terms of
the pinch-singularity surfaces (PSSs) of the corresponding massless
graphs.  The other is in terms of 
the
very different rapidity ranges of,
in this case, the exchanged gluon.  The analysis in terms of rapidity
is closely related to the use of angular ordering in leading logarithm
approximations.

\begin{figure}
  \centering
  \psfrag{kz}{$l_2^z$}
  \psfrag{k0}{$l_2^0$}
  \psfrag{A'}{}
  \psfrag{B'}{}
  \psfrag{A}{T}
  \psfrag{B}{J}
  \psfrag{S}{S}
  \includegraphics{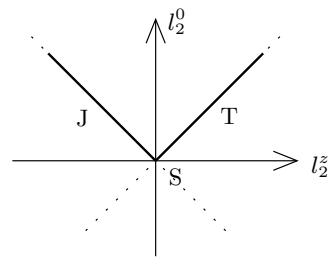}
  \caption{Pinch singularities for gluon exchange in massless theory.
    The singularities are labeled: ``S'' for soft, ``T'' for
    target-collinear and ``J'' for jet-collinear.  The solid lines
    indicate the positions of the actual pinch singularities, and the
    dots indicate their extrapolations to where there is no pinch at
    the collinear singularities.
  }
  \label{fig:PSS}
\end{figure}

For the Libby-Sterman analysis, the massless singularities,
illustrated in Fig.\ \ref{fig:PSS},  are: A soft
gluon singularity at zero gluon momentum $l_2^\mu = (0,0,{\bf 0}_\tran)$, a
target-collinear singularity at $l_2^\mu = (z_TP^+,0,{\bf 0}_\tran)$, and a
jet-collinear singularity at $l_2^\mu = (0,z_Jq^-,{\bf 0}_\tran)$.  
Here, $z_T$ and $z_J$ parameterize 
the
position along the collinear-singularity
lines in a frame-independent fashion.  
The momenta that are actually relevant are in neighborhoods that
surround the PSSs in Fig.~\ref{fig:PSS}.  
Therefore, we need to introduce variables that allow 
us to specify the proximity of the momenta to a PSS. 
For each PSS, we parameterize its neighboring
momenta by what we will term radial and angular coordinates.  Thus we
write:
\begin{itemize}
\item \underline{For the soft region} 
  \begin{equation}
  \label{eq:S.scale}
    l_2 = \lambda \big( \hat{l}_2^+, \hat{l}_2^-, \hat{l}_2^\tran \big) .
  \end{equation}
  The dimensionless ``angular'' variables $\hat{l}_2^\mu$ obey some
  moderately arbitrary normalization condition, e.g.,
  $\sum\bigl|\hat{l}_2^\mu\bigr|^2 = 1$.  These variables parameterize a
  surface of constant $\lambda$ surrounding the soft point $l=0$.  
  Then $l_2{}^2$ is of order 
  $\lambda^2$, with a coefficient bounded away from zero and infinity.  This
  is the property that enables us to estimate the errors on our
  approximations systematically.

  A formal implementation of the quasi-angular integration can be made as
  follows: 
  \begin{multline}
    \hspace*{3mm}
  \int d^4l_2   \ldots  =  \int d\lambda \, \lambda^3 \, \int d^3\hat{l}_2 \, \ldots 
\\
       \equiv  \int d\lambda \, \lambda^3  
          \int d^4l_2 \frac{1}{\lambda^3} \,
          \delta\!\left( {\textstyle \lambda - \sqrt{\ \sum_\mu|l_2^\mu|^2} }\right) 
                 \, \ldots ,
  \end{multline}
  where we use the normalization condition proposed above.  This
  formula is written in a form equally suited for virtual gluon
  exchange.  Any constraints on the invariant mass of the gluon from
  the nature of the final state are taken to be in the integrand, the
  part indicated by ``$\ldots$''.
  
  The formula obviously does not give a Lorentz-invariant decomposition
  of the integration, but is adapted to the needs of the process.  It
  is arranged so that power-counting in $\lambda$ is straightforward: The
  angular integral phase space $\int d^3\hat{l}_2$ is independent of $\lambda$.
  Moreover, it is unambiguous that a small value for $\lambda$ corresponds to a small
  neighborhood of the origin in Fig.~\ref{fig:PSS}.  This would not be
  true if we had tried to characterize the soft region by specifying
  it in terms of a Lorentz invariant quantity, i.e., $2 l_{2}^{+}
  l_{2}^{-} - l_{2,\tran}^{2}$.

\item \underline{For the target-collinear region}
  \begin{equation}
  \label{eq:T.scale}
     l_2 = \left( z_TP^+, \frac{ \lambda^2\tilde{l}_{2L} }{ z_TP^+ }, 
                 \lambda \tilde{l}_{2,\tran} \right).
  \end{equation}
  Our choice for the normalization condition of the dimensionless
  angular variables is $|\tilde{l}_{2L}| + {\tilde{l}_{2,\tran}}^2 = 1$.  The
  asymmetric scaling is suitable for a momentum highly boosted from a
  rest frame, and again ensures that $l_2{}^2$ is proportional to
  $\lambda^2$.  
  
  A formal definition of the quasi-angular integration is:
  \begin{align}
  \hspace*{5mm}
    \int&d^4l_2  \ldots  
  =
    \int d\lambda \, \lambda^3 \int \frac{dz_T}{z_T} \, \int d^2\tilde{l}_2 \, \ldots 
  \nonumber\\ &
       \equiv \int d\lambda \, \lambda^3 \int \frac{dz_T}{z_T} \,
           \int d^4l_2 \frac{z_T}{\lambda^3} \,
                 \delta(z_T-l_2^+/P^+) \times \,
   \nonumber\\ & ~ ~
            \times \delta\!\left( {\textstyle \lambda - \sqrt{|l_2^+l_2^-|+|l_{2,\tran}|^2} }\right) 
                 \, \ldots .
  \end{align}
  In the Libby-Sterman terminology, $z_T$ is an ``intrinsic'' variable
  for the PSS, parameterizing the position on the surface.  Then $\lambda$ can
  be treated as measuring the distance from the surface, while
  $\tilde{l}_2$ parameterizes a (2-dimensional) surface around the PSS,
  for a given value of the intrinsic variable.

\item \underline{For the jet-collinear region}
  \begin{equation}
  \label{eq:J.scale}
     l_2 = \left( \frac{ \lambda^2\bar{l}_2^L }{ z_Jq^-  }, z_Jq^-,
                 \lambda \bar{l}_2^T \right) ,
  \end{equation}
  exactly similarly to the the situation for the target-collinear region.

\end{itemize}
For any of these cases, when the four-momentum has a virtuality of
order a typical hadronic scale $\Lambda^2$, then the radial variable $\lambda$ is
itself of order $\Lambda$, and this can be regarded as the canonical size of
$l_2$ for the region.  This gives a basic intuition about the meaning
of collinear and soft momenta.  (Here we temporarily assume that the
$z_{T/J}$ variables are of order unity.)  But we integrate over all
accessible momenta 
(or up to some limit of order $Q$), so it is
important to treat $\lambda$ as ranging from 0 to order $Q$.

The approximations we use will have typical fractional errors
suppressed by a power of the various small mass scales, e.g., $\lambda/Q$,
$\Lambda/Q$, $m/Q$.  As we move to larger $\lambda$ than the ``canonical'' value,
the errors become larger.  But at the same time, as we will see, the
approximations for larger regions become useful, and the overall
effect in the subtraction formalism will be that the total error in
the sum of all the approximated contributions will be of higher twist.
That is, the fractional error will be a power of a fixed hadronic mass
scale divided by a large scale like $Q$.  (Logarithmic corrections
will slightly weaken an initially determined power-law suppression.)

We can now compare the Libby-Sterman analysis and the rapidity
analysis. With the above definitions, there is a large rapidity
difference between the two collinear regions and this will be
sufficient for us to obtain appropriate approximations.  The
Libby-Sterman analysis further requires $z_J$ and $z_T$ to be of order
unity, so that the large component of a collinear momentum is of order
the hard scale $Q$.  While this is important for discussing hard
scattering, it complicates the treatment of the important subcase
where one of these $z$ variables goes to zero.  In that case when $\lambda$
for a collinear region is sufficiently small (of order a mass times
$z$), the gluon is then simultaneously collinear, by the rapidity
criterion, and soft by the criterion of small size.  But the gluon is
not collinear by the Libby-Sterman criterion of energy being of order
$Q$.  This case is only significant for a massless gluon.  To avoid a
proliferation of special cases, we unify this case as much as possible
with the collinear case.  For our initial all-orders treatment in the
present paper, we will cut off this region by the use of a gluon mass,
as announced in the introduction.  But we will not need to do this
just yet.

Related issues have arisen in the literature in the form of a
distinction between a soft and a supersoft region \cite{supersoft}.
In the language of this section, they are distinguished by the
numerical values of $\lambda$.  When the components of $l_{2}$ are of order
$\Lambda$ we are in the conventional soft region; when they are of order
$\Lambda/Q^{2}$ we are in the supersoft region.  (This will be discussed
further in Sect.~\ref{accuracy}.)  Again we will unify as much as
possible the soft and supersoft regions.

\subsection{Unapproximated graph}

Before any approximations, the formula for Fig.~\ref{soft} is
\begin{multline}
\Gamma^{(R)}
= \frac{e_j^2 P_{\mu\nu}}{4\pi} 
    \int \frac{d^4l_2}{(2\pi)^4} \int \frac{d^4k}{(2\pi)^4}  
\\\times 
   {\rm Tr}\! \left[ \gamma^{\nu}  \bar{\jetbub}^\kappa(k+q,l_2)
                   \gamma^{\mu} \bar{\pdfbub}^\rho(k,l_2,P) \right]
   J_{g;\kappa\rho}(l_{2}). 
\label{eq:sigma2}
\end{multline} 
Here the superscript ${}^{(R)}$ denotes graphs for emission of a real
dressed
gluon, and, as before, $P_{\mu\nu}$ represents a projection for a
particular chosen structure function.
We use bars on $\bar{\pdfbub}(k_{1},l_{2})$ and
$\bar{\jetbub}(k_{1},l_{2})$ to distinguish bubbles with an extra external
gluon from those in Fig.~\ref{LOdiags}(b).  \emph{The indices
  $\kappa$ and $\rho$ on $\bar{\jetbub}$ and $\bar{\pdfbub}$ are now for the coupling to
  the gluon, not for a decomposition over Dirac matrices.} 
The final-state bubble for the gluon in Fig.~\ref{soft} is denoted by $J_{g;\kappa\rho}$.
In the approach with standard kinematical approximations this gluon would be put on-shell,
and therefore we would have $J_{g;\kappa\rho} \propto \sum_{\rm pol} \epsilon_{\kappa} \epsilon_{\rho}
\delta(l_2^2)$, where $\epsilon$ is a gluon polarization vector.
Since we keep exact kinematics we do not perform these approximations.  
Finally, we have left
implicit the quark-flavor label $j$ on these quantities.

Our strategy is as follows: We start with the smallest region, the
soft region, and construct an approximator $T_{\rm S}$ that is
(a) accurate in the soft region, and (b) suitable for the use of a Ward
identity argument to bring the total soft contribution, $\sum_\Gamma C_{\rm
  S}\Gamma^{(R)}$, into a factorized form.  The subtraction method requires
us to extend the integration in the soft term to larger regions.  Then
we follow similar steps to construct approximators $T_{\rm J}$ and $T_{\rm T}$ for the
collinear regions.  Again, these have to be compatible with Ward
identity arguments.  
Application of the methods of Sec.\ \ref{sec:sub} will provide
subtractions that compensate double counting between the terms for
different regions, so as to ensure that the sum 
of these terms,
$C_{\rm S}\Gamma^{(R)}+C_{\rm T}\Gamma^{(R)}+C_{\rm J}\Gamma^{(R)}$ gives an accurate
approximation for the union of the regions, including all intermediate
cases.  This therefore deals with all
regions involving low transverse momentum for $l_2$, i.e., for
$l_{2,\tran}\ll Q$, with relative errors being approximately of order
$l_{2,\tran}/Q$.

\subsection{Soft Region}
\label{sec:soft}

We now define our approximation for the soft region.  The method is
that of Grammer and Yennie~\cite{Grammer:1973db}, as applied in
factorization proofs (e.g., \cite{TMD}).  

In the Breit frame, the target is boosted to have a large plus
component of momentum, of order $Q$, while the outgoing jet is boosted
to have a large minus component of momentum, also of order $Q$.
Therefore, for the coupling of the jet and target bubbles, $\bar{\jetbub}^\kappa$
and $\bar{\pdfbub}^\rho$, to the exchanged gluon, we may characterize the sizes
of the vector components by their transformations under boosts.  The
largest components have $\kappa={-}$ and $\rho={+}$ respectively.  Relative to
the largest components, the smaller components have sizes
\begin{align}
    \frac{ \bar{\jetbub}^+ }{ \bar{\jetbub}^- }
    = \mathcal{O}\!\left( \frac{\Lambda^2}{Q^2} + \frac{\lambda}{Q} \right) ,
\qquad &
    \frac{ \bar{\jetbub}_\tran }{ \bar{\jetbub^-} }
    = \mathcal{O}\!\left( \frac{\Lambda}{Q} + \frac{\lambda}{Q} \right) ,
\\
    \frac{ \bar{\pdfbub}^{-} }{ \bar{\pdfbub}^{+} }
    = \mathcal{O}\!\left( \frac{\Lambda^2}{Q^2} + \frac{\lambda}{Q} \right) ,
\qquad &
    \frac{ \bar{\pdfbub}_\tran }{ \bar{\pdfbub}^{+} }
    = \mathcal{O}\!\left( \frac{\Lambda}{Q} + \frac{\lambda}{Q} \right) .
\end{align}
These power laws result both from the size of the components of the
momentum $l_2$ of the exchanged gluons and from the sizes of the
components of the collinear momenta, which are boosted from their rest
frame.  For the moment, we treat the collinear momenta as having
transverse momenta of order a normal hadronic mass scale $\Lambda$.  At
first sight, the Lorentz boosts to get collinear momenta indicates
that the non-leading longitudinal components, $\bar{\jetbub}^+$ and
$\bar{\pdfbub}^-$ for the collinear subgraph, would be of order $\Lambda^2/Q^2$
relative to the large components.  This would in fact be correct if
the minus component of the injected soft momentum $l_2$ were
sufficiently small, i.e., if $\lambda$ were less than about $\Lambda^2/Q$.  But
when it has a larger value, e.g., the ``natural'' value for a soft
momentum $l_2^- \sim \Lambda$, some lines of $\bar{\pdfbub}$ acquire this size for
their minus momentum.  Correspondingly $\bar{\pdfbub}^-/\bar{\pdfbub}^+$ increases
to a size $\lambda/Q$.

To leading power, 
(in either $\lambda/Q$ or $\Lambda/Q$)
we need to keep only the leading polarization
components, and we make the following string of approximations to the
product of bubbles in Eq.~(\ref{eq:sigma2}):
\begin{widetext}
\begin{align}
\bar{\jetbub}^\kappa(k+q,l_2) ~ J_{g; \kappa\rho}(l_{2}) ~ \bar{\pdfbub}^\rho(k,l_2,P) 
 &\simeq   \bar{\jetbub}^-(k+q,l_2) 
      ~ J_g^{+-}(l_2)
      ~ \bar{\pdfbub}^+(k,l_2,P)
\nonumber\\
  &\simeq   \bar{\jetbub}(k+q,l_2) \cdot l_2
      ~ \frac{1}{ l_2^+ }
      ~ J_g^{+-}(l_{2})
      ~ \frac{1}{ l_2^- } 
      ~ l_{2} \cdot \bar{\pdfbub}(k,l_2,P)
.
\label{grammer1}
\end{align}
Here, our aim is to obtain a form in which the gluon momentum $l_2$
is contracted with each jet factor, a situation in which we can apply
a Ward identity.  The critical step is in the last line, where we
use the fact that in the soft region $l_2 \cdot \bar{\pdfbub} \simeq l_2^-\bar{\pdfbub}^+$
and $\bar{\jetbub} \cdot l_2 \simeq \bar{\jetbub}^-l_2^+$.  This step requires that the
longitudinal components of $l_2$ be comparable to each 
other, which
in turn requires that the rapidity of $l_2$ be small.  It also
requires that $l_2$ be outside the Glauber region, as is always true
for real gluon emission. (Recall that the Glauber region is where
$|l_2^+l_2^-| \ll l_{2,\tran}^2$.)

Within the soft region, the relative error in
the approximation is then of order $\lambda/Q$.  However, in the
factorization formula, we will integrate over the whole accessible range
of $l_2$.  This will of course take us outside the soft region where the
above approximations are accurate.  By itself this is no problem,
since such a contribution
will eventually be accommodated by proper double-counting subtractions in the treatment
of other regions.  
But, particularly when we apply the same soft
approximation to virtual gluons, the denominators $l_2^+$ and $l_2^-$
will create rapidity divergences that are completely unphysical.  The
simplest solution is to replace these denominators by dot products with
$n_{\rm J}$ and $n_{\rm T}$, the vectors defined in Eq.\ (\ref{primedun}):
\begin{align}
\bar{\jetbub}^\kappa(k+q,l_2) ~ J_{g; \kappa\rho}(l_{2}) ~ \bar{\pdfbub}^\rho(k,l_2,P) 
  &\simeq  \bar{\jetbub}(k+q,l_{2})  \cdot l_{2} 
      ~ \frac{ n_{\rm J}^\kappa J_{g;\kappa\rho}(l_{2}) n_{\rm T}^\rho }
             { (l_{2} \cdot n_{\rm J} -i\epsilon) \, (l_{2} \cdot n_{\rm T}+i\epsilon) } 
      ~ l_{2} \cdot \bar{\pdfbub}(k,l_2,P) \,
.
\label{grammer}
\end{align}
\end{widetext}
In the soft region, $l_2 \cdot n_{\rm J} \simeq l_2^+$, and $l_2 \cdot n_{\rm T} \simeq l_2^-$, so
that the accuracy of the approximation is unimpaired.  Beyond the soft
region, these replacements provide cutoffs in an integral over the
rapidity of $l_2$.  We also make a corresponding change in the $J_g$
part of the
numerator, so that after we apply a Ward identity to sum over
all attachments of the gluon to $\bar{\pdfbub}$ and $\bar{\jetbub}$, we obtain
exactly a term where the gluon attaches to a Wilson line operator.  
Finally, we introduce the $i\epsilon$ prescriptions appropriate to the directions of
the Wilson lines determined in \cite{CM}.

\begin{figure}
\centering
\psfrag{coll. jet}{coll.\ jet}
\psfrag{coll. targ.}{coll.\ targ.}
\psfrag{soft region}{soft region}
\psfrag{y_J}{$y_{\rm J}$}
\psfrag{y_Q}{$y_Q$}
\psfrag{y_T}{$y_{\rm T}$}
\epsfig{file=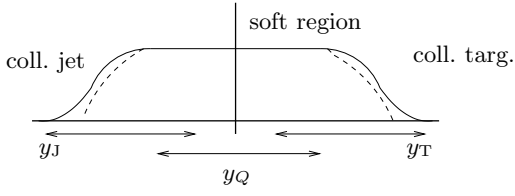,scale=0.55}
\caption{Distribution of gluon rapidity.  The solid line is the exact
  distribution, whereas 
the dotted line represents the distribution obtained from the soft
approximator, $T_{S}$.} 
\label{rapdist}
\end{figure}

Let us choose the rapidities $y_{\rm T}$ and 
$y_{\rm J}$ that define the vectors $n_{\rm T}$ and $n_{\rm J}$ to match
the target and jet rapidities.  Then there is a natural 
correspondence between the
rapidity cutoff provided in the soft approximation and that provided
by the $\bar{\pdfbub}$ and $\bar{\jetbub}$ subgraphs before approximation.  This
is illustrated in Fig.~\ref{rapdist}.

\begin{figure}
  \centering
  $\raisebox{-0.3\totalheight}{\includegraphics[scale=0.6]{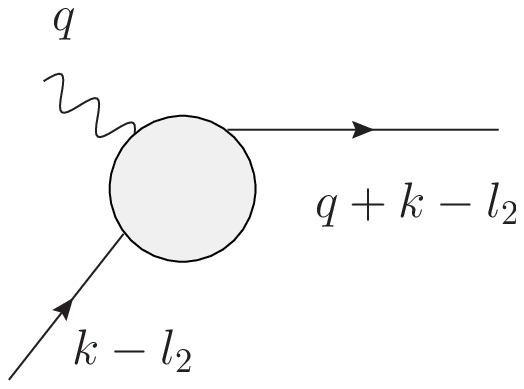}}
  ~ \mapsto ~
   \raisebox{-0.3\height}{\includegraphics[scale=0.6]{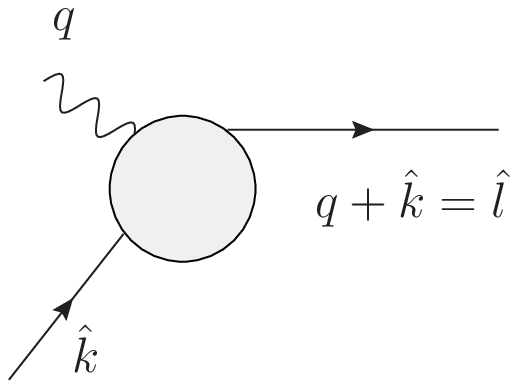}}
  $
  \caption{Corrections to current vertex, and kinematic projection for
  hard scattering.}
  \label{fig:vertex.corr}
\end{figure}

One further refinement in the exact definition of the approximation is
needed to ensure that it works suitably when there are higher-order hard
corrections at the electromagnetic vertex, the left-hand-side of Fig.\
\ref{fig:vertex.corr}.  The momentum $l_2$ flows through the vertex,
so that the hard factor can vary with $l_2$.  In the soft region 
(of the $l_2$ integral) this
is an unimportant power-suppressed effect, but in the 
complete
integral over
all $l_2$ it can create a big effect.  When multiple soft gluons are
exchanged, we will be likely to find that the definition of the soft
factor needs to be changed from Eq.\ (\ref{softdef}) in an
inconvenient way: instead of Wilson lines joining at two point
vertices at 0 and $w$, we will have nonlocal vertices, with the
nonlocality governed by detailed properties of the hard scattering. To
avoid this issue unambiguously, we define the hard scattering subgraph
to be evaluated at suitably projected momenta that stay fixed as $l_2$
varies.  This is readily done by requiring from the beginning that the
external momenta of the hard subgraph always be projected down to the
parton model values, as in Fig.\ \ref{fig:vertex.corr}.  There we use
the momenta $\hat{k} = (-q^+,0,{\bf 0}_\tran)$ 
and $\hat{l} = q+\hat{k}=(0,q^-,{\bf 0}_\tran)$ that we defined earlier.  
Thus the incoming and outgoing momenta of the approximated vertex are
exactly massless, on-shell and independent of $l_2$.  Naturally, we
also need to apply the projections ${\cal P}_{\rm T}$ and ${\cal P}_{\rm J}$ in
the Dirac algebra, exactly as in the basic parton model.

Putting these elements together gives the definition of the
approximator for the soft region for Eq.~(\ref{eq:sigma2}):
\begin{widetext}
\begin{equation}
T_{\rm S} \Gamma^{(R)}\, =
  \frac{e_j^2 P_{\mu\nu}}{4\pi} 
  \int \frac{d^4l_{2}}{(2\pi)^4} \int \frac{d^4k}{(2\pi)^4}  
{\rm Tr} \left[ {\cal P}_{\rm J} \gamma^{\nu} {\cal P}_{\rm J} ~ l_{2} \cdot \bar{\jetbub}(k+q,l_{2})  ~
                {\cal P}_{\rm T} \gamma^{\mu} {\cal P}_{\rm T} ~
                l_{2} \cdot \bar{\pdfbub}(k,l_2,P) \right] 
      \, \frac{ n_{\rm J}^{\kappa} J_{g;\kappa\rho}(l_{2}) n_{\rm T}^\rho }
             { (l_{2} \cdot n_{\rm J} - i\epsilon) \, (l_{2} \cdot n_{\rm T} + i\epsilon) }. 
\label{softapprox}
\end{equation}
Now apart from the explicit gluon line, the only off-shell external
lines of $\bar{\pdfbub}$ and $\bar{\jetbub}$ are the quark lines at the photon
vertex.  So an application of a Ward identity to the contraction of
$l_2$ with these factors, summed over graphs, takes the gluon line
outside of $\bar{\pdfbub}$ and $\bar{\jetbub}$, to give
\begin{align}
\sum_\Gamma T_{\rm S}\Gamma^{(R)} & = 
   \frac{e_j^2 P_{\mu\nu}}{4\pi} 
   \int \frac{d^4l_{2}}{(2\pi)^4} \, 
   \int \frac{d^4k}{(2\pi)^4} \, 
{\rm Tr} \left[ {\cal P}_{\rm J} \gamma^{\nu} {\cal P}_{\rm J} ~ \jetbub(k+q-l_2)  ~
                {\cal P}_{\rm T} \gamma^{\mu} {\cal P}_{\rm T} ~
                \pdfbub(k,P) \right] 
      \frac{ -g^{2} C_F \, n_{\rm J}^\kappa J_{g;\kappa\rho}(l_{2}) n_{\rm T}^\rho }
             { (l_{2} \cdot n_{\rm J} - i\epsilon) \, (l_{2} \cdot n_{\rm T} + i\epsilon) }.
\label{sigsimple4}
\end{align}
Notice how the bubbles are replaced by those that were used in the
parton model, in Eq.\ (\ref{pm2}), but with the momentum in $\jetbub$
shifted by $-l_2$.  Also, there is an overall minus sign, and a factor
$C_F$ that arises from the color matrices on the quark lines.  
As for the Ward identity, its derivation is well-known, for the case
of the model Abelian gauge theory that we use at the moment --- see
e.g., \cite[p.\ 339]{Sterman}.  We review the derivation in
App.~\ref{sec:WI} in the context of graphs such as Fig.~\ref{soft} and
Eq.~(\ref{sigsimple4}).
Complications arise in the non-Abelian case which does not
yet have a complete treatment.
We postpone the issue of the corresponding derivation in a non-Abelian
theory (i.e., QCD) to later work.

\begin{figure}
\centering
\epsfig{file=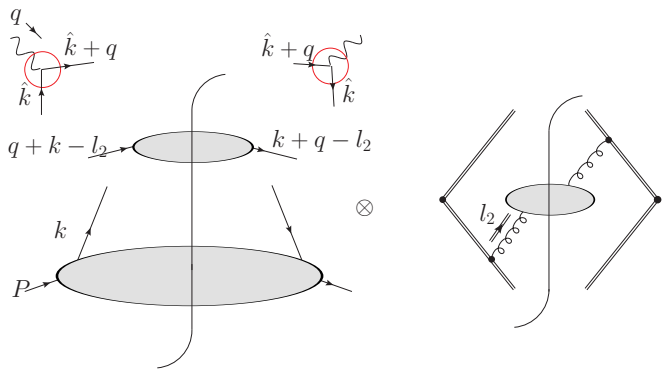,scale=0.45}
\caption{Factorization of one soft gluon. }
\label{softfactorized}
\end{figure}%

Finally we rewrite this result using the notation we used for the case
of the pure parton model:
\begin{equation}
\label{sigsimple7}
    \sum_\Gamma T_{\rm S} \Gamma^{(R)}
= \frac{1}{2 \pi} \int \frac{d^{4} l_{2}}{(2 \pi)^{4}} 
  \int \frac{d^4k}{(2 \pi)^{4} (-q^+q^-) } 
         \big| \hardfact_0(q,\hat{k}) \big|^2 \, \pdfbub^+(k,P) \, \jetbub^{-}(k+q-l_2) \,
 \soft^{(R,1)}(l_{2},y_{\rm T},y_{\rm J}) .
\end{equation}
\end{widetext}
which we notate diagrammatically in Fig.\ \ref{softfactorized}.  The
indices ``$+$'' on $\pdfbub^+$ and ``$-$'' on $\jetbub^-$ now denote, not Lorentz
indices
for a gluon, but the same projections concerning the leading part of the
Dirac matrix structure that we used in the parton model in Sec.\
\ref{sec:basic.approx}.  We define the one-loop real-gluon
contribution to the soft factor as
\begin{equation}
\label{softcont}
  \soft^{(R,1)}(l_{2},y_{\rm T},y_{\rm J})
= 
      \frac{ -g^{2} C_F \, n_{\rm J}^\kappa J_{g;\kappa\rho}(l_{2}) n_{\rm T}^\rho }
           { (l_{2} \cdot n_{\rm J} - i\epsilon) \, (l_{2} \cdot n_{\rm T} + i\epsilon) } ,
\end{equation}
and as indicated in Fig.\ \ref{softfactorized}, this is obtained from
the Feynman rules from our definition of the soft factor in Eq.\
(\ref{softdef}).  The lowest-order hard-scattering factor 
squared $\big| \hardfact_0 \big|^2$ is
given by Eq.\ (\ref{eq:H0}), so it is
\emph{exactly} the same as we found in the parton model
approximation.  This is necessary
if factorization is to hold; the hard scattering does not depend on
how many gluons attach to Wilson lines in the soft factor or on
their momenta.

\subsection{Accuracy and limits of soft approximation}
\label{accuracy}

After application of a Ward identity, we get a simple soft factor in
Fig.\ \ref{softfactorized}, with one graph for the $l_2$-dependent
soft factor.  However, depending on the precise size and rapidity of
$l_2$, different kinds of graph will dominate in Fig.\ \ref{soft},
before the Ward identities are used.  
This leads to complications if 
one wishes to set up factorization by considering individual graphs 
with on-shell final state partons; as we will now show, different 
types of graphs dominate depending on how soft the radiated gluon is.  
Therefore, the unified treatment
of the soft region discussed in the previous subsection, with all 
graphs implicitly included in the final state bubbles, has the 
notable advantage of dealing with all types of soft gluon behavior at
once.

\begin{figure*}
  \centering
  \begin{tabular}{c@{\hspace*{5mm}}c}
    \multicolumn{2}{c}{
        \includegraphics[scale=0.4]{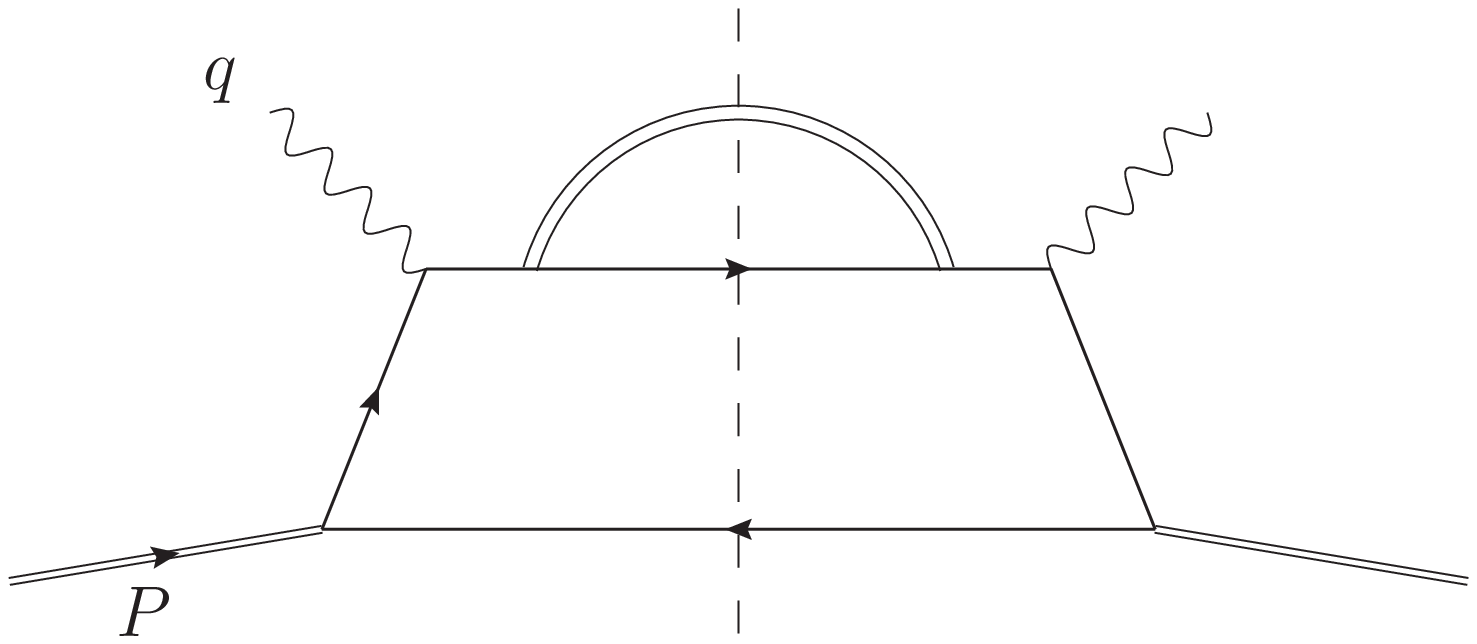}
     }

  \\
     \multicolumn{2}{c}{(a)}
  \\[5mm]
     \includegraphics[scale=0.45]{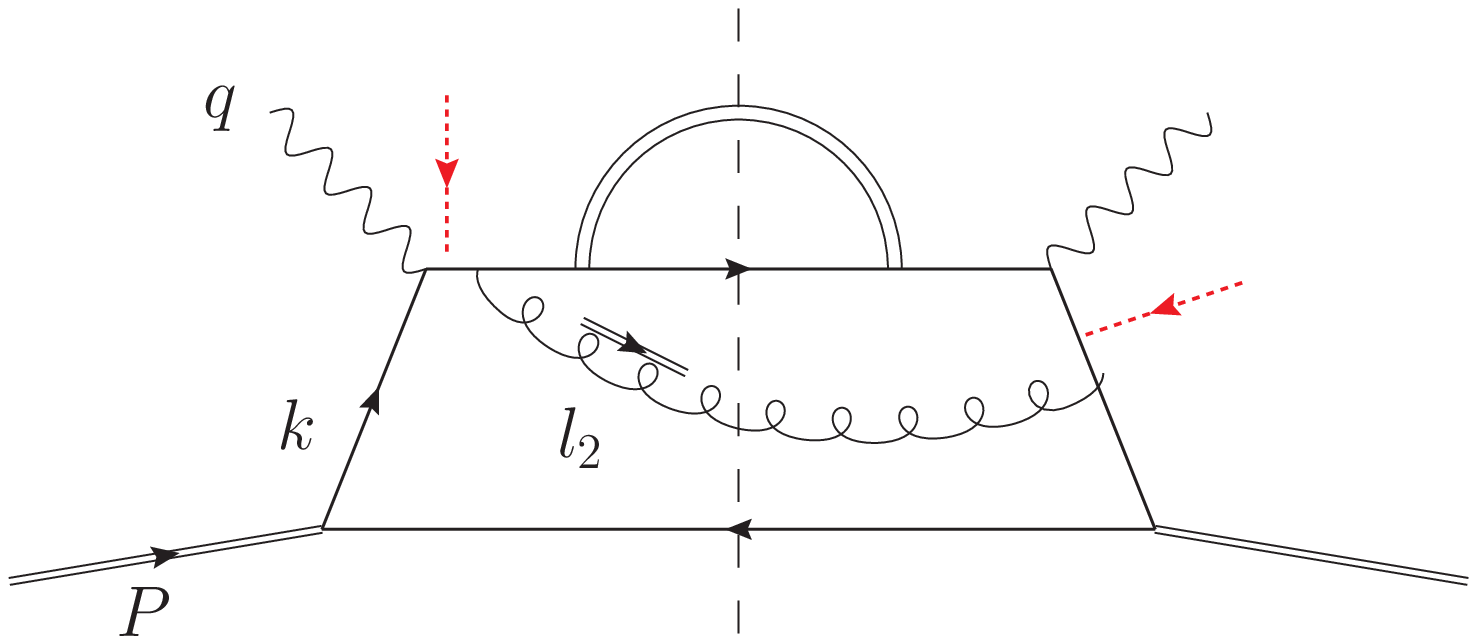}&
     \includegraphics[scale=0.45]{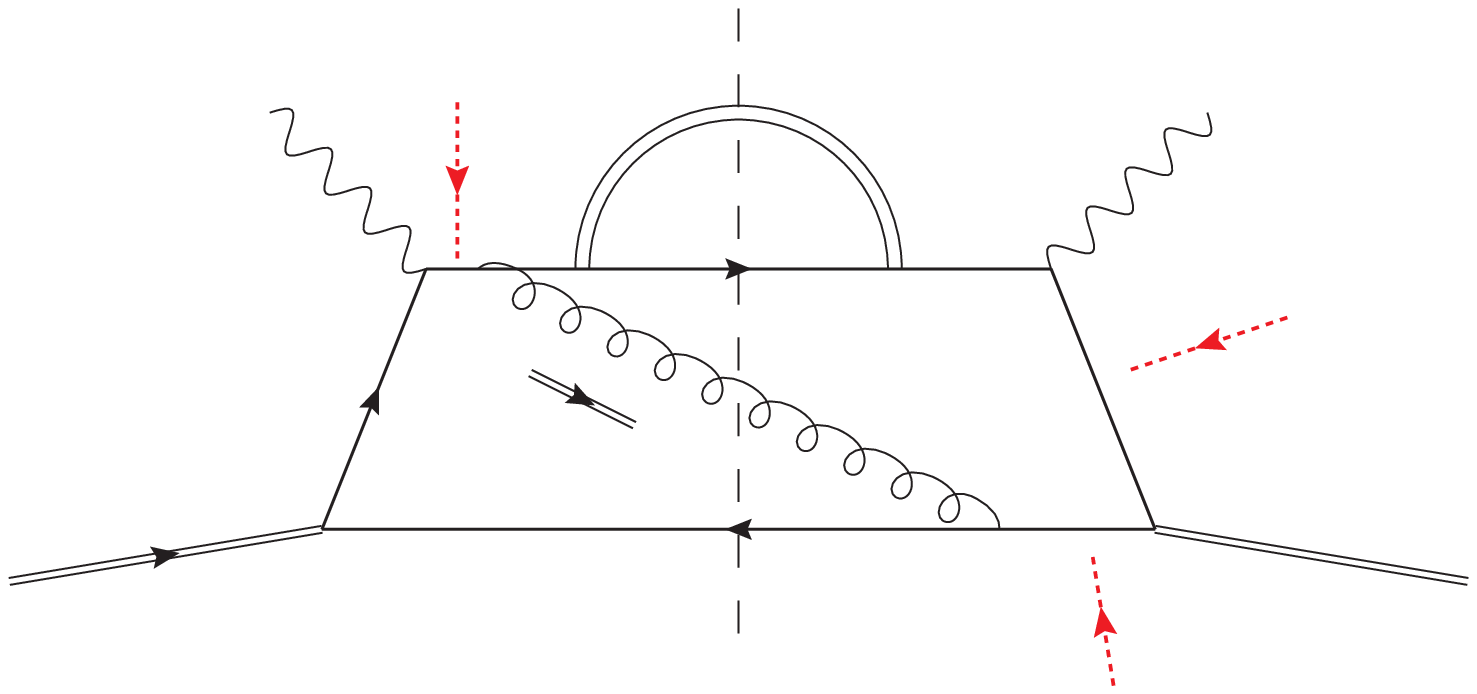}
  \\
     (b) & (c) 
  \\[5mm]
     \includegraphics[scale=0.45]{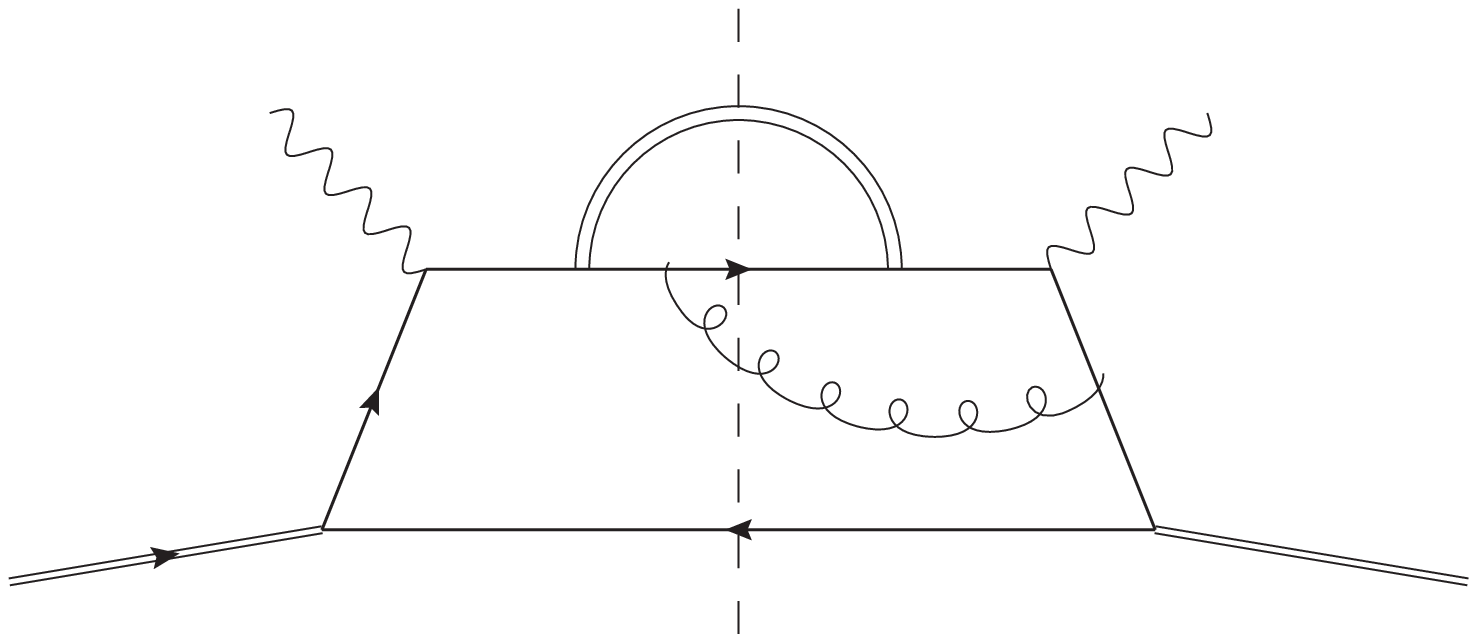}&
     \includegraphics[scale=0.45]{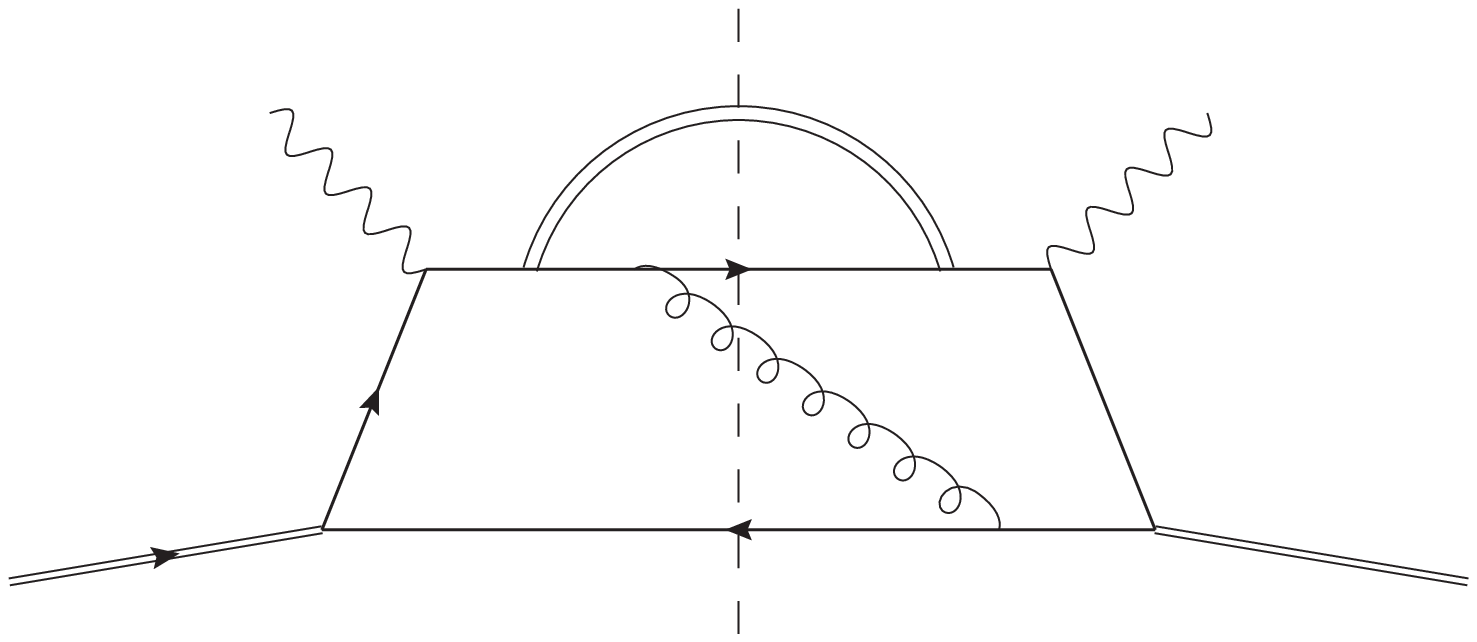}
  \\
     (d) & (e)
  \end{tabular}
  \caption{Illustrating the different detailed leading regions for an
      exchanged gluon in a spectator model.  The double lines
      are for a color-singlet scalar field treated as an analog of a
      hadron.  
      The dashed arrows in graphs (b) and (c) indicate the lines that
      are pushed 
      far off-shell by the inclusion of the extra gluon.
  } 
  \label{fig:model}
\end{figure*}

We illustrate this by examining the graphs in Fig.\ \ref{fig:model} in
a model field theory.  For this we use an Abelian gauge theory
supplemented by a color-singlet scalar field, which we treat as
describing the model's hadrons, and which we use for the measured
initial- and final-state particles.  By using an Abelian gauge theory,
we are allowed a gluon mass $m_g$, which can be zero or nonzero; we
use this to conveniently illustrate the issues associated with the
mass of the gluon.

First, without gluons we have graph (a), which gives a parton-model
description of DIS with a parton density and a fragmentation function
to describe an observed hadron in the jet --- the target splits into a
quark and an antiquark, each with a large plus component of momentum,
and the incident virtual photon scatters from the quark.  

Then we add one extra emitted real soft gluon.  To keep the discussion
simple, we restrict our attention to the situation where the rest of
the final state is near the parton model region.  That is, in the
final state, the transverse momenta (in the Breit frame) are limited
to at most order $\Lambda$, and the hadron and quark fields have masses of
order $\Lambda$.  The struck quark has a plus momentum of order $Q$, and
minus and transverse components less than or of order $\Lambda$.  Later we
will weaken these conditions.

There are graphs in which the gluon attaches only to the target or jet
parts of the graph; these simply provide corrections to the parton
density and to the fragmentation.  There remain 4 graphs of interest,
(b) to (e) in Fig.~\ref{fig:model} (plus their 
Hermitian conjugates).

We first treat the case that the extra emitted gluon has small
rapidity and has transverse momentum of order $\Lambda$.  Then it pushes
some of the quark lines off-shell.  For example, in graph (b), 
these requirements and the other 
requirements listed above on the spectator antiquark and the outgoing hadron-quark
system
imply that $(k+q)^{2}$ and $(k - l_{2})^{2}$ are of order $\Lambda
Q$:
\begin{widetext}
\begin{align}
\label{eq:denom.b}
  \frac{1}{ (q+k)^2 -m^2 } 
& = \frac{1}{(q + k - l_{2})^{2} - l_{2}^{2} - m^{2} + 2 l_{2} \cdot (k+q) }
 \sim \frac{1}{\Lambda Q}, 
\\
  \frac{1}{ (k-l_2)^2 -m^2 }
 &= \frac{1}{ k^2-m^2 + l_2^2 - 2 l_2 \cdot k }
 \sim \frac{1}{\Lambda Q}. 
\end{align}
\end{widetext}
Here we have used the fact that $(k+q)^- \sim Q$ and $k^+ \sim Q$.
Although in the upper part of the graph we have routed $l_2$ towards
the final state on the quark line of momentum $q+k-l_2$, it is the
line $q+k$ that goes off-shell.  The two lines listed above are the
only such far off-shell propagators in graph (b), so that the two
$1/Q$ factors are compensated by a factor of $Q^2$ from the numerator.
Thus there is no suppression of graph (b)

In contrast, there are at least three 
far
off-shell propagators in graphs
(c) through (e).  
Therefore, graph (b) dominates, and the others are
power suppressed.  In graph (c), the lines that are far off-shell are
indicated with 
dashed arrows.  

As long as the mass of the gluon is of order $\Lambda$, and the transverse momenta
in the target and in the jet system are also of this magnitude then we
can regard graph (b) as the only important graph at this order and we
are finished.  

But if 
we break with the requirements listed above by allowing the gluon 
mass to approach zero,
then the transverse momentum
of the gluon can go much smaller while keeping leading power
contributions 
in graphs (c) through (e).   
Alternatively, we could raise the transverse momentum
in the target and/or jet systems, for example, to order $\sqrt{\Lambda Q}$,
while still preserving the essential collinearity.  
In these situations the relative importance of 
graphs (b) through (e) in Fig.~\ref{fig:model} changes.

Let us use a zero gluon mass and continue to keep its rapidity
central.  When its transverse momentum is reduced to about $\Lambda^2/Q$,
there is no longer a penalty from 
off-shell propagators: all the
denominators are of order $\Lambda^2$.  When the gluon's transverse momentum
is reduced much further, graph (b) becomes unimportant, because the
decreasing denominators in (\ref{eq:denom.b}) plateau at order $\Lambda^2$.
For such supersoft gluons to contribute we must have denominators that
continue to decrease.  This happens only for graph (e), where the
gluon attaches directly to final-state colored lines.  This is in fact
just the ordinary IR divergence that 
appears when a massless gauge
boson is emitted from an outgoing on-shell fermion.  In a suitable sum
over final states, the ordinary IR divergences from vanishing
transverse momentum cancel against the graphs for virtual gluon
emission.

All of these cases are covered by our general Ward identity argument.
We just let a soft gluon couple in all possible ways to the collinear
subgraphs.  Some of these are smaller than others, but that does not
matter.

\begin{figure*}
  \centering
  \psfrag{ln kT}{$\ln(l_2^T/\Lambda)$}
  \psfrag{ln Q}{$\ln(Q/\Lambda)$}
  \psfrag{ln Q}{\llap{$l_2^T\sim Q$}}
  \psfrag{-ln Q}{\llap{$l_2^T\sim\Lambda^2/Q$}}
  \psfrag{y}{$y$}
  \psfrag{yT}{}
  \psfrag{yJ}{}
  \begin{tabular}{c@{\hspace*{5mm}}c}
      \hspace*{1cm}
      \includegraphics[scale=0.8]{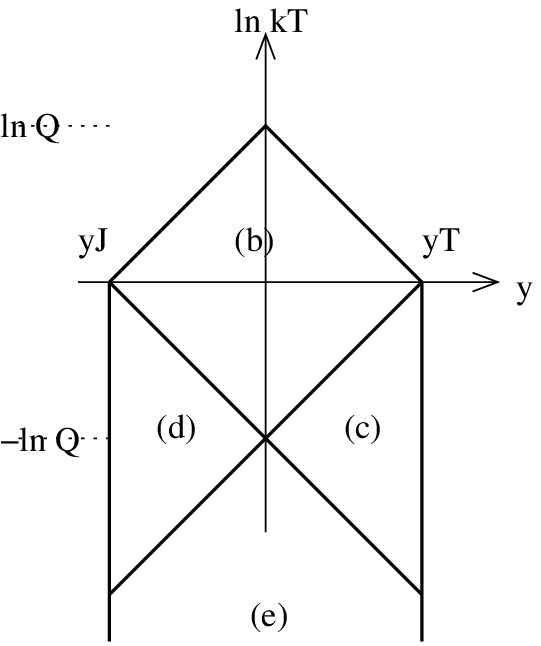}
      \hspace*{1cm}
  & 
      \includegraphics[scale=0.8]{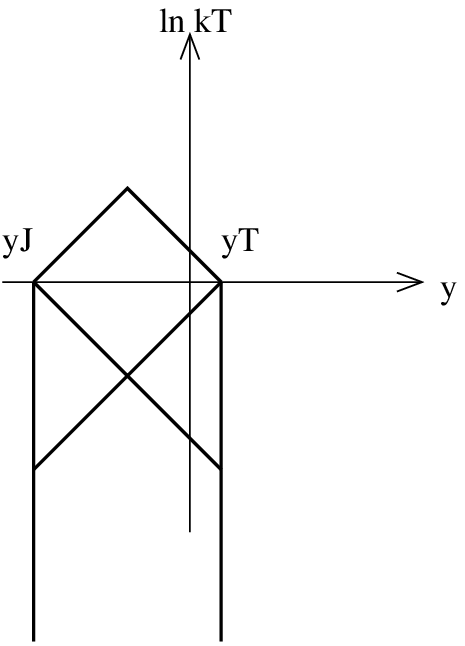}
  \\ 
     (i) & (ii)
  \end{tabular}
  \caption{(i) Ranges of gluon transverse and rapidity where the different
    graphs in Fig.\ \ref{fig:model} dominate. We use a logarithmic
    scale for $l_2^T$.  The upper diagonal lines correspond to where the
    energy of the gluon in the Breit frame is of order $Q$, and the
    vertical lines correspond to the rapidities of the jet and the
    target. 
    (ii) The case that the gluon attaches to the target remnant and
    that $x_{\rm Bj}$ is close to unity.}
  \label{fig:soft.split}
\end{figure*}

These results can be generalized to include also collinear gluons,
i.e., of large rapidity.  This gives Fig.\ \ref{fig:soft.split}(i),
where we plot the regions in gluon transverse momentum and rapidity
where the different graphs dominate.  
The figure is labeled by letters that indicate which of the 
single gluon radiation graphs 
in Fig.\ \ref{fig:model} dominate in different kinematical regimes. 
Naturally, on a boundary between
the regions the graphs associated with both sides of the boundary are
important.  The soft region corresponds to the part of the graph near the
vertical axis, the target-collinear region to the part at positive
rapidity, and the jet-collinear region to the part at negative
rapidities.

We now describe Fig.\ \ref{fig:soft.split}(i) in more detail.
We start by
describing the top diamond-shaped area where graph (b) gives an important
contribution.  For a massless real gluon in the final state, we may
write its momentum as
\begin{equation}
\label{bound1}
  l_2 = \left( 
           e^y \, \frac{|l_{2,\tran}|}{\sqrt{2}},  ~
           e^{-y} \, \frac{|l_{2,\tran}|}{\sqrt{2}}, ~
           {\bf l}_{2,\tran} 
       \right).
\end{equation}
Then the condition that $l_2^+/Q \lesssim 1$ gives one bound on the rapidity
for the area where (b) contributes: $y \lesssim \ln (Q/l_2^T)$.  Likewise,
the condition $l_2^-/Q$ gives a bound $y \gtrsim - \ln (Q/l_2^T)$.  These
bounds give the top two diagonal lines for the boundary of the
graph-(b)-dominant area in Fig.\ \ref{fig:soft.split}(i).

Furthermore, the propagator $q+k-l_2$ should have virtuality at least
of order $\Lambda^2$, otherwise we gain by going to graph (d) with the gluon
attaching directly to the final-state quark.  This gives a bound $y \gtrsim
-\ln(Ql_{2,\tran}/\Lambda^2)$, the lower left edge of the diamond.  The
corresponding bound, $y \lesssim \ln(Ql_{2,\tran}/\Lambda^2)$, on the target side gives
the remaining side of the diamond.  When $l_2$ strongly violates any
of these bounds, there is a power-law suppression relative to the
largest contribution.

Next we obtain the triangular area where graph (c)
becomes important.
For graph (c), the upper end of the gluon line still connects to an
internal quark line, so the bound $y \gtrsim -\ln(Ql_{2,\tran}/\Lambda^2)$ also applies
to graph (c)'s area, i.e., the continuation of the lower left boundary
of the region for graph (b).  But on the target side, we need $y \gtrsim
\ln(Ql_{2,\tran}/\Lambda^2)$, to avoid gaining two far off-shell propagators.  In
addition, there is a suppression of the contribution whenever $y$ is
bigger than the target rapidity $y_T$, since then the previous
dominance of the $2l_2^-(P^+-k^+)$ term in the $P-k+l_2$ propagator is
cutoff by the term $2l_2^+(P^--k^-)$.  This gives the right-hand
vertical line bounding the area for graph (c).

Exactly similar considerations give the area where graph (d) is
important.

The bounds relative to target and jet rapidity, $y_J \lesssim y \lesssim y_T$,
actually apply to all the graphs, but are automatically implied by the
other bounds except for two cases: the target rapidity bound for graph
(c) and the jet rapidity bound for graph (d).

Finally, in graph (e), the gluon connects to both of the final-state
quarks.  Whenever $l_2$ goes far outside the bounds given by the
lowest two diagonal lines in Fig.\ \ref{fig:soft.split}(i), pairs of
lines on the target or jet side are made off-shell by much more than
$\Lambda^2$, and we get a suppression.  In addition, the rapidity is
restricted to be between the jet and target rapidities.  Therefore,
graph (e) dominates in the area beneath the wedge shape at the bottom
of Fig.\ \ref{fig:soft.split}(i).

Notice that if the gluon had a mass of order $\Lambda$, there would be a
suppression of the region with $k_\tran$ much less than $\Lambda$.  Then graph
(b) would dominate in the soft region, and graphs (c) and (d) would
only contribute in the target- and jet-collinear regions, while graph
(e) would always be power suppressed.  This shows some of the
essential complications caused by the masslessness of the gluon in
Feynman graph calculations, even though the regions of supersoft gluon
momenta are presumably cut off by non-perturbative confinement in real
QCD.

Our definition of the soft approximation, adapted \cite{JCC.Sud} from
the Grammer-Yennie \cite{Grammer:1973db} paper, is arranged to avoid
any need to discuss the details of which graphs are important in
different ranges of soft $l_2^T$.  The only approximations are made in
the numerator coupling each jet factor to the soft gluon $l_2$.  These
approximations become $100\%$ inaccurate when either the energy of the
gluon is of order the jet energy or when the rapidity of the gluon is
comparable to the jet rapidity.  For the coupling to the target, the
magnitude of the fractional error is then the maximum of $l_2^+/Q$,
$\Lambda/Q$, and $e^{-(y_{\rm T}-y)}$, where $y_{\rm T}$ is the target
rapidity.  For the coupling to the jet, the fractional error is
similarly the maximum of $l_2^-/Q$, $\Lambda/Q$, and $e^{-(y - y_{\rm J})}$.
(The calculation is to be done in the Breit frame, and we have assumed
$x_{\rm Bj}$ is not close to unity, otherwise the limits are decreased
for accuracy of the coupling to $\bar{\pdfbub}$.)

These error estimates are arranged in a form that is equally suitable
for the case that $l_2$ is collinear to the jet or target.  
For the
soft approximation, we need to apply all the error estimates
simultaneously and then we find its fractional error is the maximum of
$\lambda/Q$, $\Lambda/Q$, $e^{-(y_{\rm T}-y)}$, and $e^{-(y - y_{\rm J})}$,
where $\lambda$ is defined by Eq.\ (\ref{eq:S.scale}).
That
is the soft approximation is valid when $\lambda\ll Q$, $e^{y_{\rm T}} \ll e^y \ll
e^{y_{\rm J}}$, and of course when $\Lambda$ and the transverse momentum
scale of the jets is much less than the hard scattering scale $Q$.
Thus the gluon has to have low momentum and have central rapidity.

Observe that a complication ensues when $x_{\rm Bj}$ gets close to
unity.  
For small or moderate $x_{\rm Bj}$, we expect the outgoing jet to have
large positive rapidity as in Fig.~\ref{fig:soft.split}(i).  But, for
$x_{\rm Bj} \sim 1$,
the  target remnant has low energy and rapidity, so that
the range of applicability of the soft region is very restricted on
the target side, as shown in Fig.\ \ref{fig:soft.split}(ii): The target
region
has a wide range of rapidity.

\subsection{Target-Collinear Region}
\label{sec:is.coll}

Now that we have characterized the smallest region, a treatment of the 
collinear regions follows naturally 
with
the aid of the subtraction
method described in Sect.~\ref{sec:sub}.
For the contribution of the 
target-collinear gluons to a graph,
such as the single gluon emission graph of Fig.~\ref{soft}.
we now construct an approximator, $T_{\rm T}$. To avoid double counting, the subtraction
formalism requires the contribution of the 
target-collinear
region to be obtained by applying $T_{\rm T}$ to the graph minus its
soft approximation:
\begin{equation}
C_{\rm T}\Gamma^{(R)} = T_{\rm T} (\Gamma^{(R)} - T_{\rm S} \Gamma^{(R)}). 
\label{coll}
\end{equation}

At the top end of the $l_2$ gluon, it attaches to a subgraph where the
rapidity is much more negative, when the gluon is in the
target-collinear region.  All the same issues about leading
polarizations apply as for the case that the gluon is soft, so at this
end of the gluon the approximator is very similar to the soft
approximator $T_{\rm S}$.

As we discussed earlier, there are two ways we could define the
target-collinear regions.  One is that the gluon has low transverse
momentum (e.g., of order $\Lambda$) and its plus momentum is of order $Q$ in
the Breit frame.  In the case of the gluon-exchange graphs in
Fig.\ \ref{fig:model} only graphs (b) and (c) are then important;
graphs (d) and (e) are suppressed because they have extra off-shell
propagators.  There is one far off-shell line: the quark $k+q$ between
the gluon attachment to the upper subgraph and the hard vertex has
virtuality of order $Q^2$.  We could correctly consider this line as
part of the hard subgraph.  (Note that although $l_2$ is not
routed through this line, we have imposed a low-mass requirement on
the hadron-quark part of the final state, immediately to its right.
So the large plus momentum of $l_2$ actually flows on the $k+q$ line.)

However, when the gluon is massless, its transverse momentum can be
very small, and then it can be useful to define the target-collinear
regions by the gluon rapidity being comparable to the target rapidity,
$e^y \sim e^{y_{\rm T}}$.  As shown in Fig.\ \ref{fig:soft.split}(i),
which graphs contribute depends on exactly how small the transverse
momentum is.  In all cases, the same method of approximation applies
for the coupling of the gluon to the upper subgraph $\bar{\jetbub}$.  The
Ward identity argument for summing the contributions will work; it
will extract the gluon from the upper subgraph and convert to
couple to a Wilson line independently of which 
of 
the graphs are
involved. 

In view of these issues, it is not totally obvious where to apply the
projection on the Dirac matrices.  So we formulate the approximator
in two stages.  The first stage just involves writing a
Ward-identity-compatible form for the coupling to $\bar{\jetbub}$, just as
in Eq.\ (\ref{grammer}), to obtain
\begin{widetext}
\begin{align}
T_{\rm T} \Gamma^{(R)}\!
   = \frac{ e_j^2 P_{\mu\nu} }{ 4\pi }
     \int \frac{d^4 l_{2}}{(2\pi)^4}  \int \frac{d^4 k}{(2\pi)^4} 
   {\rm Tr}
   \left[\gamma^{\nu} l_{2} \cdot \bar{\jetbub}(k+q,l_{2}) \gamma^{\mu} \bar{\pdfbub}^{\rho}(k,l_2,P) \right] 
    \frac{n_s^\kappa}{l_{2} \cdot n_s -i\epsilon}
    J_{g;\kappa\rho}(l_{2}). 
\label{collapprox}
\end{align} 
Here we use the vector $n_s$ of 
nearly 
zero rapidity, rather than the
vector $n_{\rm J}$ of very negative rapidity that we used in the
corresponding part of the soft approximation.  This provides a cutoff
at central rapidities, as appropriate for a collinear region.  When
$l_2$ has a large positive rapidity, both vectors agree in projecting
out the plus component: $l_2\cdot n_s \simeq l_2\cdot n_{\rm J} \simeq l_2^+$.

Again, as before, we apply a Ward identity, but now only on the jet side,
to obtain
\begin{align}
\sum_\Gamma T_{\rm T} \Gamma^{(R)} 
   = \frac{P_{\mu\nu}}{4\pi} \int \frac{d^4k}{(2\pi)^4}  \int \frac{d^4l_2}{(2\pi)^4}
   {\rm Tr} \left\{ \gamma^{\nu} \jetbub(k+q-l_2) \gamma^{\mu} 
   \left[
   \frac{ g \, n_s^\kappa J_{g;\kappa\rho}(l_{2}) }
        {l_2 \cdot n_s -i\epsilon}
    \bar{\pdfbub}^\rho(k,l_2,P)
 \right] \right\}.
\label{eq:coll3}
\end{align}
We identify the square bracket factor in Eq.~(\ref{eq:coll3}) as a
contribution to the numerator of the PCF defined in Eq.~(\ref{pcf2}) with the gluon
coupling to the Wilson line.  The remaining factor in the integrand is
exactly the same as the jet factor in the ordinary handbag diagram,
Eq.\ (\ref{pm2}), aside from a different labeling of the loop
momenta.  

\begin{figure}
\centering
    \epsfig{file=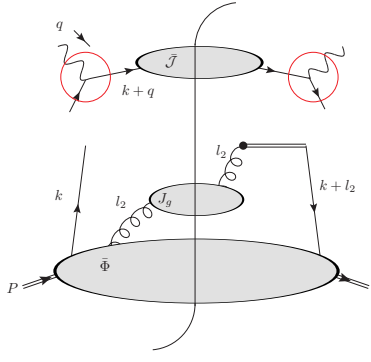,scale = 0.55}
\caption{The approximated graph of Eq.~(\ref{eq:coll3}). }
\label{initcoll}
\end{figure}

The definition of the collinear approximator is completed by applying
the parton-model approximator at the electromagnetic vertex, as
illustrated in Fig.\ \ref{initcoll}.  We also make a shift of
integration variable, replacing $k$ by $k-l_2$, so as to make the
correspondence with the simple handbag diagram clear.
\begin{equation}
\sum_\Gamma T_{\rm T} \Gamma^{(R)} 
= \frac{1}{2\pi} 
   \int \frac{d^4 k}{(2\pi)^4} 
   \frac{1}{-q^+q^-}
   \, \jetbub^{-}(k +q) \, \big| \hardfact_0(q,\hat{k}) \big|^2 
\left\{ \int \frac{d^4l_2}{(2\pi)^4}
        J_{g;\kappa\rho}(l_2) 
    \frac{ g \, n_s^\kappa }
         { l_2 \cdot n_s  - i\epsilon }
    {\rm Tr} \left[ \frac{\gamma^+}{4} \bar{\pdfbub}^\rho(k + l_{2},l_2,P) \right]
\right \}.
\label{eq:coll5}
\end{equation}
Here $\jetbub^-$ 
and $\big| \hardfact_0 \big|^2$ are exactly the same jet factor and hard
scattering coefficient as in the simplest handbag diagram 
Fig.~\ref{LOdiags}(a), and the
factor in braces corresponds to the numerator in the definition of the
PCF Eq.~(\ref{pcf2}) with one gluon connecting to the Wilson line.  

Finally we need the double-counting-subtraction term in
Eq.~(\ref{coll}).  The new item needed is the result of first applying
the soft approximator $T_{\rm S}$ and then the collinear approximator
$T_{\rm T}$.  The collinear approximator simply modifies the upper
vertex of $l_2$ by replacing the vector $n_{\rm J}$ by the vector $n_s$:
\begin{align}
\label{eq:TS}
  T_{\rm T} T_{\rm S} \bar{\jetbub}^\kappa(k+q,l_2)
& = 
  T_{\rm T}  \bar{\jetbub}(k+q,l_2) \cdot l_2 
         ~ \frac{ n_{\rm J}^\kappa } { l_2 \cdot n_{\rm J} -i\epsilon }
\nonumber\\
& = 
  \bar{\jetbub}(k+q,l_2) \cdot l_2 
   ~ \frac{ n_{\rm J} \cdot l_2 } { l_2 \cdot n_{\rm J} -i\epsilon }
   ~ \frac{ n_s^\kappa } { l_2 \cdot n_s -i\epsilon }
\nonumber\\
 & = 
  \bar{\jetbub}(k+q,l_2) \cdot l_2 
   ~ \frac{ n_s^\kappa } { l_2 \cdot n_s -i\epsilon }
\end{align}
Applying Ward identities now gives us a term just like that for the
soft approximation Fig.\ \ref{softfactorized}, except with the change
from $n_{\rm J}$ to $n_s$.  However, we now must identify the eikonal factor
as part of the PDF.  

We therefore write the contribution of the target-collinear region as
\begin{equation}
\sum_\Gamma C_{\rm T} \Gamma^{(R)} \! 
=
 \frac{1}{2 \pi} \int \frac{d^4k}{(2\pi)^4}  \frac{1}{-q^+q^-}
   \, \big| \hardfact_0(q,\hat{k}) \big|^2 \, \jetbub^{-}(k+q)
    F_{(R,1)}(k,P), 
\label{eq:coll8}
\end{equation}
where
\begin{equation}
   F_{(R,1)}(k,P) 
   \equiv 
\int \frac{d^4l_2}{(2\pi)^4}
  \Biggl\{ 
        J_{g;\kappa\rho}(l_2) 
    \frac{ g \, n_s^\kappa }
         { l_2 \cdot n_s - i\epsilon }
    {\rm Tr} \left[ \frac{\gamma^+}{4} \bar{\pdfbub}^\rho(k,l_2,P)  \right]
  - 
  \pdfbub^+(k+l_2,P) 
  \soft^{(R,1)}(l_2,n_{\rm T},n_s) \Biggr\} .
\label{eq:coll7}
\end{equation}
We now recognize $F_{(R,1)}(k,P)$ as a part of the PCF due to
one-gluon exchange with the Wilson lines in its definition, Eq.\
(\ref{pcf2}).  The one-gluon exchange term is found from expanding the
Wilson line operators in powers of the coupling.  Then the first term
in (\ref{eq:coll7}), as we have already noted, is from the numerator
of Eq.~(\ref{pcf2}).  The second term arises from the $O(g^2)$ term
in the denominator.  In coordinate space, this multiplies the lowest
order term in the numerator.  Fourier transformation gives the
convolution product in the second term in Eq.\ (\ref{eq:coll7}), as
illustrated in Fig.~\ref{pdfsubtracted}.

These results support the
correctness of Eq.~(\ref{pcf2}) as the definition of the PCF. 

\begin{figure*}
\centering
$  F(k,P) =
   \raisebox{-0.5\totalheight}{\includegraphics[scale=0.55]{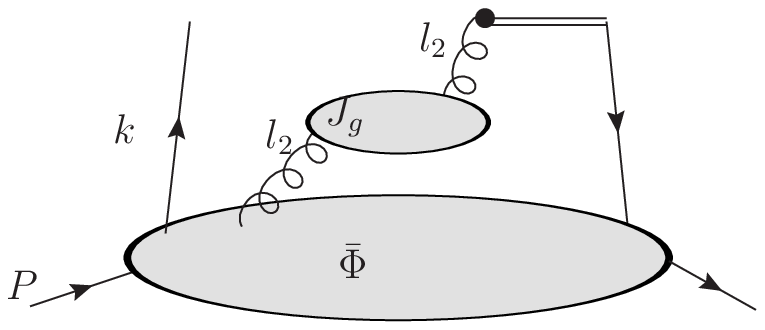}}
   ~-~ \raisebox{-0.5\totalheight}{\includegraphics[scale=0.55]{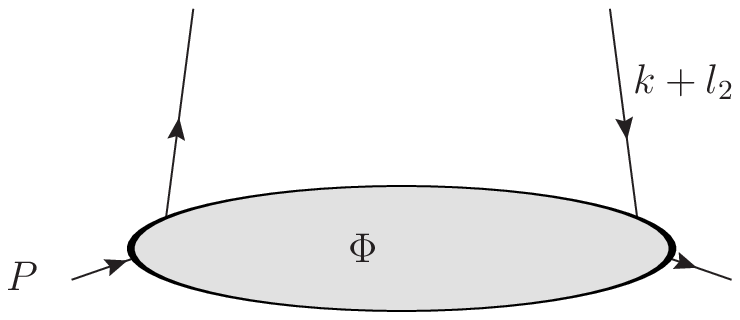}}
   \otimes \raisebox{-0.5\totalheight}{\includegraphics[scale=0.55]{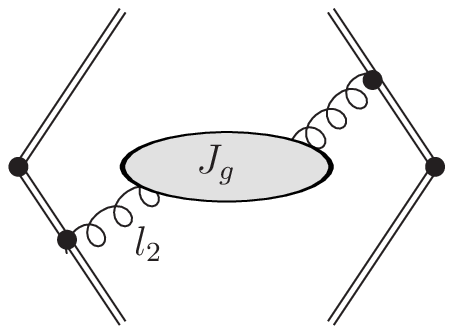}}
$
\caption{The target-collinear contribution for the gluon in Fig.\
  \ref{soft} gives this factor that is a contribution to the fully
  unintegrated PCF.  See Eq.~(\ref{eq:coll7}).} 
\label{pdfsubtracted}
\end{figure*}

\subsection{Jet-collinear region}
\label{sec:fs.coll}

The treatment of the remaining region, the jet-collinear region, is
very similar --- one follows the same steps as in the target collinear case.  
For completeness we state the result here.
The result for the contribution of this region is
\begin{equation}
\sum_\Gamma C_{\rm J} \Gamma^{(R)} 
\equiv \sum_\Gamma T_{\rm J} (\Gamma^{(R)} - T_{\rm S} \Gamma^{(R)})
=
   \frac{1}{2 \pi} \int \frac{d^4k}{(2\pi)^4}  
   \frac{1}{-q^+q^-}
   \big| \hardfact_{0}(q,\hat{k}) \big|^2
   \jet_{(R,1)}(k+q) \pdfbub^+(k,P), 
\label{eq:coll15}
\end{equation}
where the one-gluon-exchange contribution to the fragmentation PCF is
\begin{equation}
\jet_{(R,1)}(k+q) 
= 
    \int \frac{d^4l_2}{(2\pi)^4} 
    \Biggl\{ 
          {\rm Tr}\left[ \frac{\gamma^-}{4} \bar{\jetbub}^\kappa(k+q,l_2) \right]
         \frac{ -g J_{g;\kappa\rho}(l_2) n_s^\rho }{ l_2\cdot n_s -i\epsilon }
     - 
        \jetbub^-(k+q-l_2) \, 
        \soft^{(R,1)}(l_2,n_{\rm T},-n)
    \Biggr\}. 
\label{eq:coll16}
\end{equation}
In obtaining this by modifying the derivation for the target-collinear
region, we replaced $n_s$ by $-n_s$, in accordance with the results of
\cite{CM} and the discussion in sect.~\ref{sec:defns}.

\subsection{Hard Gluons}
\label{hard}

Although the issue of extracting the NLO hard scattering component is 
beyond the scope of this article, it is worth describing the
general method we would follow.
The NLO hard contribution from the single gluon emission diagrams will 
follow naturally in the subtraction formalism:
\begin{equation}
\label{hardsub}
\Gamma^{R, NLO}_{H}(k,l_{2}) = \Gamma^{(R)}(k,l_{2}) 
     - \left[ \Gamma^{(R)}_S(k,l_{2}) + \Gamma^{(R)}_T(k,l_{2}) + \Gamma^{(R)}_J(k,l_{2}) \right].
\end{equation}
\end{widetext}
Here, $\Gamma^{(R)}_{S,T,J}(k,l)$ represents the result of applying 
the soft, hard, or jet approximator to the unapproximated graph.

\section{Virtual corrections}
\label{sec:virt}

In this section we analyze in detail the diagrams with virtual
corrections for the case of one gluon attachment.

First, we note that all situations with gluons entirely confined to
the jet or target subgraphs are already covered by the basic
parton-model argument in Sec.\ \ref{sec:basic.approx}.  These gluons
are already included in what we mean by the jet and target subgraphs.

\begin{figure}
\centering
\epsfig{file=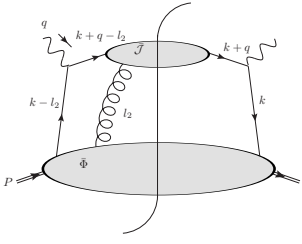,scale=0.45}
\caption{Virtual correction graph.}
\label{fig:virt_vc}
\end{figure}

Therefore we need to analyze graphs with a gluon exchanged between the
jet and target subgraphs, Fig.~\ref{fig:virt_vc}.  The general
procedure follows that for real gluon emission.  The expression for
the unapproximated graph is
\begin{multline}
  \Gamma^{(V)}_{\rm vc} =   \frac{e_j^2 P_{\mu\nu}}{4\pi}     
  \int \frac{d^4l_2}{(2\pi)^4} \int \frac{d^4k}{(2\pi)^4}  \,
\\\times 
   {\rm Tr}\! \left[ \gamma^{\nu} \,  \bar{\jetbub}^{\kappa}(k+q,l_2) \,
                   \gamma^{\mu} \, \bar{\pdfbub}^{\rho}(k,l_2,P) \right]
   G_{\kappa\rho}(l_2) \, ,
\label{eq:virt_vc_1}
\end{multline}
where $\bar{\pdfbub}^{\rho}$ and $\bar{\jetbub}^{\kappa}$ are the bubbles with one extra
gluon attached.  These bubbles are the same as for real gluon
emission, \emph{except that the gluon is on the left of the
  final-state cut in both bubbles}. The subscript `${\rm vc}$' means
virtual correction.  This formula is just like (\ref{eq:sigma2})
except that the cut dressed gluon propagator is replaced by an uncut
propagator:
\begin{equation}
J_{g;\kappa\rho}(l_2)  \longrightarrow G_{\kappa\rho}(l_2).
\end{equation}

\subsection{Soft region}
We start with the smallest region, the soft one, which was
appropriately defined in Sec.~\ref{sec:real}.  Following the procedure
used for real-gluon emission, we find that the Grammer-Yennie method
applied to (\ref{eq:virt_vc_1}) gives
\begin{multline}
\Gamma^{(V)}_{\rm vc,soft} =
   \frac{e_j^2 P_{\mu\nu}}{4\pi}     \int \frac{d^4l_2}{(2\pi)^4} 
   \int \frac{d^4k}{(2\pi)^4}  \,
\\
   {\rm Tr}\! \left[ {\cal P}_{\rm J} \gamma^{\nu} {\cal P}_{\rm J} \,  \bar{\jetbub}(k+q,l_2) \cdot l_2 \,
                   {\cal P}_{\rm T} \gamma^{\mu} {\cal P}_{\rm T} \, 
                   \bar{\pdfbub}(k,l_2,P)\cdot l_2 \right] \times 
\\
\times \,    \frac{ n_{\rm J}^\kappa G_{\kappa\rho}(l_2) n_{\rm T}^\rho }
             { (n_{\rm J}\cdot l_2-i\epsilon)(n_{\rm T}\cdot l_2+i\epsilon) } .
\label{eq:virt_vc_2}
\end{multline}
which exactly corresponds to Eq.\ (\ref{softapprox}) for the
real-gluon case.  Unlike the real-gluon case, we now have to worry
about the Glauber region, $|l_2^+l_2^-| \ll l_{2,\tran}^2$, since the
integration over a virtual gluon momentum includes arbitrarily small
values of $l_2^+$ and $l_2^-$.  To obtain the soft approximation we
needed the assertions $l_2 \cdot \bar{\pdfbub} \simeq l_2^-\bar{\pdfbub}^+$ and $\bar{\jetbub} \cdot
l_2 \simeq \bar{\jetbub}^-l_2^+$.  The first assertion becomes invalid when
$l_2^-$ is too small, and the second assertion becomes invalid when
$l_2^+$ is too small.  The conditions for the validity of both parts
of the soft approximation can be deduced from Sec.\ \ref{sec:soft}:
\begin{gather}
\label{eq:soft.rap.cond}
 \frac{\Lambda^2}{(q^-)^2} \ll  \left| \frac{l_2^+}{l_2^-} \right| 
  \ll \frac{(P^+)^2}{\Lambda^2},
\\
\label{eq:soft.kt.cond2}
  \left| \frac{l_2^+}{l_{2,\tran}} \right| \gg \frac{\Lambda}{q^-},
\qquad
  \left| \frac{l_2^-}{l_{2,\tran}} \right| \gg \frac{\Lambda}{P^+}.
\end{gather}
These estimates assume that the transverse momentum in the jet and
target subgraphs are of order $\Lambda$.  The first pair of conditions
simply state that the rapidity of the gluon must be well inside the
range between the jet and target rapidities, as is natural for the
soft region.  The second pair of conditions are that the longitudinal
components of $l_2$ should not be too much smaller than the transverse
momentum; from them can be deduced that we need
\begin{equation}
\label{eq:soft.nonglauber}
    \left| l_2^-l_2^+ \right| \gg l_{2,\tran}^2 \frac{\Lambda^2}{Q^2}.
\end{equation}

A breakdown only of the conditions on the rapidity of $l_2$ simply
takes us to one of the collinear regions, and that need not concern us
here since we will treat the collinear regions separately.  However a
breakdown of the other conditions is problematic.  We see from
(\ref{eq:soft.nonglauber}) that such a breakdown brings us to the
Glauber region, i.e., to $|l_2^+l_2^-| \ll l_{2,\tran}^2$.  

\begin{figure}
\centering
  \psfrag{l2+}{$l_2^+$}
      \includegraphics[scale=0.4]{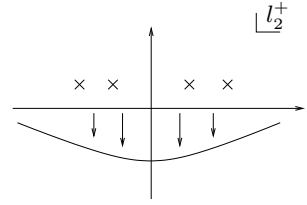}
  \caption{The deformation of the contour for the integral over
$l_2^+$ in Eq.\ (\ref{eq:virt_vc_2})}
  \label{fig:contourdef}
\end{figure}

As explained in \cite{CM,diff.hs}, we can apply a contour deformation
to get out of the Glauber region.  The contour deformation is to be
applied to $l_2^+$ only, since the only significant dependence on
$l_2^+$ in the Glauber region is in the jet subgraph.  All the
relevant singularities are final-state singularities, and are
therefore all in the upper half-plane; thus the same deformation works
for all graphs.  
This is shown graphically in the complex $l_2^+$ plane shown 
in Fig.~\ref{fig:contourdef} where the crosses represent the final
state poles.
An attempt to apply a corresponding argument to the
other longitudinal component $l_2^-$ would fail, because there can be
both initial- and final-state singularities for $l_2^-$ in the target
subgraph.  For the application of a Ward identity, it is essential to
have a single contour deformation applied to $l_2$ for every graph
that is summed by the Ward identity.  The choice of $i\epsilon$ in the
eikonal denominators, particularly $n_{\rm J}\cdot l_2-i\epsilon$, was determined by
compatibility with the contour deformation of $l_2^+$ away from
final-state poles in the jet subgraph.
Depending on the size of $l_2^+$, the contour deformation may take
$l_2$ to the conventional soft region or to the target-collinear
region.  In either case we have a situation for which we have an
applicable technique.

Now that we have deformed the contour integration out of the Glauber region, the
soft approximation is valid over the whole of the soft region, as
defined by the rapidity condition.  Therefore we can apply Ward
identities to the sum over graphs, just as with real-gluon emission,
to obtain
\begin{multline}
\Gamma^{(V)}_{\rm vc,soft} =   \frac{e_j^2 P_{\mu\nu}}{4\pi}
   \int \frac{d^4k}{(2\pi)^4}  \,
\\
   {\rm Tr} \left[ {\cal P}_{\rm J} \gamma^\nu {\cal P}_{\rm J} \,  \jetbub(k+q)  \,
                   {\cal P}_{\rm T} \gamma^\mu {\cal P}_{\rm T} \, \pdfbub(k,P) \right] \times \\
\times \,    \int \frac{d^4l_2}{(2\pi)^4}\, 
      \frac{ g^2\, C_F\, n_{\rm J}^\kappa G_{\kappa\rho} n_{\rm T}^\rho }
           { (n_{\rm J}\cdot l_2-i\epsilon) (n_{\rm T}\cdot l_2+i\epsilon) }  ,
\label{eq:virt_vc_3}
\end{multline}
where we have performed summation over colors and the strong coupling
constant has been taken out of the vertices $\bar{\jetbub}^{\kappa}$ and
$\bar{\pdfbub}^{\rho}$.  The soft gluon has factorized from the rest of the
graph.  Effectively the soft gluon only sees the total color charge
and direction of the target and jet lines, and has no sensitivity to
the details.

The expression in the last line of (\ref{eq:virt_vc_3})
\begin{multline}
  \tilde{\soft}^{\rm (V,1)}(y_{\rm T}-y_{\rm J}) 
\\ 
  =  g^2\, C_F \, \int \frac{d^4 l_2}{(2\pi)^4}
      \frac{ n_{\rm J}^\kappa G_{\kappa\rho} n_{\rm T}^\rho }
           { (n_{\rm J}\cdot l_2-i\epsilon) (n_{\rm T}\cdot l_2+i\epsilon) }  ,
\label{virtual_soft}
\end{multline}
is the lowest order term of the vacuum expectation value of two Wilson
lines, exactly corresponding to the relevant term in our definition of
the soft factor in Eq.\ (\ref{softdef}) when the intermediate state
between the left three Wilson lines and the right three is the vacuum
state.

The use of non-lightlike lines, in directions $n_{\rm T}$ and $n_{\rm J}$ cuts off
the rapidity divergences that would occur if light-like lines were
used \cite{Collins:1989bt}.  We could follow an alternative procedure,
suggested in \cite{Collins:1989bt} of using light-like Wilson lines,
but with extra generalized renormalization factors in the definition
of the soft factor to cancel the rapidity divergences.  But we will
not follow this idea here.

We do remark that that Eq.\ (\ref{virtual_soft}) does have a UV
divergence.  We define it to be removed by applying renormalization,
as usual; this is an ordinary UV divergence associated with the cusp
joining the two segments of the Wilson line.

\subsection{Target-collinear region}

The treatment of the target-collinear region for $l_2$ works exactly
as in the real emission case.  
But we must apply the approximators and the subtraction to the 
graph with the contour deformed (to avoid the Glauber region), in order 
for the double-counting subtraction for the soft region to be correct.
This requires that in the Grammer-Yennie
approximation for the collinear region we apply the $i\epsilon$ prescription
to the eikonal denominator that is compatible with the contour
deformation:
\begin{multline}
\gamma^{\nu} \, \bar{\jetbub}^{\kappa}(k+q,l_2) \, \gamma^{\mu} \, \bar{\pdfbub}^{\rho}(k,l_2,P)
\,G_{\kappa\rho}
\\
 \simeq \gamma^{\nu} \, \bar{\jetbub}(k+q,l_2)  \cdot l_2 \, \gamma^{\mu}
   \,\frac{n_s^{\kappa}G_{\kappa \rho} }{l_2 \cdot n_s-i\epsilon} \,  \bar{\pdfbub}^{\rho}(k,l_2,P) \; .
\end{multline}
This leads to a first form of the approximator:
\begin{multline}
T_{T1} \Gamma^{V} \; = \; \frac{e_j^2 P_{\mu\nu}}{4\pi} \, \int \frac{d^4k }{(2\pi)^4}
\int \frac{d^4 l _2 }{(2\pi)^4} 
\\
{\rm Tr}\! \left[\gamma^{\nu} \, \bar{\jetbub}(k+q,l_2)  \cdot l_2 \, \gamma^{\mu} \,  
     \bar{\pdfbub}^{\rho}(k,l_2,P) \right] 
\frac{ n_s^\kappa G_{\kappa \rho} }{ l_2 \cdot n_s-i\epsilon } \; .
\end{multline}
Naturally, we must use the same eikonal denominator for both real and
virtual gluon emission in order that all the contributions to the
target PCF arise from the same Wilson line.

The remaining steps follow exactly as for real emission: (i)
Subtraction of a soft term as in Eq.\ (\ref{coll}) to compensate
double counting with the smaller region.  (ii) Application of a
leading-power approximation to the hard scattering.  (iii) Use of Ward
identities.  The result is:
\begin{multline}
\sum_\Gamma C_{\rm T} \Gamma^{(V)} \! 
=
 \frac{1}{2 \pi} \int \frac{d^4k}{(2\pi)^4}  \frac{1}{-q^+q^-}
\\
   \, \big| \hardfact_0(q,\hat{k}) \big|^2  \, \jetbub^{-}(k+q)
    F_{(V,1)}(k,P) \,   ,
\label{eq:virt_coll8}
\end{multline}
with the 1-gluon virtual gluon contribution to the parton correlation
function defined as
\begin{align}
 {F}_{\rm (V,1)}(k,P)
={}& \int \frac{d^4 l_2}{(2\pi)^4} \, 
   \left[ \bar{\pdfbub}^{\rho}(k,l_2,P) \, 
          \frac{ -g \,C_F \,n_s^\kappa G_{\kappa \rho} }{ l_2\cdot n_s -i\epsilon}
   \right]
\nonumber\\
   & - \pdfbub^+(k,P) \, \soft^{\rm (V,1)}(y_{\rm T}-y_s)\; .
\end{align}
This is exactly the sum of the contributions caused by one virtual
gluon coupling to the Wilson line(s) in the target PCF.

\begin{figure}
\centering
\epsfig{file=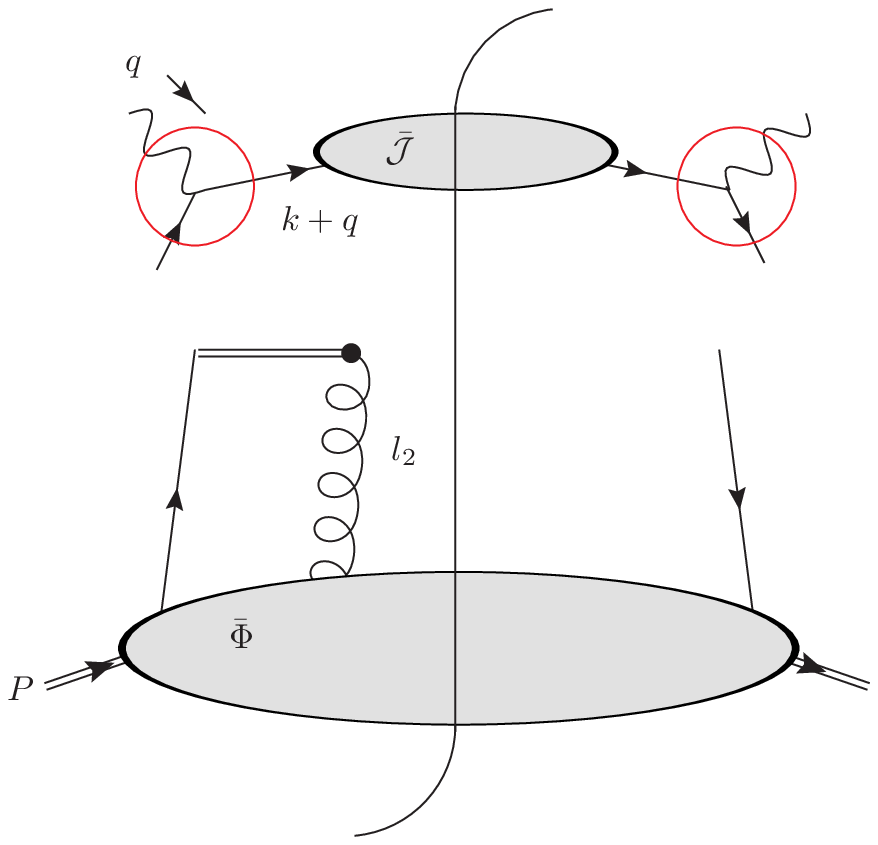,scale=0.5}
\caption{Graphical representation of the target-collinear contribution
  for one virtual collinear gluon, as in (\ref{eq:virt_coll8}). From
  this is 
  to be subtracted a soft term. }
\label{fig:initcoll_virt_fact}
\end{figure}

\subsection{Jet-collinear region}

The case in which the gluon is collinear to the outgoing jet can be
worked out in complete analogy with the previous cases.  The result is
\begin{multline}
\sum_\Gamma C_{\rm J} \Gamma^{(V)} \!
=
 \frac{1}{2 \pi} \int \frac{d^4k}{(2\pi)^4}  \frac{1}{-q^+q^-}
\\
   \, \big| \hardfact_0(q,\hat{k}) \big|^2 \, \jetbub_{(V,1)}(k+q)
    \pdfbub^+(k,P) \,   ,
\label{eq:virt_coll8_1}
\end{multline}
where
\begin{align}
 \jet_{\rm (V,1)}(k,P)  
 = {}& \int \frac{d^4 l_2}{(2\pi)^4} \, 
    \bar{\jetbub}^{\kappa}(k,l_2,P) ~
    \frac{ -g \, C_F \, G_{\kappa \rho}n_s^\rho }{ l_2\cdot n_s - i\epsilon}
\nonumber\\
  & - \jetbub^{-}(k,P) \, \soft^{(V,1)}(y_s-y_{\rm J})\; .
\end{align}
To summarize, a graphical depiction of the resulting factorization 
for the virtual contribution is shown in Fig.~\ref{fig:initcoll_virt_fact}.

\subsection{Hard vertex correction}
We have simplified our work by restricting to final states without
particles or jets of extra high transverse momentum.  Thus we did not
need to treat real corrections where extra partons are emitted from
the hard scattering.  We treated collinear and soft gluons, but not
hard gluons.

\begin{figure}[b]
  \centering
  \includegraphics[scale=0.5]{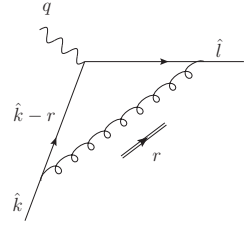}
  \caption{One loop correction to current vertex.}
  \label{fig:vc}
\end{figure}

For virtual corrections, the situation is different: loop momenta are
not restricted by the external states, and can be arbitrarily large.
However, in the graphs entailed by Fig.\ \ref{fig:virt_vc}, the only
one in which a hard gluon gives a leading-power contribution is in the
vertex graph of Fig.\ \ref{fig:vc}.  This graph is, of course, to be
considered as a particular subgraph for Fig.\ \ref{fig:virt_vc}.

So we now present the correction to the hard scattering coefficient,
associated with the vertex graph.  To it is 
to be added the complex
conjugate contribution from the same correction at the current vertex
on the right of the final-state cut.  Let us use $V$ to denote the
vertex graph.  Its contributions where the gluon is soft or collinear
have already been allowed for, so there remains only the contribution
from the hard region.  In accordance with our subtraction formalism,
this contribution is
\begin{widetext}
\begin{align}
\label{eq:vcH}
  C_H V = {}& T_H\left[ 1 - T_S - T_{\rm T} (1 - T_S) - T_{\rm J}(1 - T_S) \right] V
\nonumber\\
  = {}& T_H\left[ 1 - T_{\rm T} - T_{\rm J}  - (1 - T_{\rm T} - T_{\rm J}) T_S \right] V
\nonumber\\
  ={}& \frac{ -ig^2 }{ (2\pi)^4 } \int d^4r \frac{1}{r^2+i\epsilon} \,
      {\cal P}_{\rm T} 
      \Bigg\{ 
         \frac{ \gamma^\alpha \, \gamma\cdot(\hat{l}-r) \, \gamma^\mu \, \gamma\cdot(\hat{k}-r) \, \gamma_\alpha
         }
         { [(\hat{l}-r)^2+i\epsilon] \, [(\hat{k} -r )^2+i\epsilon] }
\nonumber\\
   & \hspace*{1cm}
       -
         \frac{ \gamma^\mu \, \gamma\cdot(\hat{k}-r) \,\gamma\cdot n_s
         }
         { (-r\cdot n_s +i\epsilon) \, [(\hat{k} -r )^2+i\epsilon] }
        -
         \frac{ \gamma\cdot n_s \, \gamma\cdot(\hat{l}-r) \, \gamma^\mu 
         }
         { [(\hat{l} -r )^2+i\epsilon] \, (-r\cdot n_s +i\epsilon) }
      \Bigg\} 
     {\cal P}_{\rm T} 
\nonumber\\
    & + \mbox{UV counterterm}.
\end{align}
Our normal recipe for a hard scattering 
required us to insert the
projection matrix ${\cal P}_{\rm T}$ at each side, and to use the massless
on-shell external momenta $\hat{k}$ and $\hat{l}$ that were defined
earlier. We also used the fact that in Feynman gauge the three
subtraction terms involving $T_S$ exactly cancel.  These terms $- T_H
(1 - T_{\rm T} - T_{\rm J}) T_S V$ are like those in Eq.\ (\ref{eq:vcH}) but with
the factor in braces replaced by
\begin{align}
         - \frac{ u_{\rm J}\cdot u_{\rm T} }
         { (-u_{\rm J} \cdot r + i\epsilon) \, (-u_{\rm T} \cdot r - i\epsilon) }
        + \frac{ n_s \cdot u_{\rm T} }
         { (-n_s \cdot r + i\epsilon) \, (-u_{\rm T} \cdot r - i\epsilon) }
        + \frac{ u_{\rm J}\cdot n_s }
         { (-u_{\rm J} \cdot r + i\epsilon) \, (-n_s \cdot r + i\epsilon) },
\end{align}
\end{widetext}
which is exactly zero.  Since we take the massless limit in the hard
scattering, we have replace the non-lightlike vectors $n_{\rm T}$ and $n_{\rm J}$
by their lightlike counterparts $u_{\rm T}$ and $u_{\rm J}$.

\section{Full factorization}
\label{sec:factorization}
We now have enough 
techniques to obtain a full factorization formula,
valid at the leading power in $Q$ including all logarithmic
corrections.  However, we will restrict our derivation to a model
theory with massive Abelian gluons.  This avoids certain complications
with Ward identities in a non-Abelian theory and with actual IR
divergences associated with the masslessness of the gluon.  
These complications we leave
to later work.  The model exhibits the issues of gluons
coupling 
subgraphs with different kinds of momenta and the need for
appropriate Wilson lines in the definitions of the PCFs.  Even so, the
formulation of factorization, with the definitions of the PCFs already
exhibited, is equally applicable to QCD.

As we have done throughout in this paper, we also restrict to final
states with low transverse momenta in the Breit frame.  Thus the
leading regions do not involve extra 
groups (or jets) of 
collinear partons emitted from
the hard scattering.  Since this is a restriction on the final state,
and our approximation methods leave unaltered all final state momenta,
this is a safe restriction appropriate for exhibiting the simplest
and some of the most important cases where PCFs are important.

\begin{figure*}
  \centering
  \begin{tabular}{c@{\hspace*{10mm}}c}
    \includegraphics[scale=0.45]{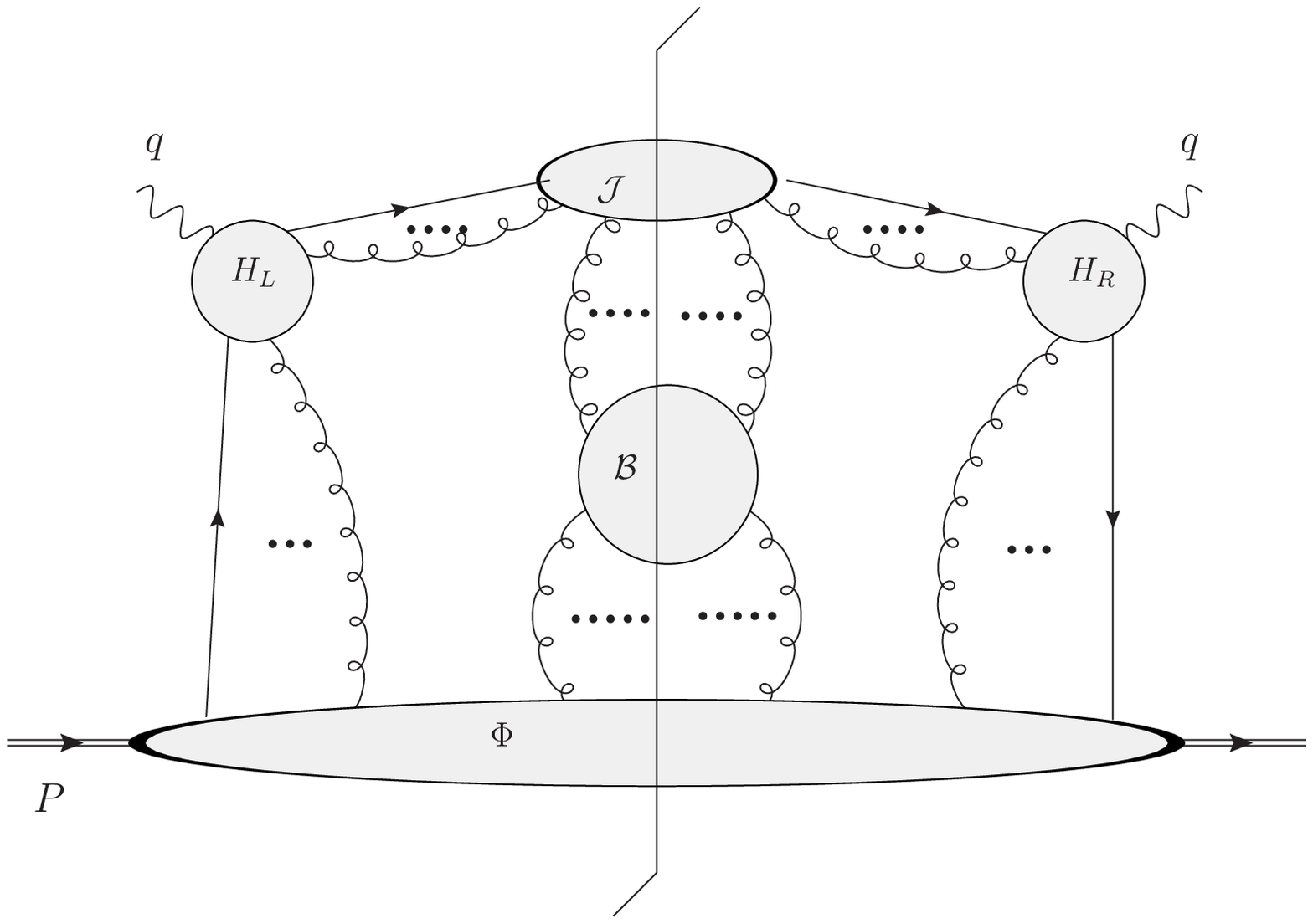}
    &
    \includegraphics[scale=0.45]{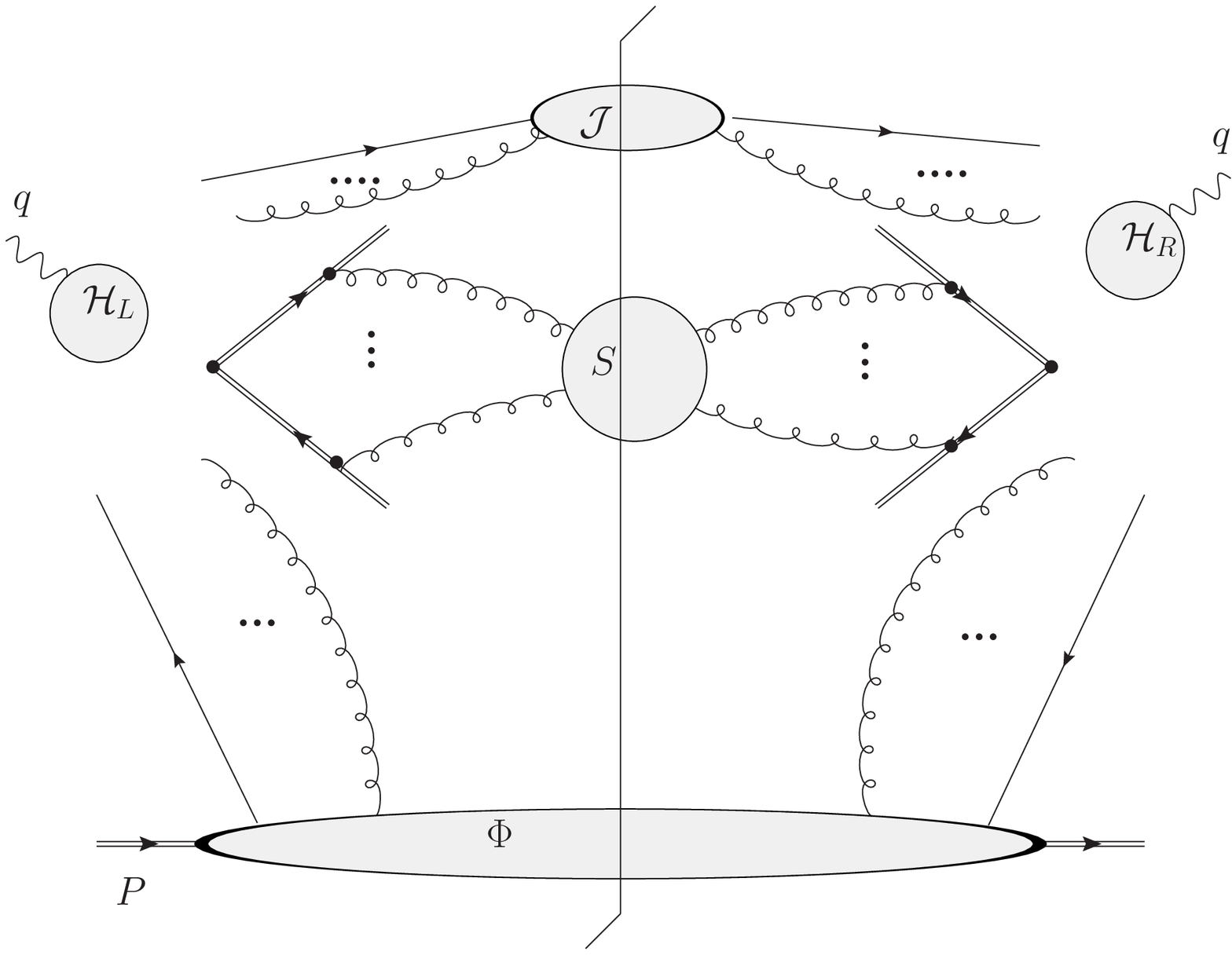}
  \\
  (a) & (b)
  \\[10mm]
  \multicolumn{2}{c}{
    \includegraphics[scale=0.45]{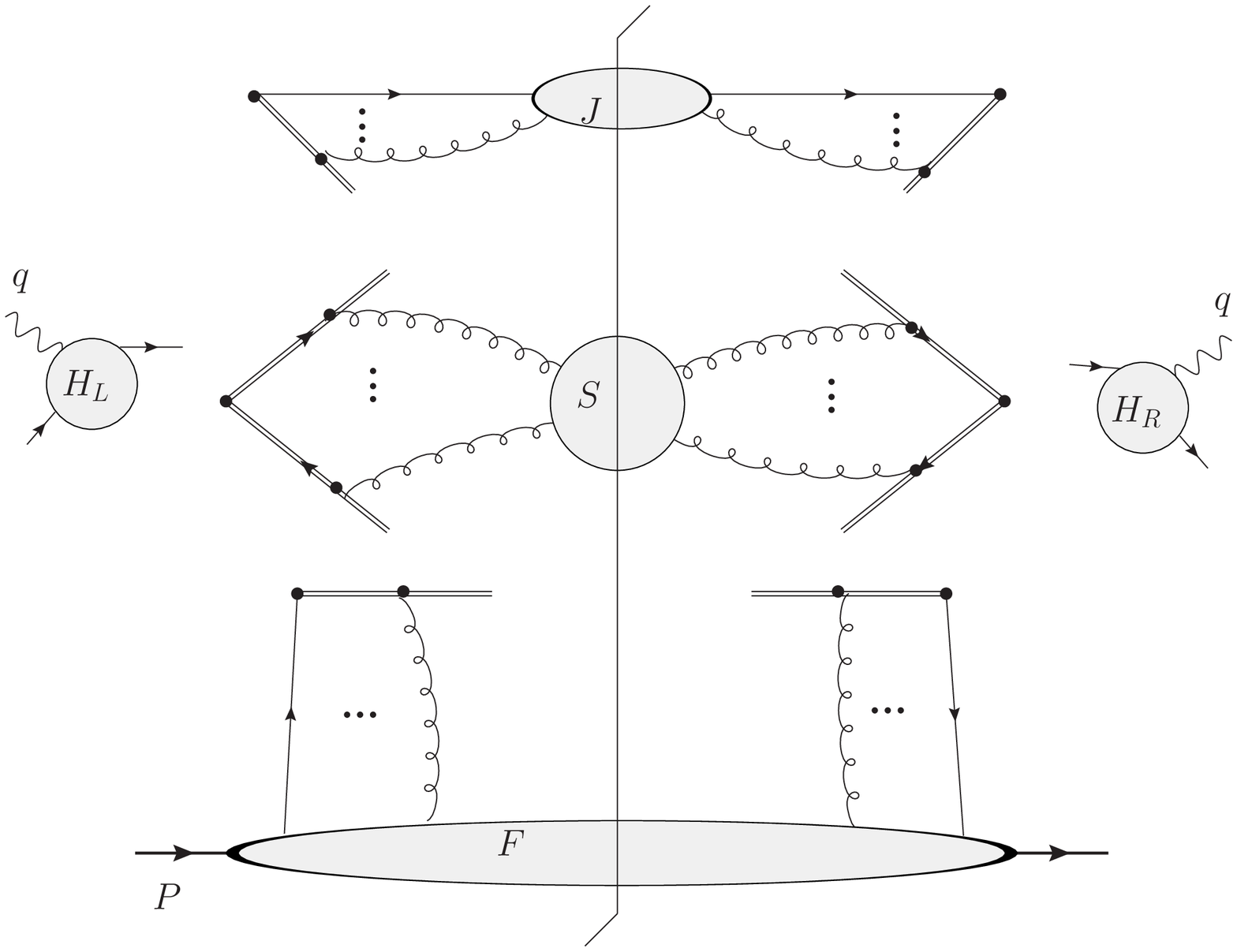}
  }
  \\
  \multicolumn{2}{c}{(c)}
  \end{tabular}
  \caption{Conversion to factorized form: (a) All leading reduced
    graphs with generalized handbag structure.  (b) After application
    of Ward identities to connection of gluons from soft subgraphs to
    collinear subgraphs.  (c) After application of Ward identities to
    connection of collinear subgraphs to hard subgraphs.}
\label{fig:WI.appl}
\end{figure*}

In Fig.\ \ref{fig:WI.appl}, we give a graphical overview of the proof:
\begin{itemize}
\item In graph (a) we symbolize the most general leading region.  It
  has a connected hard subgraph associated with each of the external
  vertices for electromagnetic current.  It has a connected collinear
  subgraph for the target and for the jet, each of which has, in
  addition to the standard quark connection, arbitrarily many gluons
  connecting it to the hard subgraphs.  It has a soft subgraph
  connected by gluons to both collinear subgraphs.  The soft subgraph
  has arbitrarily many connected components (including zero as a
  possibility), and each component must couple to both collinear
  subgraphs.

  Let $R$ denote any such region of a general graph $\Gamma$ for DIS.

\item We then apply the operation $C_R\Gamma$ to obtain the contribution
  associated with this region.  We can now understand graph (a) to
  denote $C_R\Gamma$ with a sum over $R$ and $\Gamma$ to give the complete
  leading-power contribution.  Subtractions, as defined in Eq.\
  (\ref{eq:CR}), ensure that the contributions from smaller regions
  are suppressed.  That is, when the loop momenta are close to the
  defining surface of a smaller region $R'$, there is an actual
  power-law suppression, and when the loop momenta are far from this
  surface, there is a subtraction to prevent double counting between
  $C_R\Gamma$ and $C_{R'}\Gamma$. 

\item In accordance with our earlier discussions, we do not apply any
  restriction on the internal momenta of the graph; any suppression
  comes purely from overall momenta conservation, from the nature of
  the subtractions, and/or from any restrictions explicitly imposed in the
  definition of the 
  approximators.  

\item Then we apply Ward identities to the connections of the soft
  subgraph to the collinear subgraphs.  
  One of the reasons we have been insisting on \emph{not} applying
  restrictions to the internal momenta in a graph, is that this is necessary in 
  order that Ward identities work exactly.  The derivations of Ward identities involve shifts
  of loop momenta, so that any internal restriction on loop momenta is
  liable to violate the Ward identity:  A shift of an integration
  variable near a boundary can move a momentum across the boundary.

  This contrasts with the situation in Soft-Collinear Effective Theory
  (SCET) \cite{Bauer:2002nz}, and other methods related to a Wilsonian
  renormalization group.  In SCET, fields are defined with explicit
  restrictions to particular localities in momentum space, and it is
  not at all obvious how this is to be actually implemented,
  at least not with the exact preservation of Ward identities.
  
  In the present context of deriving a proper factorization formula,
  with the anticipated power suppressed corrections, a non-exact Ward 
  identity presents a serious problem.  In a particular region, one may 
  be tempted to apply an approximate Ward identity to the sum over
  graphs.
  Close to
  the particular region under investigation, the terms that violate the 
  exact Ward identity may be power suppressed.  However,  within the subtraction formalism
  (which allows for a proper derivation of a factorization formula) we encounter 
  equations like~(\ref{eq:CR}), where the approximated graphs are used far 
  outside their corresponding regions, where the leftover terms in the 
  Ward identities are not small.
 
\item
  In an Abelian theory, application of a Ward identity is
  entirely straightforward, and after the sum over all relevant graphs
  the gluons from the soft subgraph are attached to vertices for
  Wilson lines, one Wilson line for each quark at the hard scattering.
  We must define the approximator for the region $R$ so that the
  approximated hard subgraphs are exactly independent of the soft
  momenta.  Then the Wilson lines meet at a point and we have
  separated out a soft factor --- Fig.\ \ref{fig:WI.appl}(b).  The
  soft factor is a PCF defined by Eq.\ (\ref{softdef}).

  There are some subtleties in defining the approximators so that the
  hard subgraphs become exactly independent of the soft momenta; we
  will return to this issue.

\item
  The Ward identities apply most directly to the approximated graph
  $T_R\Gamma$; 
  then the collinear subgraphs are just ordinary Green functions.
  However, the double counting subtractions for smaller regions ---
  Eq.\ (\ref{eq:CR}) --- change this situation, and we must also
  ensure, as we will do later, that the double counting subtractions
  are compatible with the Ward identities.

\item Finally, we apply Ward identities to the gluons connecting the
  collinear subgraphs to each of the two hard subgraphs.  This gives
  Fig.\ \ref{fig:WI.appl}(c), giving the jet and target PCF factors.
  We will have to investigate more carefully the subtractions for
  smaller regions, and we need to show that these produce the
  denominators in the definitions (\ref{pcf2}) and (\ref{pcf3}).

\item The hard subgraphs are now restricted to having the same
  external lines as in the parton model.  Loop graphs have
  subtractions for non-hard regions, so that we have a properly
  defined hard coefficient.

\end{itemize}

\subsection{Factorization formula}
The resulting factorization formula is
\begin{widetext}
\begin{multline}
  P_{\mu \nu} W^{\mu\nu}
  = 
  \int \frac{ d^4k_{\rm T} }{ (2\pi)^4 }
    \frac{ d^4k_{\rm J} }{ (2\pi)^4 }
    \frac{ d^4k_S }{ (2\pi)^4 }
    (2\pi)^4 \delta^{(4)}(q+P-k_{\rm T}-k_{\rm J}-k_S)
\times \\ \times 
    |\hardfact(Q,\mu)|^2
    ~ \soft(k_S,y_{\rm T},y_{\rm J},\mu) 
    ~ \pdf_{\rm mod}(k_{\rm T},y_p,y_{\rm T},y_s,\mu)
    ~ \jet_{\rm mod}(k_{\rm J},y_{\rm J},y_s,\mu),
\end{multline}
where $\soft$, $\pdf_{\rm mod}$ and $\jet_{\rm mod}$ are the Fourier 
transforms [Eq.\
(\ref{pcf4})] into momentum
space of the soft, target, and jet PCFs defined in Eqs.\
(\ref{softdef}), (\ref{pcf2}), and (\ref{pcf3}).  
The variables $k_{\rm T}$, $k_{\rm J}$, and $k_S$ are the momenta
of the final states in the target, jet, and soft PCFs.

This formula can be simplified.  
The target (and jet) PCFs have two rapidity arguments for Wilson
lines, which is in contrast to the situation in the CS formalism.
This can be changed by moving the denominators from the target and jet
PCFs to the soft PCF, so that the soft PCF is redefined to 
\begin{equation}
  \tilde{\soft}_{1}(w,y_{\rm T},y_{\rm J},y_s,\mu) 
= \frac{ \tilde{\soft}(w,y_{\rm T},y_{\rm J},\mu) }
       { \tilde{\soft}(w,y_{\rm T},y_{s},\mu) ~ \tilde{\soft}(w,y_{s},y_{\rm J},\mu) }.
\label{softdef.new}
\end{equation}
In this soft factor, the denominators remove the contributions from
large positive and negative rapidities.  We expect to be able to take
the limits that the vectors $n_{\rm T}$ and $n_{\rm J}$ become light-like:
\begin{equation}
  \tilde{\soft}_{2}(w,y_s,\mu) 
= \tilde{\soft}_{1}(w,+\infty,-\infty,y_s,\mu).
\label{softdef.new1}
\end{equation}
This corresponds to the soft factor defined by Collins and Soper
\cite{Collins:1981uw}. 

Then the factorization formula becomes
\begin{multline}
  P_{\mu \nu} W^{\mu\nu}
  =
  \int \frac{ d^4k_{\rm T} }{ (2\pi)^4 }
    \frac{ d^4k_{\rm J} }{ (2\pi)^4 }
    \frac{ d^4k_S }{ (2\pi)^4 }
    (2\pi)^4 \delta^{(4)}(q+P-k_{\rm T}-k_{\rm J}-k_S)
\times \\ \times 
    |\hardfact(Q,\mu)|^2
    ~ \soft_2(k_S,y_s,\mu) 
    ~ \pdf(k_{\rm T},y_p,y_s,\mu)
    ~ \jet(k_{\rm J},y_s,\mu),
\end{multline}
\end{widetext}
where we use the target PCF $\pdf$ defined by Eq.\ (\ref{pcfguess})
instead of $\pdf_{\rm mod}$ (and similarly for the jet PCF) defined by Eq.\ (\ref{pcf2}).
Note that there is a difference from the CS case, where the soft
factor is independent of $y_s$.  The CS soft factor is defined
appropriately for $k_\tran$ factorization, so that it is the integral over
$k_S^+$ and $k_S^-$ of the soft factor defined here, to give a
function of transverse momentum alone.  This quantity is invariant
under boosts in the $z$ direction.  In contrast, our soft factor
depends on longitudinal momenta as well.

\subsection{Momentum routing in approximators}

It is important for factorization that the hard scattering coefficient
should depend only on $Q^2$.  It should not depend on the loop momenta
in the soft factor.  Therefore in constructing the approximator for a
region, we need to arrange that the hard subgraphs are approximated as
independent of the soft momenta.  In addition, so that the hard
scattering coefficients can be treated as on-shell matrix elements
(modified by subtractions), we will approximate their external parton
momenta by on-shell massless momenta. 
So that we may be sure that this can be done in general, for a
graph with arbitrarily many soft and collinear gluons, we must be 
certain that the approximators that we use do not introduce any anomalous
dependence on soft gluon momenta, and that all approximations are 
exactly consistent with the application of Ward identity relations.

The difficulty in constructing a general prescription obeying these
requirements can be seen from the simple case of Fig.\ \ref{soft}.
There, the trouble comes from the correct choice for ``routing'' momenta.
By a choice of routing we mean the choice of which
momenta are 
to be treated as independent variables.  Different choices correspond to 
different explicit appearances of momentum variables around different loops in 
the graph.  Of course, the choice of momentum labeling is arbitrary and has 
no effect on an \emph{unapproximated} Feynman graph.  However, the approximators
are defined with respect to a certain set of variables.  Hence, the same 
instructions for approximating a graph will lead to different results for
different routings of momentum.  In other words, we can say that the approximators
are not completely defined until a choice of momentum routing is made.
   
For example, in Fig.\ \ref{soft} the soft momentum $l_2$ is routed
through the left-hand vertex. 
Thus, it appears natural to define the approximated momenta for the
hard vertex by replacing $l_2$ by zero and by then 
keeping only the minus and plus components of the quark momentum.
That is, \emph{in the 
  hard vertex} we perform the replacements $q+k-l_2 \mapsto (0,q^-+k^-,{\bf 0}_\tran)$
and $k-l_2 \mapsto (k^+,0,{\bf 0}_\tran)$ on the external momenta of the left hand
current vertex.  At the right-hand vertex there is no soft momentum,
so we simply make the replacements $q+k \mapsto (0,q^-+k^-,{\bf 0}_\tran)$ and $k \mapsto
(k^+,0,{\bf 0}_\tran)$.

Observe that the value of minus momentum on the quark line going to the
jet was changed on the left-hand side, but not on the right-hand side,
and similarly for the quark from the target subgraph.  We can reverse
this situation by simply changing the routing of $l_2$ to go through
the right-hand current vertex, for example by changing variables to
$k_1=k-l_2$.  In that case the momenta at the left-hand vertex are
$q+k_1$ and $k_1$, while the momenta at the right-hand vertex are
$q+k_1+l_2$ and $k_1+l_2$.  Thus the definition of the approximation
varies depending on the routing of the loop momenta, and the two
routings we have shown are equally legitimate.

In the true soft region, the components of $l_2$ are small compared with $Q$ and so the
difference between the definitions is a small, power-suppressed
effect.  But we perform an integral over all kinematically accessible
momenta, so the difference amounts to a genuine inconsistency.

Nevertheless, the inconsistency did not affect our treatment of Fig.\
\ref{soft}, because its hard subgraphs are single vertices and hence
independent of momentum.  But if we consider a more general situation,
as in Fig.\ \ref{fig:WI.appl}, with non-trivial hard subgraphs, the
momentum dependence of the hard subgraphs creates an inconsistency
between (at least) two possible definitions of approximation for  the
hard subgraph.

A further problem appears when we observe that the approximated minus
momentum on the $q+k-l_2$ and $q+k$ lines is not $q^-$, and that the
approximated plus momentum on the $k-l_2$ and $k$ lines is not $-q^+$.
This creates the situation that, beyond lowest order in the hard
scattering, the hard scattering coefficient does not depend just on
$Q$.  Again, in a situation of really collinear momenta this is a
small effect, but we integrate the PCFs out beyond this region.

These issues are closely related to the kinematic inconsistencies in
parton showering algorithms that were found by Bengtsson and Sj\"ostrand
\cite{BS}, and that led \cite{MRW1,MRW2,Collins:2005uv,CZ} to the
proposal to use PCFs rather than regular parton densities.

The essential difficulty in defining the approximations is that we
have to know what are the independent variables in the various
subgraphs.  However, the momenta of different lines are constrained by
momentum conservation and are not all independent.  The dangers are
made 
even worse by our pervasive use of Ward identities.  Graphical proofs
of Ward identities require shifts of integration variables.
Consistency of Ward identities with the approximations requires that
the approximations be invariant under certain reroutings of loop
momenta. 
As we will now see, a consistent prescription for labeling 
momenta involves treating \emph{all} of the outgoing parton momenta 
as independent variables, but treating the photon momentum as a 
\emph{dependent} variable.  

\begin{figure}
  \centering
  \includegraphics[scale=0.45]{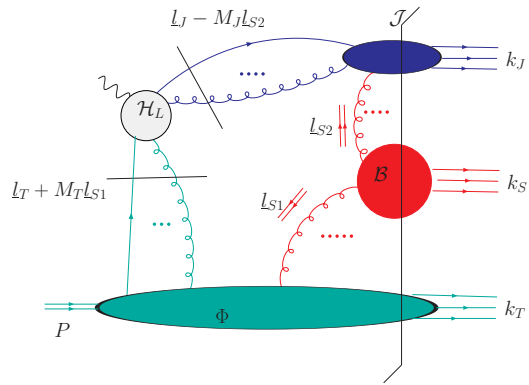}
  \caption{Definition of momentum variables for a given region.
           The subgraphs are coded by color as: red = soft, blue =
           jet-collinear, green = target-collinear, black = hard.
  }
  \label{fig:routing}
\end{figure}

Our solution is two fold: In the first step we route all the momenta
from the soft subgraph out through the hard vertex; i.e., 
the photon momentum 
is \emph{not}
treated as an independent variable, and is not fixed to $q$.
See Fig.\ \ref{fig:routing} for a graphical depiction of the basic setup.  
This choice of independent variables is
sufficient to treat both hard subgraphs the same way, and gives a routing
consistent with all the Ward identities we apply.  
The second step, after approximating
the external lines of the hard subgraph as before, is to
define a new set of variables by
rescaling the
approximated minus and plus momenta so that 
outgoing jet momenta sum to $(0,q^{-},{\bf 0}_\tran)$ and
the incoming jet momenta sum to $(-q^{+},0,{\bf 0}_\tran)$.  
We end up with a treatment of momenta that is similar to the
prescription
of 
Collins and Zu \cite{CZ},
but is more complicated because of the need to treat soft subgraphs.

Now let us go through the steps described above explicitly.
We let $\underline{l}_{S2}$ denote the \emph{collection} of momenta 
entering the jet-collinear subgraph from
the soft subgraph.  Then we write the
collection of momenta 
leaving the hard subgraph to the
jet-collinear subgraph as
\begin{equation}
  \underline{l}_{\rm J} - M_{\rm J} \underline{l}_{S2}.
\end{equation}
Here $M_{\rm J}$ is an incidence matrix.  
Thus for a \emph{particular} outgoing jet line, labeled by an index,
$j$, we have
\begin{equation}
\label{eq:soft.route.example}
l_{{\rm J} j} - \sum_k M_{{\rm J} jk} \, l_{S2 k},
\end{equation}
where the sum is over the soft lines outgoing from the subgraph $\jetbub$.
Similarly,
we write the collection of momenta entering the hard subgraph from the
target-collinear subgraph as
\begin{equation}
  \underline{l}_{\rm T} + M_{\rm T} \underline{l}_{S1}.
\end{equation}
Thus for a \emph{particular} incoming target line, we have
\begin{equation}
l_{{\rm T} j} + \sum_k M_{{\rm T} jk} \, l_{S1 k}. 
\end{equation}
where the sum is over the soft lines outgoing from the subgraph $\pdfbub$.
Momentum conservation on the final state will impose some constraints
between the momenta within $\underline{l}_{\rm J}$, within
$\underline{l}_{\rm T}$, and within
$(\underline{l}_{S2},\underline{l}_{S1})$, but this will not affect
our argument.  Within the hard subgraph, we define the 
approximator so that it makes the following replacements:
\begin{align}
\label{eq:coll.B}
  \underline{l}_{\rm J} - M_{\rm J} \underline{l}_{S2}
  & \mapsto \lambda_{\rm J} ( 0, \underline{l}_{\rm J}^-, {\bf 0}_\tran ) ,
\\
\label{eq:coll.A}
  \underline{l}_{\rm T} + M_{\rm T} \underline{l}_{S1}
  & \mapsto \lambda_{\rm T} ( \underline{l}_{\rm T}^+, 0, {\bf 0}_\tran ) .
\end{align}
That is, we project down to the minus and plus components of the
collinear momenta only. 
After this projection of momentum components, the next step is to apply a rescaling so that the momenta
sum to $q$ in the hard vertex.  More specifically,~(\ref{eq:coll.B}) and~(\ref{eq:coll.A}) mean that
for
each of the
outgoing and incoming lines, we project onto a new set of variables as follows:
\begin{align}
\label{eq:coll.B2}
  l_{{\rm J} j} - \sum_k M_{{\rm J} jk} \, l_{S2  k}
  & \mapsto \hat{l}_{{\rm J} j} = \lambda_{\rm J} ( 0, l_{{\rm J} j}^-, {\bf 0}_\tran ) ,
\\
\label{eq:coll.A2}
  l_{{\rm T} j} + \sum_k M_{{\rm T} jk} \, l_{S1  k}
  & \mapsto \hat{l}_{{\rm T} j} = \lambda_{\rm T} ( l_{{\rm T} j}^+, 0, {\bf 0}_\tran ) .
\end{align}
As before, a ``hat'' on a momentum variable indicates 
that it is an approximated momentum variable.
The scalings are determined from the condition that 
the incoming target collinear momenta add up to $(-q^{+},0,{\bf 0}_\tran)$ 
while the outgoing jet collinear momenta add up to 
$(0,q^{-},{\bf 0}_\tran)$,
\begin{align}
\label{eq:coll.B3}
  \sum_j \hat{l}_{{\rm J} j} = \sum_j \lambda_{\rm J} \left( 0, l_{{\rm J} j}^-, {\bf 0}_\tran \right)
                  = \left(0,q^{-}, {\bf 0}_\tran \right),
\\
\label{eq:coll.A3}
  \sum_j \hat{l}_{{\rm T} j} = \sum_j \lambda_{\rm T} \left( l_{{\rm T} j}^+, 0, {\bf 0}_\tran \right)
                  = \left( -q^{+},0, {\bf 0}_\tran \right).
\end{align}
Hence we get the following definitions for the scalings:
\begin{align}
  \lambda_{\rm J} &= \frac{q^-}{ \sum_j l_{{\rm J} j}^- } ,
\\
  \lambda_{\rm T} &= \frac{-q^+}{ \sum_j l_{{\rm T} j}^+ } .
\end{align}
We also define 
the approximator so that
the fermion lines are equipped with projection
matrices, ${\cal P}_{\rm T}$ in the amplitude and ${\cal P}_{\rm J}$ in the
complex conjugate. At the gluons we apply an appropriate
Grammer-Yennie-style approximation, as in Eq.\ (\ref{collapprox}).

To summarize, we have generalized (and modified) the replacement scheme
of Sect.~\ref{sec:basic.approx} and Ref.~\cite{CZ} in such a way that we now 
consistently  
deal with the presence of soft gluons.

There is one non-obvious step
for the Grammer-Yennie approximation
that arises from the replacement of exact momenta with approximated 
momenta in the
hard subgraph.  For a gluon coupling to the hard subgraph from the
target-collinear subgraph, we make the replacement
\begin{equation}
  \hardbub^\kappa(l_{{\rm T} j}, \ldots) 
  \mapsto
  \frac{n_s^\kappa}{l_{{\rm T} j} \cdot n_s -i\epsilon} ~ \hat{l}_{{\rm T} j} \cdot \hardbub (\hat{l}_{{\rm T} j}, \ldots). 
\end{equation}
Here $l_{{\rm T} j}$ denotes the momentum of the line flowing into the hard
subgraph on the gluon line.  In the hard part we replace this momentum
by its approximation defined in Eq.\ (\ref{eq:coll.A}).  So that the
relevant Ward identity is exactly valid, we have performed the same
replacement in the factor of momentum contracted with the hard
subgraph.  In contrast the denominator is left unaltered.  This
enables the $n_s$ vector to 
fulfill its purpose of cutting off the
integral over the rapidity of $l_{{\rm T} j}$.  This unaltered denominator also
ensures that we get 
exactly 
the expected Wilson line operator for the target
PCF.

Similar definitions apply to the gluonic connections from the other
collinear subgraph.  
To summarize, the sequence of replacements 
listed in this subsection defines an approximator which (a) leaves
the hard scattering subgraph independent of soft momenta, and (b) allows 
for the exact application of Ward identities.

\subsection{Ward identities for soft into collinear subgraphs}

We first examine the Ward identity argument for the connection of soft
gluons from subgraph $\softbub$ to the jet subgraph $\jetbub$ in Fig.\
\ref{fig:WI.appl}(a).  For each 
soft
gluon the Grammer-Yennie approximation
is applied, by a replacement of the form
\begin{equation}
\label{eq:S.to.B}
  \jetbub^\kappa(l_{S2 j}, \ldots) 
  \mapsto
  \frac{n_{\rm J}^\kappa}{l_{S2 j} \cdot n_{\rm J} -i\epsilon} ~ l_{S2 j} \cdot \jetbub (l_{S2 j}, \ldots) \; ,
\end{equation}
where $l_{S2 j}$ is the soft-gluon momentum oriented to enter subgraph $\jetbub$.
Now the sum over regions and graphs for the whole process can be
written as independent sums over the different subgraphs, subject to
consistency on the kinds of lines joining them.  So we now sum over
subgraphs $\jetbub$ with the other subgraphs fixed,
and also apply the same argument for the connection of the soft
subgraph to the target-collinear subgraph.
In the absence of the 
double-counting subtractions, a standard Ward identity applied to
every gluon connecting the soft subgraph to the collinear subgraphs
would result in Fig.\ \ref{fig:WI.appl}(b).  

However, to each graph is applied a series of subtractions
from smaller regions, 
as in Eq.~(\ref{eq:CR}),
so we must
examine their compatibility with the Ward identities. So let $R$
denote a particular region of the form of Fig.\ \ref{fig:WI.appl}(a),
for which we wish to construct its contribution as in Eq.\
(\ref{eq:CR}).  The direct Ward identity argument applies to the term
$T_R\Gamma$.  Subtractions concern all possible strictly smaller regions
$R'<R$. 
Each $R'$ assigns a momentum category (hard, collinear of some
type, soft) to each line, and implicit in the subtractions in $C_R\Gamma$
are therefore approximations appropriate to regions smaller than $R$.

The following discussion entails considering the relation between the
two different regions $R$ and $R'$ for a single graph $\Gamma$ and also
treating a sum over the possibilities for $\Gamma$, $R$, and $R'$, as in
\begin{align}
\label{eq:sum.CR}
  \sum_{\Gamma,R} C_R \Gamma & =  \sum_{\Gamma,R} T_R\Bigl( \Gamma - \sum_{R'<R} C_{R'}\Gamma \Bigr) 
\nonumber\\
  & = \sum_{\Gamma,R} \Bigl( T_R \Gamma - \sum_{R'<R} T_RT_{R'}\Gamma + \ldots \Bigr).
\end{align}
Each region
can be specified in terms of its $\jetbub$, $\pdfbub$, $\hardbub$ and $\softbub$ subgraphs, so
we will use a notation $\jetbub(R)$, $\softbub(R)$ etc for the jet and soft
subgraphs, etc, associated with the region $R$.  The region $R'$ has
its corresponding $\jetbub(R')$, $\softbub(R')$ subgraphs etc.   

\begin{figure*}
  \centering
  \psfrag{T_R}{$T_R$}
  \psfrag{T_Rp}{$T_{R'}$}
  \begin{tabular}{c@{\hspace*{1cm}}c@{\hspace*{1cm}}c}
    \includegraphics{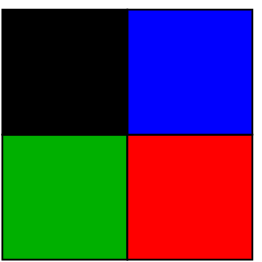}
    &
    \includegraphics{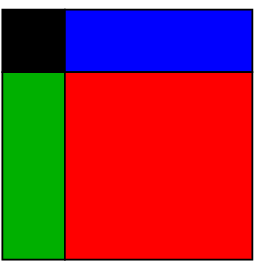}
    &
    \includegraphics{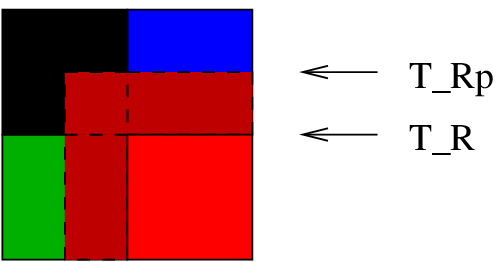}
  \\
    (a) & (b) & (c)
  \end{tabular}
  \caption{(a) Representation of subgraphs for region $R$.  (b)
    Representation for smaller region $R'$.  (c) Soft subgraph of
    $R'$ overlaid on subgraphs for $R$.  
    The color coding is the same as in Fig.\ \ref{fig:routing}. }
\label{fig:regions}
\end{figure*}

Since $R'$ is smaller than $R$, it imposes tighter constraints on the
momentum categories of the lines.  Hence, as regards the soft lines,
the set of lines categorized (or labeled) as soft for $R'$ is at least
as big as the set of lines labeled as soft for $R$.  That is, the soft
subgraph of $R'$ is at least as big as the soft subgraph of $R$, i.e.,
$\softbub(R') \supseteq \softbub(R)$, as shown in Fig.\ \ref{fig:regions}.  In forming the
term $T_RC_{R'}\Gamma$ in Eq.\ (\ref{eq:sum.CR}), first the approximator
$T_{R'}$ is applied and then $T_R$.  (There can be further
subtractions inside $C_{R'} \Gamma$, but that need not concern us here.)

In the representation
in Fig.\ \ref{fig:regions} the approximator $T_R$ applies some
operation at the boundary of $\softbub(R)$, while $T_{R'}$ applies some
operation at the boundary of $\softbub(R')$.  (There are also operations
applied at the boundaries between the collinear and hard subgraphs,
but that will be our topic later.)
Since $\softbub(R') \supseteq \softbub(R)$, the operation $T_{R'}$ has no effect on the soft
subgraph $\softbub(R)$ for the original region $R$.  But some lines of jet
subgraph $\jetbub(R)$ may be soft according to $R'$, so the approximator
$T_{R'}$ has an effect inside $\jetbub(R)$, when we form the term
$T_RT_{R'}\Gamma$.

\begin{figure*}
  \centering
  \begin{tabular}{c@{\hspace*{1cm}}c@{\hspace*{1cm}}c}
    \includegraphics[scale=0.45]{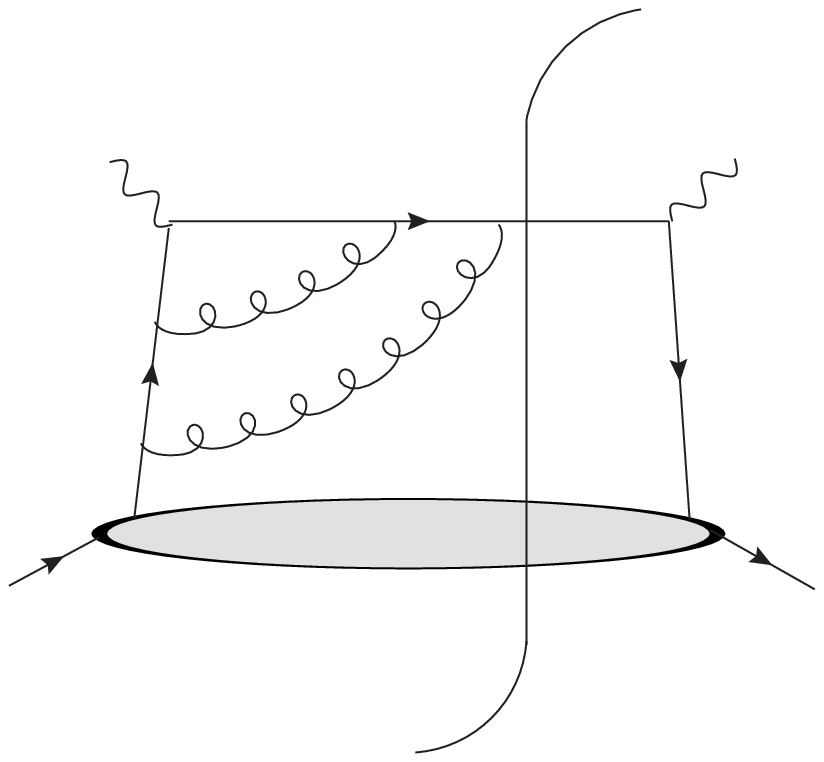}
    &
    \includegraphics[scale=0.45]{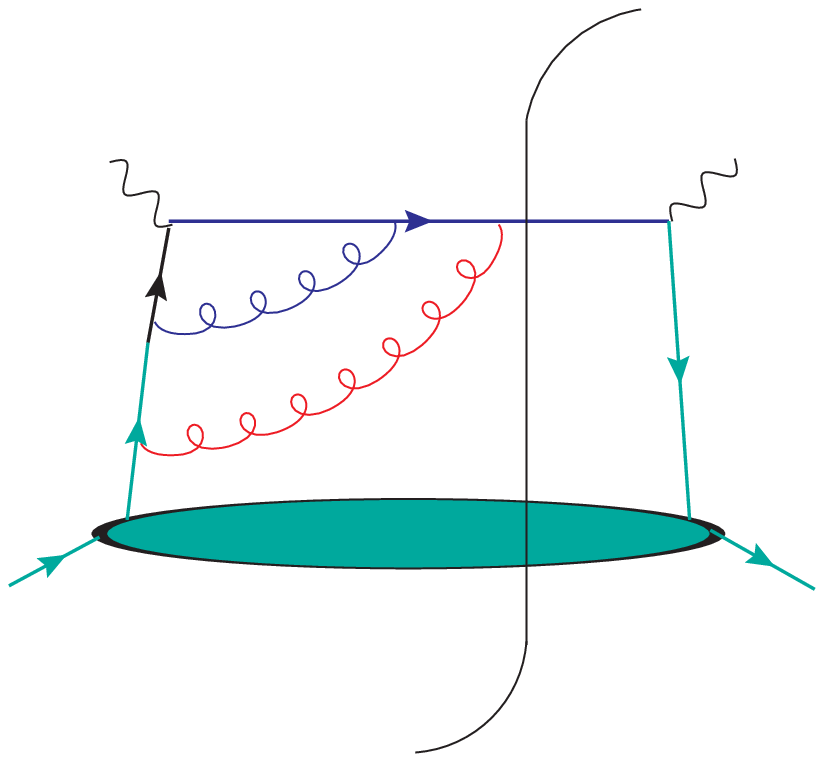}
    &
    \includegraphics[scale=0.45]{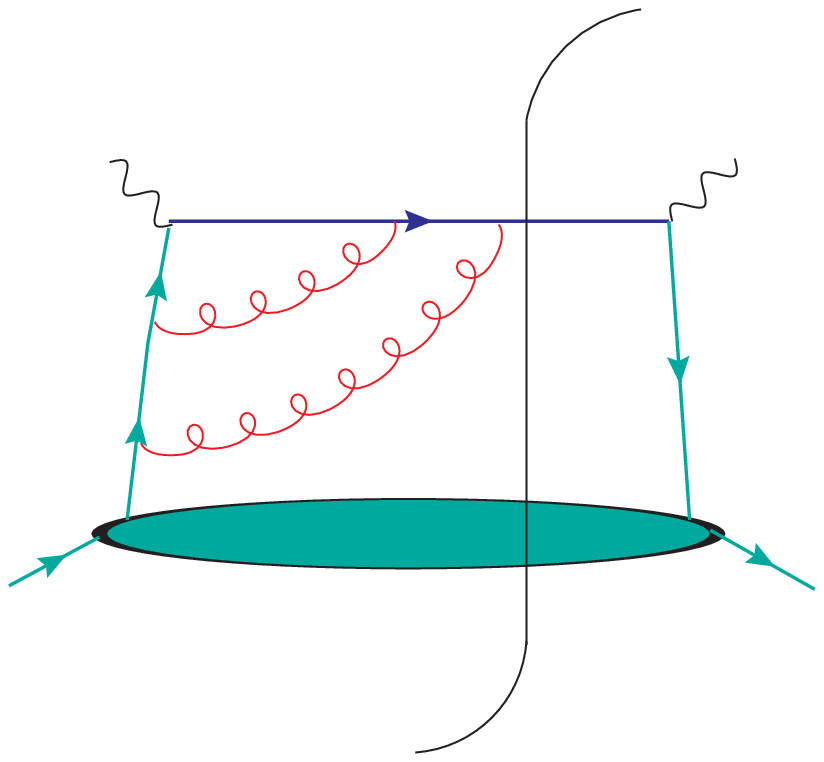}
  \\
    (a) & (b) & (c)
  \\
    \includegraphics[scale=0.45]{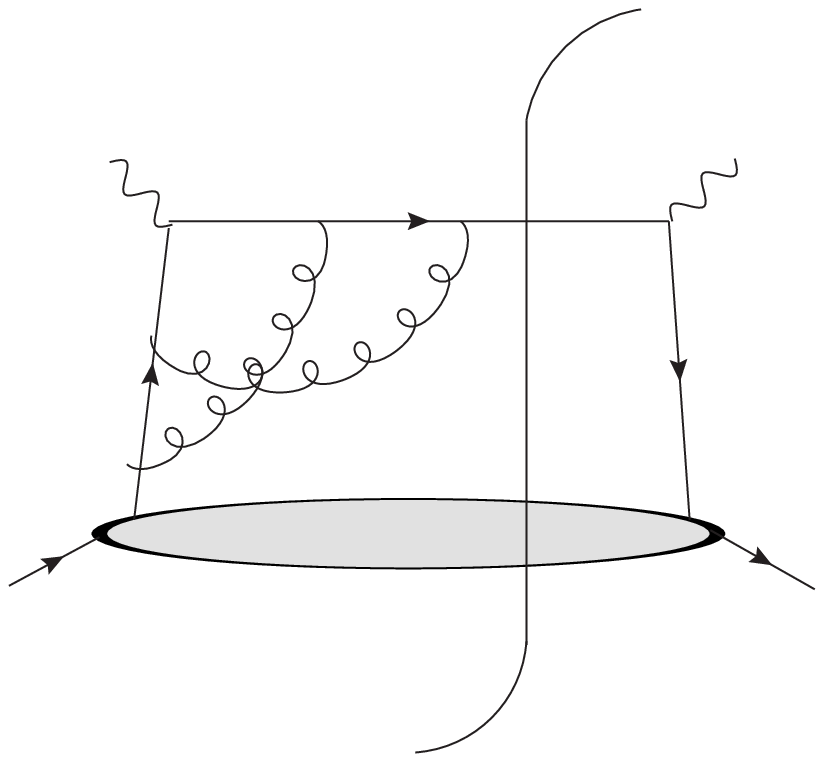}
    &
    \includegraphics[scale=0.45]{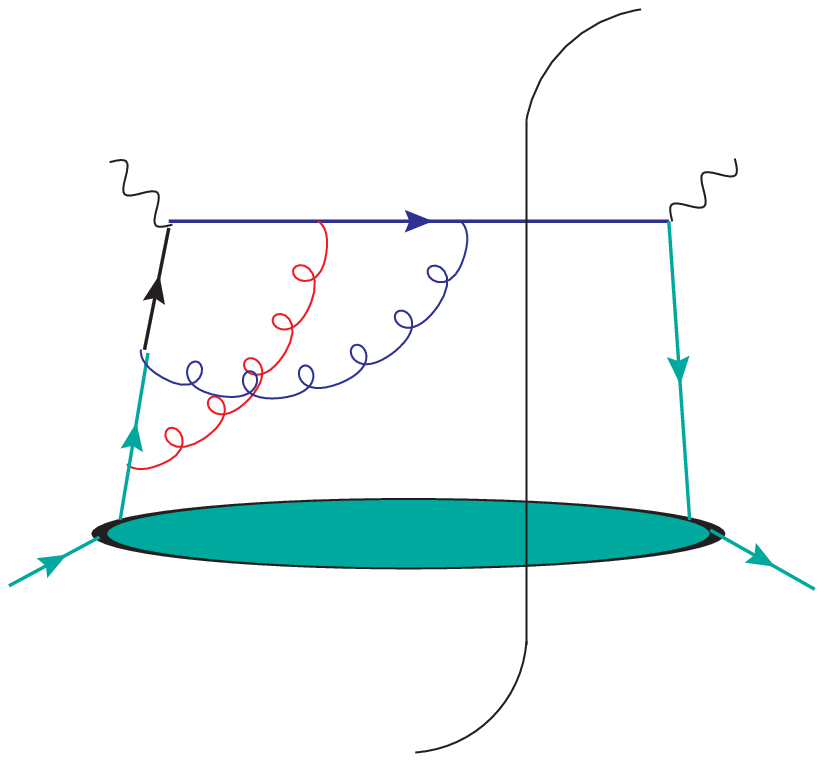}
    &
    \includegraphics[scale=0.45]{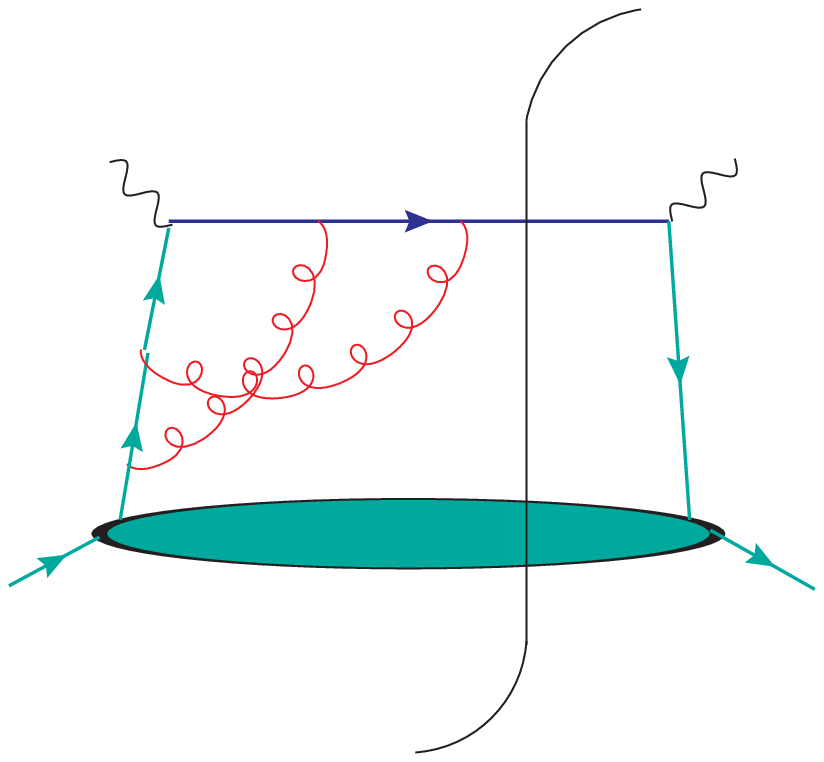}
  \\
    (d) & (e) & (f)
  \end{tabular}
  \caption{(a--c) A graph for DIS, and two of its regions.
           (d--f) Another graph, and two of its regions.
    The diagrams in the left-hand column, all in black, are particular
    Feynman graphs for the process, i.e., a candidate for $\Gamma$.
    The colored diagrams in the middle column represent a particular
    region (a candidate for $R$) for each of these graphs
    where one of the gluons is soft and one is jet-collinear.
    The diagrams in the right-hand column represent a smaller region,
    a candidate for $R'$ where both gluons are soft.
    The color coding is the same as in Fig.\ \ref{fig:routing}. }
  \label{fig:2gluon}
\end{figure*}

An example of the situation we wish to discuss is shown in Fig.\
\ref{fig:2gluon}.  For each graph $\Gamma_1$ and $\Gamma_2$ in the left-hand
column, suppose we have constructed for each graph the term
$C_{R'_j}\Gamma_j$ for the region shown in the right-hand column where the
two gluons are both soft.  Next we wish to construct the the term
$C_{R_j}\Gamma_j$ for the bigger region shown in the middle column, where
only one gluon is soft and the other is jet-collinear.  After
application of the soft approximation, a Ward identity, discussed
below, will let us combine the two ways of attaching the red gluon to
the upper quark line.  There are also subtractions to prevent
double-counting.  The nature of the soft approximation lets us combine
the subtraction terms related to the right-hand column in the same way
as in the middle column, thus ensuring that the Ward identities are
compatible with subtractions.  

Suitably viewed, this pair of graphs actually leads us to the general
case.  
First, let us examine the
standard graphical derivation of the Ward identity.  The basic step is
the following identity applied to the connection of a soft gluon to
one quark line, after the replacement (\ref{eq:S.to.B}) is applied:
\begin{widetext}
\begin{subequations}
\label{eq:WI-elem}
\begin{align}
  \frac{ -ign_{\rm J}^\kappa}{l \cdot n_{\rm J} -i\epsilon} ~
  \frac{ i }{ \slashed{k} - m  } ~
  \slashed{l} ~
  \frac{ i }{ \slashed{k} - \slashed{l} - m  }
=
  \frac{ -gn_{\rm J}^\kappa }{ l \cdot n_{\rm J} -i\epsilon } 
  \left(
     \frac{ i }{ \slashed{k} - m  }
  -
     \frac{ i }{ \slashed{k} - \slashed{l} - m  }
  \right),
\intertext{which can be written graphically as}
  \includegraphics[scale=0.45]{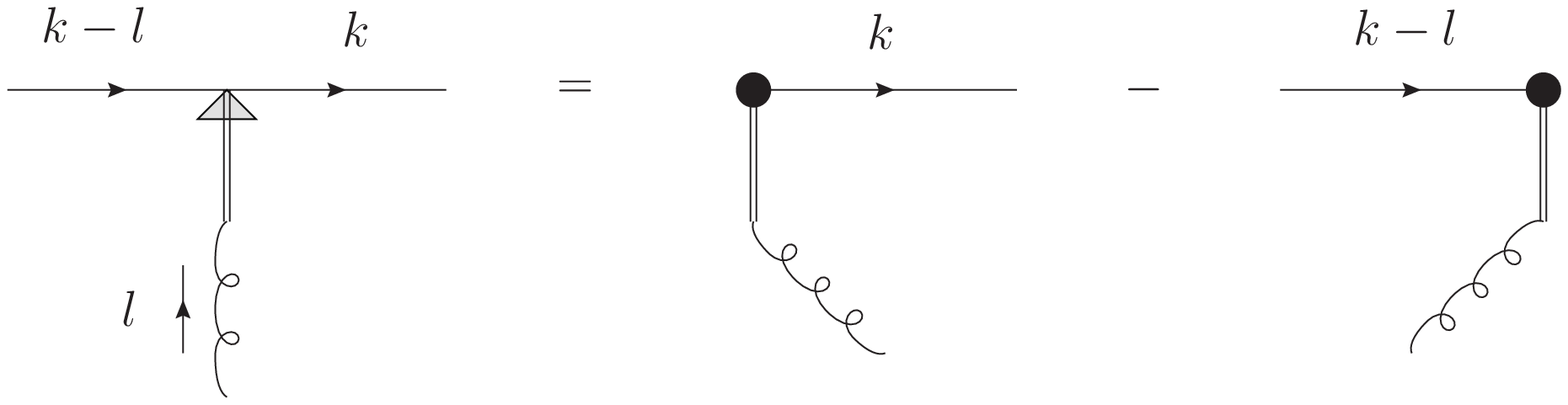}
\end{align}
\end{subequations}
On the left-hand side, the double line and the arrow at the end of the
gluon denote, respectively, the first factor and the $l_{S2j} \cdot \, $ factor in
(\ref{eq:S.to.B}).  The double-line notation is to exhibit that its
factor corresponds to Feynman rules for a Wilson line.  
The big dots at the vertices on the right hand side of (\ref{eq:WI-elem}b) are
used to emphasize the special vertex where the Wilson-line component
attaches to the rest of the graph.

We now examine what happens when we 
take
two different graphs for $\jetbub(R)$
that are related by having the gluon $l$ attached on opposite sides of
some other vertex, with another gluon $l_1$.  
Embedded inside the subgraph $\jetbub(R)$, we have the following situation:
\begin{subequations}
  \label{eq:WI-2gluon}
  \begin{align}
  \label{eq:WI-2gluon1}
  \includegraphics[scale=0.45]{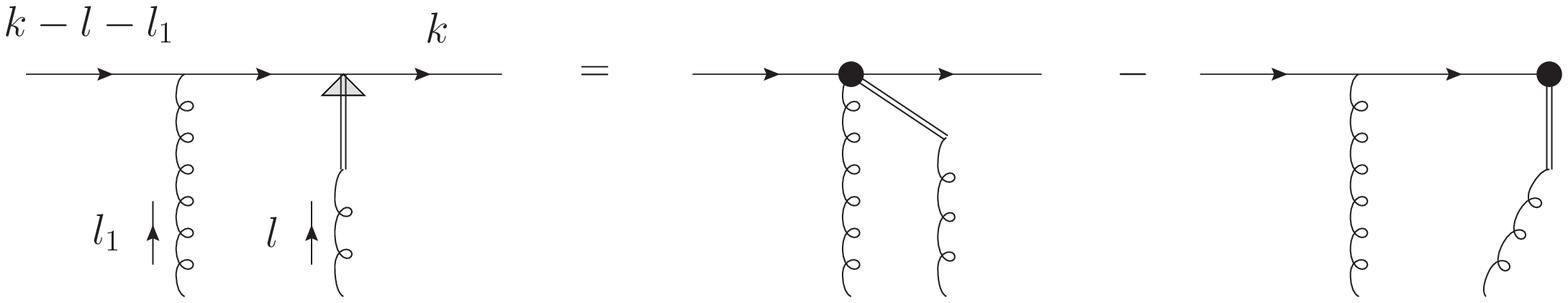}
 \\
  \label{eq:WI-2gluon2}
  \includegraphics[scale=0.45]{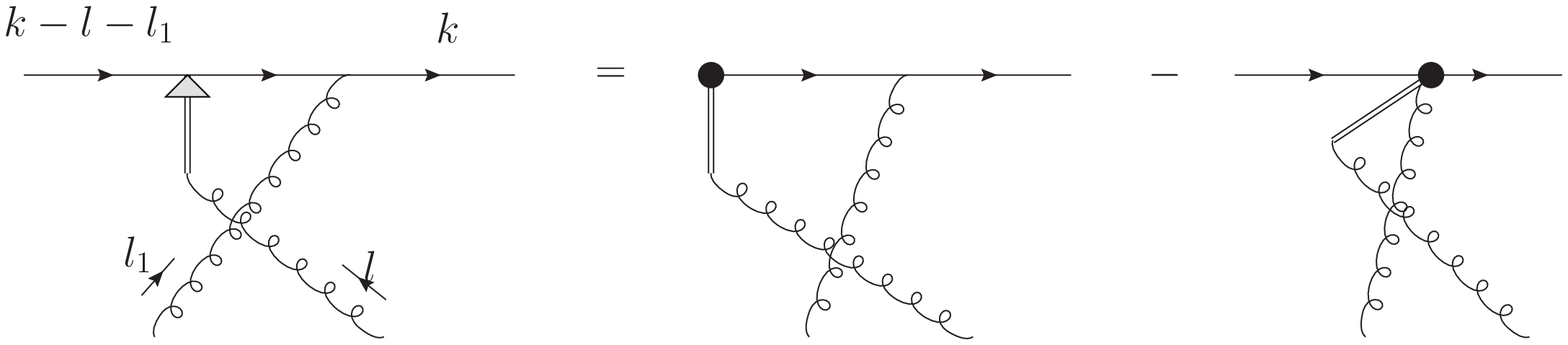}
  \end{align}
\end{subequations}
\end{widetext}
An example of this situation is provided by Fig.\ \ref{fig:2gluon},
with the gluon $l$ being the red gluon in the middle column.

After we sum over the two graphs, on the left-hand-side, there is a
cancellation between the two terms where the special vertex is at the
second gluon, i.e., of the first term on the right of Eq.\
(\ref{eq:WI-2gluon1}) with the second term on the right of Eq.\
(\ref{eq:WI-2gluon2}):
\begin{equation}
  \label{eq:WI-cancel}
  \includegraphics[scale=0.42]{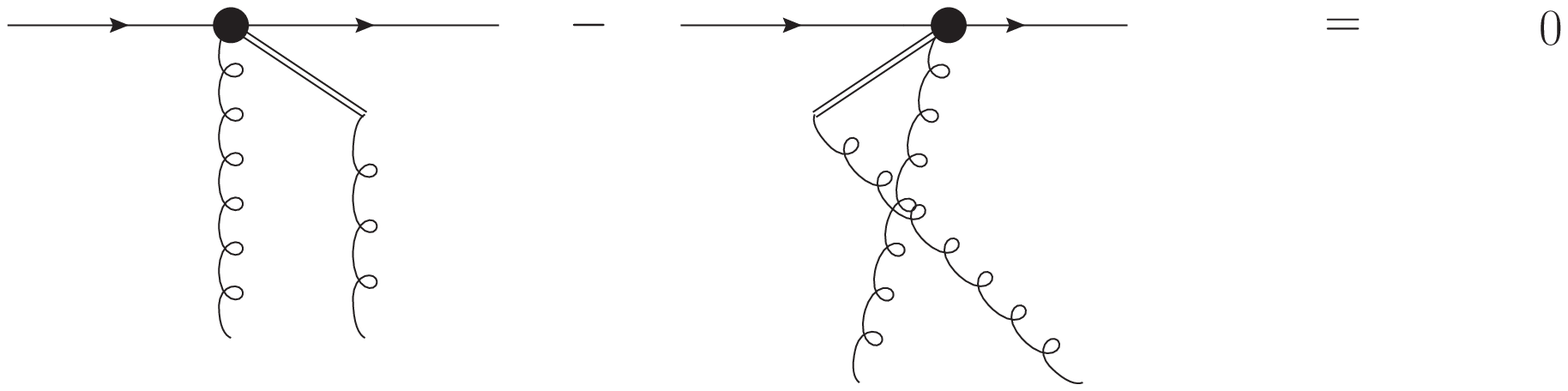}
\end{equation}
Then follows a cascade of such cancellations when we sum over all ways
of attaching gluon $l$ to the unsubtracted jet-collinear subgraph,
i.e., when we sum over the relevant possibilities for $\jetbub(R)$, e.g.,
over the the middle column in Fig.\ \ref{fig:2gluon}.

There remain
only terms at the end of fermion lines, and those at on-shell
ends give zero.  Summing over all graphs for $\jetbub(R)$ gives
\begin{equation}
  \includegraphics[scale=0.43]{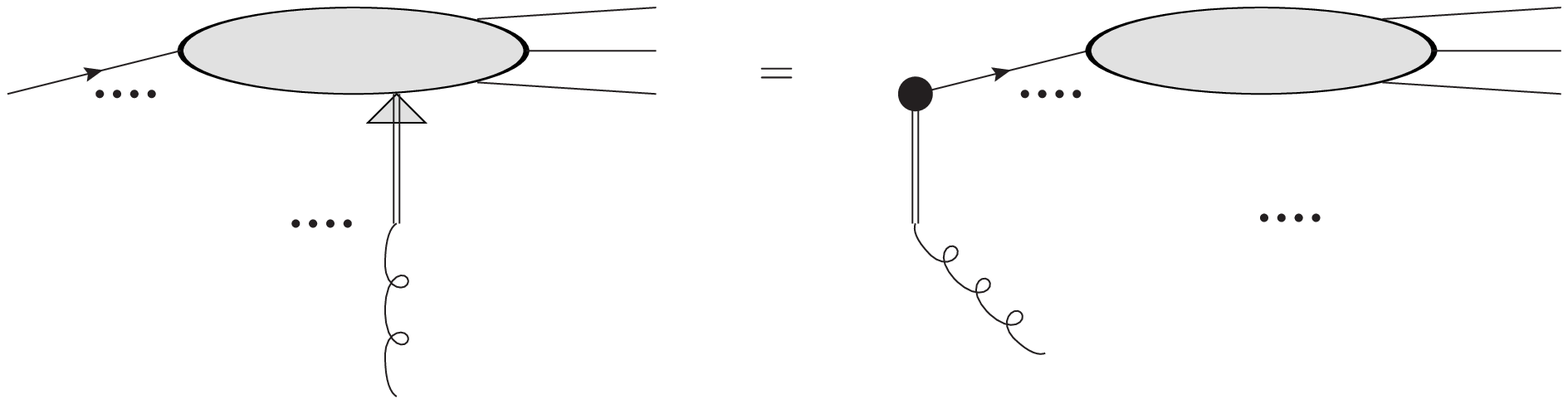}
\end{equation}
where the dots indicate arbitrarily many gluon connections, both
between $\jetbub(R)$ and the hard subgraph, and between $\jetbub(R)$ and the soft
subgraph 
$\softbub(R)$.  Applying the same argument to all the gluons from $\softbub(R)$ to
$\jetbub(R)$, and then summing over all cases for $\softbub(R)$, gives 
accumulated Wilson line components at the left which exactly
correspond to the Feynman rules for the $n_{\rm J}$ Wilson line in the
definition of the soft PCF.  

But for each term in the sum over $\jetbub(R)$, there is a sum over
subregions $R'<R$.  For the cancellation (\ref{eq:WI-cancel}) to work in
the presence of subtractions, we make a 1-to-1 correspondence between
the regions $R$ and $R'$ appropriate to the two graphs in
(\ref{eq:WI-2gluon}), considered as embedded in graphs for $\jetbub(R)$. 

For example, in Eq.~(\ref{eq:WI-2gluon1}) we can consider two regions:
The larger region we call $R_a$ for the case that the other gluon
$l_1$ is inside $\jetbub(R_a)$, i.e., is collinear to the outgoing jet ---
we refer to $R_a$ as the outer region.  There can be a subregion $R'_a$
for which this gluon is labeled soft, but such that the fermions are
still labeled collinear.  The subtraction for this region involves
applying a Grammer-Yennie approximation at the end of gluon $l_1$.
But when we exchange the two gluons, so that we are instead
considering Eq.~(\ref{eq:WI-2gluon2}), the same assignment of momentum
types still corresponds to a leading region.  Thus, we have a outer
region $R_b$ and a subregion $R'_b$ for Eq.~(\ref{eq:WI-2gluon2})
analogous to $R_a$ and $R'_a$ for Eq.~(\ref{eq:WI-2gluon1}).  In both
of the cases shown in Eq.~(\ref{eq:WI-2gluon}) we have subregions
$R'_j$ (embedded in some graph for $\jetbub(R_j)$), where the fermions are
collinear and $l_1$ is soft, and exactly the same approximation is
applied at the end of the gluon $l_1$ in the second line.  Note that
the outer regions, $R_a$ and $R_b$, are not literally the
same because they apply to a different overall graphs for $\jetbub(R_j)$;
the same applies to the subregions, of course.  

Of course, we have to sum over all possibilities for leading-power
subregions.  There is only a very limited set of cases, and the same
argument applies to all the other possibilities.  Hence the
cancellation (\ref{eq:WI-cancel}) also applies in the presence of
subtractions.  This means that, the Ward identities apply both to the
unsubtracted graphs and the subtracted graphs, given our definitions
of the approximators.

\subsection{Ward identities for collinear into hard subgraphs}

The situation is somewhat more complicated with the Ward identities for
collinear gluons connecting to a hard subgraph.  

\begin{figure}
  \centering
  \includegraphics[scale=0.45]{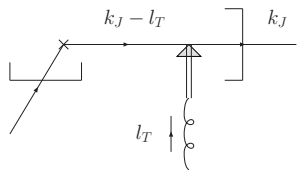}
  \caption{Lowest-order connection to hard subgraph of
    target-collinear gluon. The hooked lines on the outermost fermion
    lines indicate where fermion momenta are replaced by massless
    on-shell approximations. }
  \label{fig:T.to.H}
\end{figure}

\begin{figure}
  \centering
  \begin{tabular}{c@{\hspace*{3mm}}c}
    \includegraphics[scale=0.45]{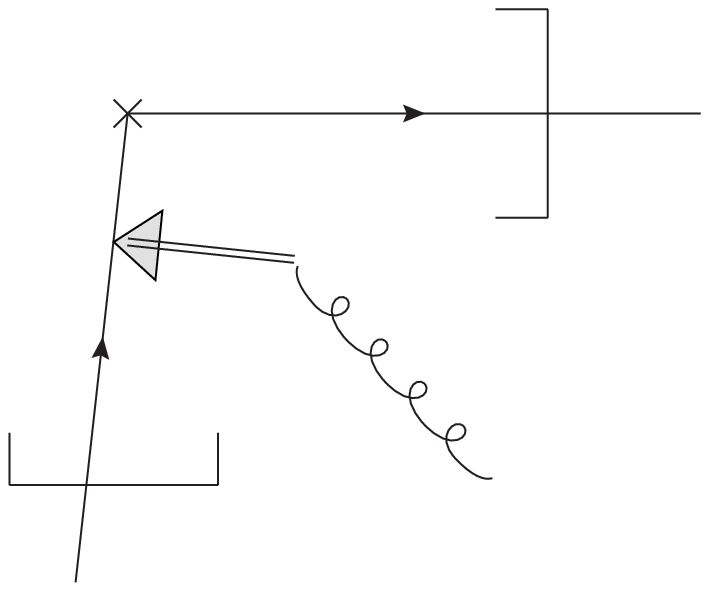}
  &    
    \includegraphics[scale=0.45]{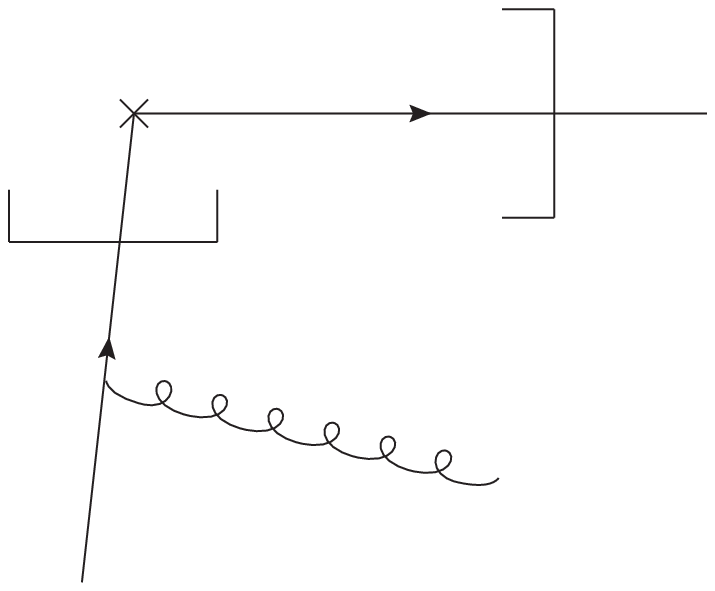}
  \\
     (a) & (b)
  \end{tabular}

  \caption{(a) This is \emph{not} an allowed 
    connection of a target-collinear gluon
    to the hard subgraph, since the gluon and the attached quark are
    both target-collinear: the intermediate quark is therefore
    collinear, not hard. (b) The corrected treatment of the graph: The
    hard subgraph is just the vertex for the electromagnetic current.}
  \label{fig:T.to.H.not}
\end{figure}

First let us examine a lowest-order connection of a target-collinear
gluon to the hard subgraph, Fig.\ \ref{fig:T.to.H}.  The vector in the
approximation is now $n_s$ instead of $n_{\rm J}$.  Of the two terms in the
elementary Ward identity (\ref{eq:WI-elem}), only the first survives.
The second term is zero because in the hard subgraph the quark $k_{\rm J}$
is replaced by a massless 
one with an
on-shell momentum, and the associated
Dirac-matrix projector ${\cal P}_{\rm T}$ is equivalent to using an on-shell
Dirac wave function.  Although the hard subgraph is like an on-shell
matrix element, one graph is missing, Fig.\ \ref{fig:T.to.H.not}(a), which
has the target-collinear gluon connecting to the target-collinear
quark.  Thus the sum over graphs --- really just one graph here ---
gives the appropriate Wilson line factor for the target PCF.  This
procedure readily generalizes to all the
 \emph{unsubtracted}
graphs.

The only subregion and hence the only subtraction for Fig.\
\ref{fig:T.to.H} is where the gluon is soft.  As can be seen from Eq.\
(\ref{eq:TS}), this has no effect on the approximator as applied at
the hard subgraph (the hard vertex), so the subtraction leaves the Ward
identity unaltered.

\begin{figure}
  \centering
  \begin{tabular}{c@{\hspace*{3mm}}c}
    \includegraphics[scale=0.43]{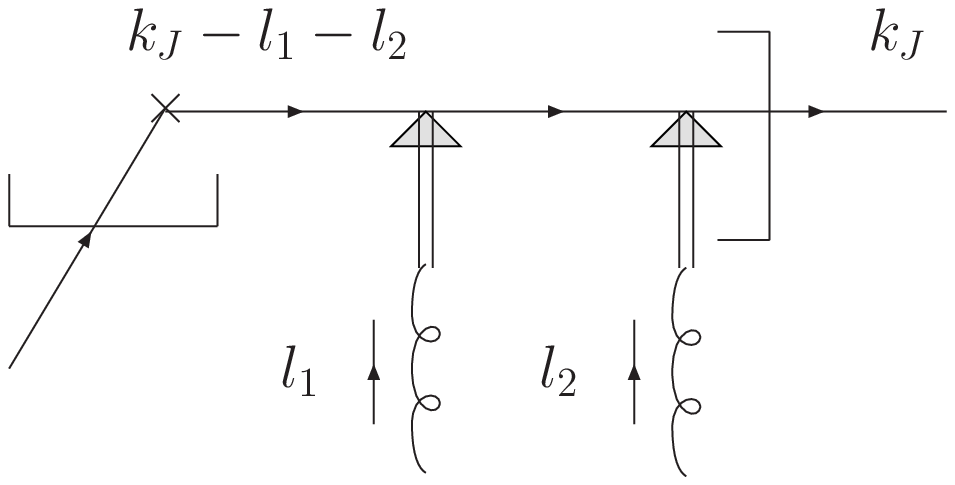}
  &
    \includegraphics[scale=0.43]{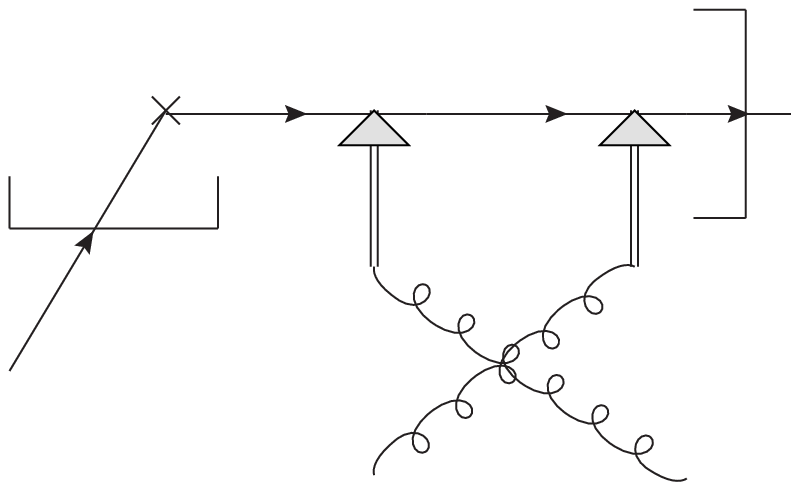}
  \\
     (a) & (b)
  \\
  \multicolumn{2}{c}{
    \includegraphics[scale=0.43]{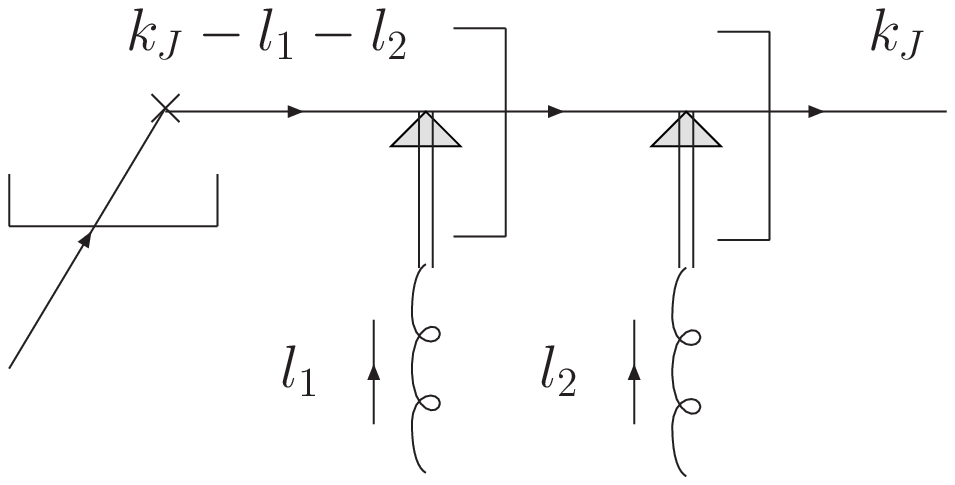}
  }
  \\
  \multicolumn{2}{c}{(c)}
  \end{tabular}
  \caption{(a) and (b) The lowest-order possibilities for two
    target-collinear gluons attaching to the hard subgraph. (c)
    Subregion with $l_1$ collinear and $l_2$ soft is allowed only for
    graph (a) and not graph (b), but in the subtraction the
    intermediate quark is on-shell.}
  \label{fig:T2.to.H}
\end{figure}

\begin{figure*}
  \centering
  \begin{equation*}
    \includegraphics[scale=0.6]{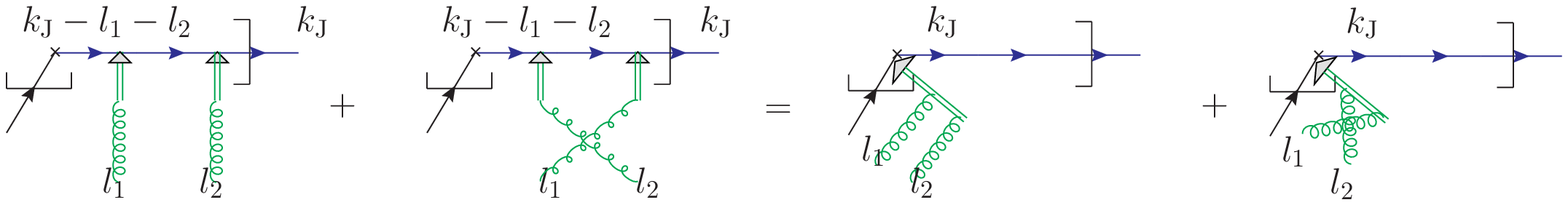}
  \tag{a}
  \end{equation*}
  \begin{equation*}
    \includegraphics[scale=0.6]{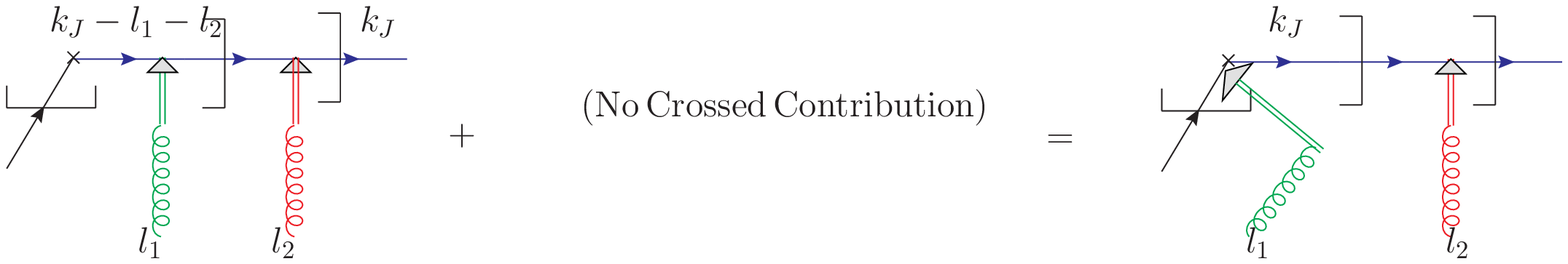}
  \tag{b}
  \end{equation*}
  \caption{(a) The application of the Ward identity when both gluons
    are collinear. (b) The application of the Ward identity when one
    of the gluons is soft.}
  \label{T2equationcons}
\end{figure*}

A complication concerning subtractions does arise because there is no
longer always a 1-to-1 correspondence between 
the subregions $R'$ used in the subtractions for different terms in
the sum (over $R$ and $\Gamma$) to which we wish to apply a Ward identity
argument, of the kind we applied to soft gluons in
Eq.~(\ref{eq:WI-2gluon1}). 
To illustrate the situation explicitly, we analyze the case 
of two gluons, labeled $l_1$ and $l_2$ attaching to the jet quark line
--- the two graphs in Fig.\ \ref{fig:T2.to.H}.
For each 
of these graphs, we will
consider two \emph{regions}.  
One is 
the region where both gluons are target collinear.  We will call
this region $R_{CC}$ (the outer region).  
The second, smaller region 
is where 
$l_1$ remains target collinear, but $l_2$ becomes soft.  We will call 
this region $R_{CS}$, and we will need to take this region into account 
in setting up subtractions.  To connect with the previous discussion, we 
note that $R_{CC}$ corresponds to the outer region, $R$, and $R_{CS}$ 
corresponds to the smaller region, $R^{\prime}$. 
(Of course, there are other subregions we could consider.  We consider 
this collection of regions for the purpose of illustration.)

Looking first at the outer region, $R_{CC}$, we can immediately 
sum over the
two possible graphs (a) and (b), giving the appropriate two-gluon
coupling to the Wilson line for the target PCF.  
This is shown graphically
in Fig.~\ref{T2equationcons}(a).
Note that in the hard
subgraph, the approximator replaces the collinear momenta by their
large components.  Thus $k_{\rm J}-l_1-l_2$ is replaced by
$(-l_1^+-l_2^+,k_{\rm J}^-,{\bf 0}_\tran)$ 
(with some possible scaling as well).
Next we consider the smaller region, $R_{CS}$, which will be 
relevant for subtractions.
We will refer to the contribution from a particular graph to $R_{CS}$ 
as the subregion for that graph.
For the subtractions obtained from subregions for graph (a) the one possibility
we consider here,
$R_{CS}$, is that gluon $l_1$ is collinear and gluon $l_2$ is soft,
and this generates a subtraction for graph (a).  But for graph (b)
this region is not leading, and there is no subtraction.  Thus the
subregions do not correspond between graphs.  (A similar situation
occurs with the graphs and momentum assignments reversed.)  
However, as far 
as
the exact application of the Ward identity is concerned, the
contribution from graph (b) is not needed in the subtraction terms.
This is because in the subtraction term for graph (a), depicted in
graph (c), the approximator $T_{R_{CS}}$ for the subregion sets the
intermediate quark line, 
that propagates between the two gluon attachments,
on-shell, i.e.,  
$k_{\rm J}-l_2$ 
is replaced by $(0,k_{\rm J}^-,{\bf 0}_\tran)$. 
(Note the extra ``hook'' that appears 
on the intermediate line connecting the two gluon attachments in graph (c).)  
In order for the Ward identity to work, 
there is, therefore, 
no need to also consider graph (b) in the subtractions.
In other words, after the application of $T_{R_{CS}}$, graph (c) is 
already in the desired form.
The application of the Ward identity in the smaller region gives 
the graphical equation in Fig.~\ref{T2equationcons}(b).

For the subregion subtraction, we also need to apply Ward identities
to the gluon $l_2$ that is labeled soft in the subregion.  Without
the approximator $T_{R_{CS}}$, we would need a cancellation with
something related to the crossed graph (b).  But since the
approximator 
discussed in the previous section 
makes the inner subgraph exactly independent of $l_2$,
this does not matter.  
It should be recalled that we specifically chose the 
definitions of our soft approximators so 
that these steps would work, i.e., so that 
Figs.~\ref{T2equationcons}(a) and~\ref{T2equationcons}(b)  
would \emph{both} be exact.

A fully detailed and explicit treatment is left to a future work.

\section{Conclusions, Outlook, and Future Work}
\label{sec:conclusion}
In this paper, we have set up a factorization framework for deep inelastic scattering 
in QCD in which the exact kinematics of the partons 
is conserved.
We have shown that this seemingly simple requirement leads to a nontrivial 
and significant conceptual shift even at the lowest, parton model level.
The requirement of exact kinematics means that one has to abandon 
standard integrated parton densities and fragmentation functions,
and instead use the parton correlation functions.

The exact treatment of kinematics imposes particularly stringent
requirements on the methods by which factorization is formulated and
derived in the presence of gluon exchanges between the different
subgraphs (collinear and hard) with different kinds of parton
momentum.  We have shown in detail how this works for one gluon
exchange.  Then we extended the results quite generally, to obtain a
factorization property with definite gauge invariant operator
definitions for the PCFs as well as for the soft factor.  These
definitions posses additional parameters (related to rapidity) which
are introduced via non-light-like Wilson lines, comparable to those in
the Collins-Soper-Sterman formalism for TMD distributions. 

So far, the factorization formula we have derived involves 
a rather trivial hard scattering coefficient.  However, having
precisely formulated the factorization for the zeroth order 
hard scattering, the structure of higher order hard scattering
coefficients can be readily determined using the subtraction formalism.
A calculation of this type is in progress.

We expect that a reasonably straightforward generalization of the
methods of Ref.~\cite{CZ} will be suitable.  Thus the hard scattering
will be obtained from on-shell partonic cross sections with
subtractions that compensate double counting with regions associated
with lower-order hard scattering.  The subtractions remove singular
contributions where some of the partons in the hard scattering become
collinear or soft.  Because of our use of exact parton kinematics, the
subtractions will be applied point-by-point in parton momentum, and
will result in hard scattering coefficients that are ordinary
integrable functions of external parton momenta.  Thus they will
correctly represent the corrections to parton probabilities
differential in parton kinematics.
In contrast, standard methods of collinear factorization
 result in hard-scattering coefficients
that contain non-trivial generalized functions, e.g., the well-known
plus distributions.  Such distributions are very singular and cannot
represent the detailed differential distribution of parton
kinematics.  They only give correct results for cross sections that
involve a broad average over parton kinematics.

There is still much work to do.  First, the evolution equations for
the PCFs need to be derived, presumably a natural generalization of
the the CSS equations.  Then we need to obtain methods for
higher-order corrections to the hard scattering matrix elements.  We
need to extend the derivations to a non-Abelian gauge theory.  In
addition, we need to determine if and how this formalism can be recast
in terms of PCFs defined with light-like Wilson lines, but with
factors that cancel light-cone divergences, as in \cite{C03}.

\section*{Acknowledgments}
We would like to thank Andreas Metz for discussions.  
We thank Markus Diehl for proof reading.
Feynman diagrams 
where made using JaxoDraw~\cite{Binosi:2003yf}.
A.M.S. 
was
supported by the Polish Committee for Scientific Research grant
No.\ KBN 1 P03B 028 28.  All the authors were supported by the
U.S. D.O.E. under grant number DE-FG02-90ER-40577.


\appendix
\section{Projection matrices}
\label{sec:projection}
The projection matrices are defined as follows:
\begin{equation}
{\cal P}_{\rm T} = \frac{1}{2} \gamma^- \gamma^+ \; ,
\qquad
{\cal P}_{\rm J} = \frac{1}{2} \gamma^+ \gamma^- \; , 
\end{equation}
where
\begin{equation}
\gamma^-=\frac{1}{\sqrt{2}}(\gamma^0 -\gamma^3),
\qquad
\gamma^+=\frac{1}{\sqrt{2}}(\gamma^0+\gamma^3) \; .
\end{equation}

The projectors satisfy the following relations
\begin{align}
{\cal P}_{\rm T}+{\cal P}_{\rm J} & =  \one \; ,\\
 {\cal P}_{\rm J} \, \gamma^- = \gamma^+ \, {\cal P}_{\rm J} & = 0 \; ,\\
{\cal P}_{\rm T} \, \gamma^+= \gamma^- \, {\cal P}_{\rm T} & = 0 \;, \\
\overline{{\cal P}}_{\rm J} = \gamma^0 {\cal P}_{\rm J}^{\dagger} \gamma^0 & = {\cal P}_{\rm T} \; ,\\
{\cal P}_{\rm T} \gamma^{-} {\cal P }_{\rm J} &= \gamma^- ,\\
{\cal P}_{\rm J} \gamma^{+} {\cal P }_{\rm T} &= \gamma^+ \\
{\cal P}_{\rm T}^2 & = {\cal P}_{\rm T} \; ,\\
{\cal P}_{\rm J}^2 & = {\cal P}_{\rm J} \; .
\end{align}

\begin{widetext}
\section{The elementary Ward identity}
\label{sec:WI}
We have made frequent use of Ward identities 
to disentangle soft and collinear gluons from 
the soft and collinear bubbles such as those 
in Fig.~\ref{LOdiags}(c).  In this appendix we
review the derivation of the elementary Ward identity
for an Abelian gauge theory.  
We treat the case where
a soft gluon attaches to the outgoing jet
bubble -- the upper bubble in Fig.~\ref{soft}.
Analogous results hold for the other situations we have considered that
require Ward identities (such 
as when the gluon attaches at the target bubble)
and follow from
similar arguments.

Consider the contribution to the outgoing jet bubble represented by the 
following sum of graphs:
\begin{equation}
  \label{summed_graphs}
  \includegraphics[scale=0.6]{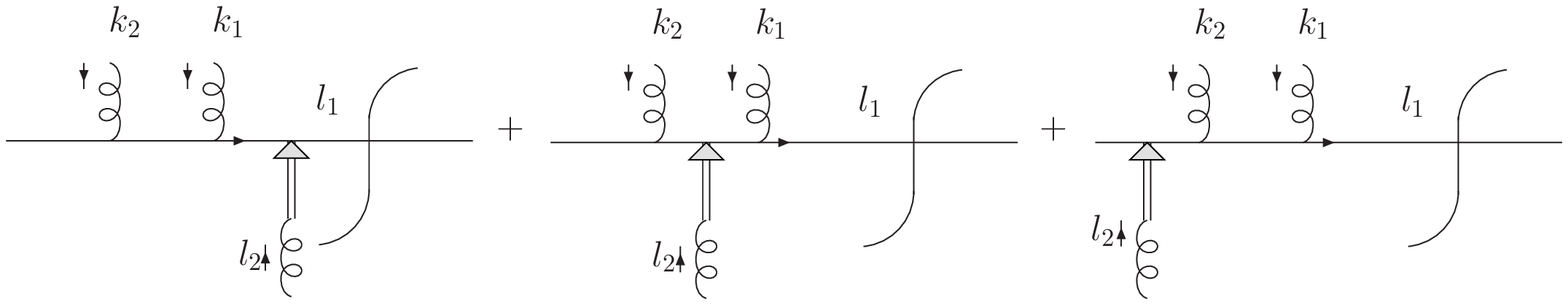}
\end{equation}
Here the outgoing quark attaches to two additional gluons as 
it enters the final state on the left side of the final state cut.  
(The other ends 
of these extra gluons may attach anywhere else in the 
jet bubble).  In this example,
we assume that the final state quark is on shell.
The soft gluon insertion is represented by the attachment with 
a double line and arrow, indicating that we have made the 
approximations discussed in the main text that lead to 
Eq.~(\ref{softapprox}).  In particular, the soft momentum, $l_2$ 
has been contracted with the soft gluon vertex, and there is a
division by $l_2 \cdot n_J$.
The sum of graphs in (\ref{summed_graphs}) contributes the 
following factor to the full graph:
\begin{multline}
\label{eq:sumgraphs}
\frac{ig^3}{l_2 \cdot n_J} \bar{u}(l_1) \slashed{l}_2 \left( \frac{i}{\slashed{l}_1 - \slashed{l}_2 - m} \right) \gamma_{\rho_1} 
\left(  \frac{i}{\slashed{l}_1 - \slashed{l}_2 - \slashed{k}_1 - m} \right) \gamma_{\rho_2} \left(  \frac{i}{\slashed{l}_1 - \slashed{l}_2 - \slashed{k}_1 - \slashed{k}_2 - m} \right) + \\ +
\frac{ig^3}{l_2 \cdot n_J} \bar{u}(l_1) \gamma_{\rho_1} \left( \frac{i}{\slashed{l}_1 - \slashed{k}_1 - m} \right) \slashed{l}_2
\left(  \frac{i}{\slashed{l}_1 - \slashed{l}_2 - \slashed{k}_1 - m} \right) \gamma_{\rho_2} \left(  \frac{i}{\slashed{l}_1 - \slashed{l}_2 - \slashed{k}_1 - \slashed{k}_2 - m} \right) + \\ +
\frac{ig^3}{l_2 \cdot n_J} \bar{u}(l_1) \gamma_{\rho_1} \left( \frac{i}{\slashed{l}_1 - \slashed{k}_1 - m} \right) \gamma_{\rho_2}
\left(  \frac{i}{\slashed{l}_1 - \slashed{k}_1 - \slashed{k}_2 - m} \right) \slashed{l}_2 \left(  \frac{i}{\slashed{l}_1 - \slashed{l}_2 - \slashed{k}_1 - \slashed{k}_2 - m} \right).
\end{multline}
Let us call the first term $A$, the second term $B$ and the third term $C$.
Now we notice that we can 
eliminate quark propagators and 
utilize the Dirac equation by substituting 
the following trivial identities for $\slashed{l}_2$:
\begin{align*}
\slashed{l}_2 =& -(\slashed{l}_1 - \slashed{l}_2 - m) + (\slashed{l}_1 - m) && \text{in term A} \\
=& -(\slashed{1}_1 - \slashed{l}_2 - \slashed{k}_1 - m) + (\slashed{l}_1 - \slashed{k}_1 - m)  &&\text{in term B} \\
=& -(\slashed{l}_1 - \slashed{l}_2 - \slashed{k}_1 - \slashed{k}_2 - m) + (\slashed{l}_1 - \slashed{k}_1 - \slashed{k}_2 - m).  &&\text{in term C}
\end{align*}
These are each of the form of a difference of the denominators for the
two quark lines next to the vertex for the $l_2$ gluon, so that
\begin{eqnarray}
A & = & \frac{g^3}{l_2 \cdot n_J} \bar{u}(l_1) \gamma_{\rho_1}  
\left(  \frac{i}{\slashed{l}_1 - \slashed{l}_2 - \slashed{k}_1 - m} \right) 
\gamma_{\rho_2} \left(  \frac{i}{\slashed{l}_1 - \slashed{l}_2 - \slashed{k}_1 - \slashed{k}_2 - m} \right), \\
B & =& 
- \frac{g^3}{l_2 \cdot n_J} \bar{u}(l_1) \gamma_{\rho_1}  \left(  \frac{i}{\slashed{l}_1 - \slashed{l}_2 - \slashed{k}_1 - m} \right)
\gamma_{\rho_2} \left(  \frac{i}{\slashed{l}_1 - \slashed{l}_2 - \slashed{k}_1
    - \slashed{k}_2 - m} \right)  + \nonumber\\
& \, & 
+ \frac{g^3}{l_2 \cdot n_J} \bar{u}(l_1) \gamma_{\rho_1} \left( \frac{i}{\slashed{l}_1 - \slashed{k}_1 - m} \right) 
\gamma_{\rho_2} \left(  \frac{i}{\slashed{l}_1 - \slashed{l}_2 - \slashed{k}_1 - \slashed{k}_2 - m} \right), \\ 
C & = & 
-\frac{g^3}{l_2 \cdot n_J} \bar{u}(l_1) \gamma_{\rho_1} 
\left( \frac{i}{\slashed{l}_1 - \slashed{k}_1 - m} \right) \gamma_{\rho_2}
\left(  \frac{i}{\slashed{l}_1 - \slashed{l}_2 - \slashed{k}_1 -
    \slashed{k}_2 - m} \right) + \nonumber\\
& \, & 
+\frac{g^3}{l_2 \cdot n_J} \bar{u}(l_1) \gamma_{\rho_1} 
\left( \frac{i}{\slashed{l}_1 - \slashed{k}_1 - m} \right) \gamma_{\rho_2}
\left(  \frac{i}{\slashed{l}_1 - \slashed{k}_1 - \slashed{k}_2 - m} \right). 
\end{eqnarray}
In the sum, all the intermediate terms cancel, to leave only the last term,
\begin{equation}
\label{eq:cancel}
A + B + C = \frac{g^3}{l_2 \cdot n_J} \bar{u}(l_1) \gamma_{\rho_1} \left( \frac{i}{\slashed{l}_1 - \slashed{k}_1 - m} \right) \gamma_{\rho_2}
\left(  \frac{i}{\slashed{l}_1 - \slashed{k}_1 - \slashed{k}_2 - m} \right).
\end{equation}
The soft gluon $l_2$ 
has been factored out of the rest of the graph leaving
only an over-all factor corresponding to the eikonal line 
propagator and eikonal vertex.  (See Ref.~\cite{CSSreview} for a review of Feynman rules involving
eikonal lines.)  

Exactly the same pattern applies no matter how many gluons attach to
the quark line.  This results in the general identity for $N_J$ extra
gluons shown graphically in Fig.~\ref{Appb_WI}.

As desired, the sum of graphs with a soft gluon 
insertion is replaced by the graph with no soft gluon, and an 
over-all factor giving the expected eikonal line.
\begin{figure}
  \centering
  \includegraphics[scale=0.6]{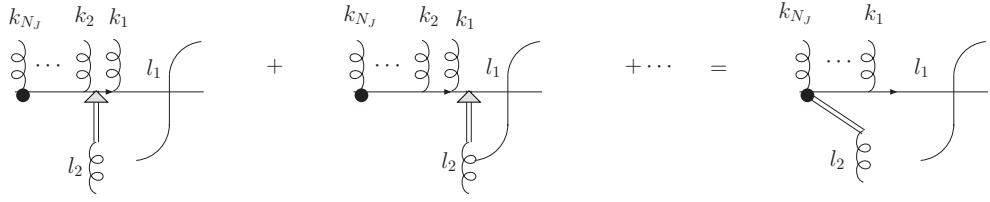}
  \caption{Applying the Ward identity to 
factorize a single soft gluon insertion.}
  \label{Appb_WI}
\end{figure}

If we repeat the above argument for multiple soft gluon insertions,
we obtain the graph shown in Fig.~\ref{Appb_WI_multa}.  
Each eikonal line gives a
propagator factor, $1/(h_j \cdot n_J)$ for $j$ running from $1$ to $N_s$
(From here on, the variables $\{h_1,h_2,\dots,h_{N_s} \}$ 
will denote the collection of soft gluon momenta). 
Figure~\ref{Appb_WI_multa} can be re-written in a way that corresponds more 
directly with the Feynman rules of Wilson lines if we note that
the product of eikonal propagators can be written as,
\begin{equation}
\label{eq:appb_sum}
\prod_j^{N_s} \frac{1}{n_J \cdot h_{N_s}} = \sum_{ \left\{ 1,2,\dots,N_s \right\} } 
\frac{1}{\left[n_J \cdot h_1 \right] \left[ n_J \cdot (h_1 + h_2) \right] 
\times \cdots \times \left[ n_J \cdot (h_1 + h_2 + \cdots + h_{N_s}) \right] }.
\end{equation}
The summation sign means that we sum over all permutations 
of the momenta $\{h_1,h_2,\dots,h_{N_s}\}$ in the denominator 
on the right side.  
That is, Fig.~\ref{Appb_WI_multa} is equivalent to summing all
possible ways of attaching the soft gluons to a single eikonal line.
To illustrate, we show this graphically in Fig.~\ref{Appb_WI_multb} for 
the simple case of two soft gluons.
\begin{figure}
  \centering
  \includegraphics[scale=0.45]{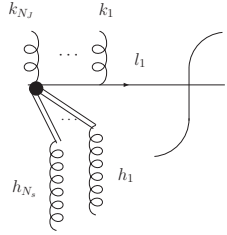}
  \caption{Result of applying the Ward identity with multiple soft gluon 
insertions}
  \label{Appb_WI_multa}
\end{figure}
\begin{figure}
  \centering
  \includegraphics[scale=0.6]{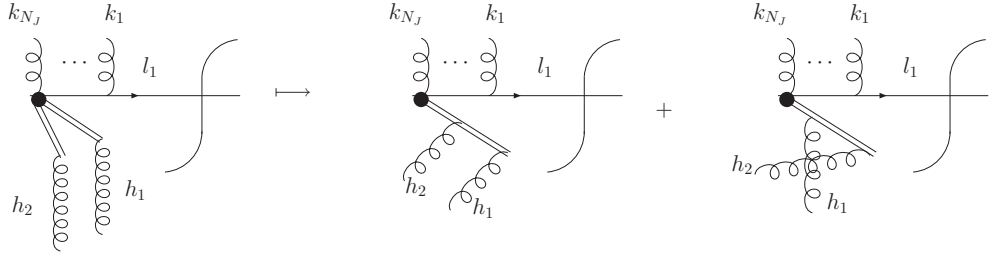}
  \caption{Using Eq.~(\ref{eq:appb_sum}) to rewrite Fig.~\ref{Appb_WI_multa}
for the case of two soft gluon insertions.}
  \label{Appb_WI_multb}
\end{figure}
\end{widetext}

\end{document}